\documentclass[a4paper,12pt]{article}

\usepackage[english]{babel}		
\usepackage{verbatim}			
\usepackage{layout}				
\usepackage{cite}				
\usepackage{amsmath}				
\usepackage{amssymb}
\usepackage{graphicx}			
\usepackage{subcaption}			

\graphicspath{{./jpg/}}		

\begin{document}

	\title{Momentum distribution functions and pair correlation functions of unpolarized uniform electron gas in warm dense matter regime}
	\author{A.S. Larkin \and V.S. Filinov \and P.R. Levashov}
	\date{28 February, 2022}
	\maketitle
	
	\pagenumbering{arabic}		
	\begin{abstract}
		In this paper we continued our research of the uniform electron gas, using the single--momentum path integral Monte Carlo method, and studied the momentum distribution functions and the pair distribution functions in the warm dense matter regime.
We discovered that the single--particle momentum distribution function deviates from the Fermi distribution and forms so-called ``quantum tails'' at high momenta, if non-ideality is strong enough in both degenerate and non-degenerate cases.
This effect is always followed by the appearance of the short--range order on the pair distribution functions and can be explained via the tunneling through the effective potential wells surrounding the electrons.
Also we calculated the average kinetic and potential energies in the wide range of states, expanding our previous results significantly.  
	\end{abstract}
	

\section{Introduction}

Over the past decade, the interest in warm dense matter (WDM) has been steadily growing.
The WDM regime can be characterized by high values of density and temperature, exceeding by several order of magnitude those in solids, so the WDM is usually non-ideal and degenerate. 
The study of the WDM is required in many astrophysical applications, such as planet interiors \cite{Knudson2012,Nettelmann2012,Mazevet2019}, brown and white dwarfs \cite{Hubbard1997,Chabrier2000}, and compact stars \cite{Haensel2007,Sharma2015}.
Also the WDM can be realized in experiments on inertial confinement fusion \cite{InertialFusion2014,InertialFusion2015} and interaction of the intense laser beams with dense plasma \cite{Roozehdar2019,Edwards2019}.

To understand this exotic state of matter, the study of it's structure and thermodynamic properties is required.
One of the most important structural quantities are the pair distribution functions (PDFs), characterizing the average spatial distribution of the particles \cite{EbelingBook2017}. 
The analysis of the PDFs gives the information about the ordering of particles and hence allows to understand the state of the system: gas--like, liquid--like or crystalline.
Besides that, the PDFs can be used for calculation of the average potential energy, which is the important quantity in thermodynamics.
The another important thermodynamic quantities are the momentum distribution functions (MDFs).
Characterizing the probability density to have certain momenta values for the particles, they are essential in calculations of many other thermodynamic quantities such as internal energy,  scattering processes, the rate constants of chemical and nuclear reactions and their thresholds \cite{Savchenko2001,Salpeter1969,Ichimaru1993,Dewitt1999}.

Due to the strong non-ideality and degeneracy, the studying of the WDM is very challenging.
The most powerful and developed analytical methods are usually based on the perturbative expansions or other asymptotic approaches.
In this way, the perturbative approximations of the MDFs for the Coulomb system leads to the ``quantum tail" exceeding the Maxwell distribution in weakly non-ideal and non-degenerate case \cite{STAROSTIN2002287,STAROSTIN1,STAROSTIN2}. 
However these analytical methods are not applicable if any appropriate small parameter does not exist, and exactly this situation is common in the case of WDM.
Therefore the numerical simulation is the most promising instrument in the studies of WDM, such as quantum Monte Carlo methods for relatively simple systems (particles in external field \cite{LarkinCPP2016,LarkinJAMP2017}, hydrogen and electron--hole plasma \cite{LarkinCPP2018,LarkinJoP2017,LarkinCPP2017,LarkinTVT2019} etc.) and the DFT--based methods for more complex and realistic materials.
However the DFT methods require the accurate knowledge about some thermodynamic quantities and structure of the electron subsystem as {\it ab initio} input data.  

One of the most actual and usable models of the electron subsystem is a well--known model of the uniform electron gas (UEG), which has also a self--consistent value of a well--known quantum analogue of one-component plasma \cite{Loos2016} and a simple model of alkali metals \cite{Mahan2000}.
It consists of electrons on a neutralizing rigid background with a uniformly distributed density of the electrical charges.
The thermodynamic state of unpolarized UEG can be described with two dimensionless parameters --- the Brueckner parameter $r_s$ and the reduced temperature $\theta$:
\begin{eqnarray}
	\label{intro_param}
	 \quad r_s = \left( \frac{3}{4\pi n a_0^3} \right)^{1/3}, \quad \theta = \frac{kT}{E_{F}},
\end{eqnarray}
where $E_{F}$ is the Fermi energy, $n$ is the electron density, $a_0$ is the Bohr radius.
Also the coupling strength $\Gamma$ and the degeneracy parameter $\chi$ are often used:
\begin{eqnarray}
	\label{intro_param1}
	\Gamma = \frac{e^2}{r_s kT}, \quad \chi = n\lambda^3,
\end{eqnarray}
where $\lambda = \sqrt{2\pi\hbar/(kT)}$ is the thermal wavelength.	

The PDFs and MDFs of weakly non-ideal case has been studied with standard perturbative methods \cite{Yasuhara1976,Kimball1975}.
In particular, the first-order approximations leads to the ``quantum tail" exceeding the Maxwell distribution in weakly non-ideal and non-degenerate case. 
However the UEG in WDM regime cannot be studied with such approaches and requires {\it ab initio} numerical simulations.
Most of the quantum Monte Carlo methods uses the coordinate representation of the partition function and do not allow to calculate the MDFs.
The recent results for the MDFs of the UEG has been obtained by using the CPIMC method \cite{Hunger2021}, based on the path integral representation in the representation of the occupation numbers.
However only the states with weak non-ideality have been considered ($r_s \lesssim 2$, $\theta \gtrsim 0.0625$), because of the poor convergence at $\Gamma \gtrsim 1$.

To overcome these difficulties, we have developed the single--momentum path integral Monte Carlo  (SMPIMC) method, based on the path integral representation of the Wigner formulation of quantum mechanics.
Dealing with the phase space, it allows one to calculate the MDFs, PDFs and other thermodynamic quantities directly from their definitions.
The method has already been applied to many quantum systems: the particle in external potential fields \cite{LarkinJAMP2017}, dense hydrogen plasma and electron--hole plasma \cite{LarkinCPP2018,LarkinTVT2019}.
In particular, we have shown that the MDFs of the plasma differs from the Fermi distribution and have the ``quantum tail'' in the strongly non-ideal and degenerate case.
Also we have calculated the average kinetic, potential and exchange--correlation energies in our recent paper \cite{LarkinPoP2021}.

In this paper we continue our research of the UEG, started in \cite{LarkinPoP2021}, and study the MDFs and the PDFs in a wide range of the parameters $0.5 \le \theta \le 4$, $0.2 \le r_s \le 36$ covering the thermodynamic states from almost ideal gas to deep WDM regime.
We have analyzed the dependence of the MDFs and the PDFs on the parameters and discovered that in the regime of strong non-ideality, when the short--range order appears and the UEG becomes liquid--like, the MDFs exceed the Fermi distribution at high momenta and have the distinct ``quantum tails''.
Also we have expanded our table from \cite{LarkinPoP2021} for the average kinetic and the potential energy on the higher values of $r_s$ up to $36$.

\section{Theoretical part}

\subsection{Uniform electron gas}
Let us consider an UEG with $N$ electrons at temperature $T$, contained in the quadratic cell with volume $V$.
We assume that the system is unpolarized, and do not take into account spin interactions, so the numbers of electrons with two different spin projections are equal and constant: $N_{u} = N_{d} = N/2$.
Also we do not taken into account any spin interactions. 
In order to reduce the finite--size effects, we apply the periodic boundary conditions (PBC).
In addition to the Coulomb interactions in the main cell, one has to take into account their interactions with all periodic images using the Ewald summation technique \cite{TOUKMAJI199673}.
As a result, the effective Hamiltonian of the UEG is as follows \cite{Dornheim2018}:
\begin{eqnarray}
	\label{eq_hamiltonian}
	\hat{H} = \sum_{a=1}^{N} \frac{\hat{\mathbf{p}}_a^2}{2m} + \sum_{a=1}^{N} \sum_{b=a+1}^{N} e^2 \Psi\left( \mathbf{q}_a, \mathbf{q}_b \right) + \frac{1}{2} e^2 N \xi_{M},
\end{eqnarray}
where $m$ and $e$ are the electronic mass and charge, $\hat{\mathbf{p}}_a$ and ${\mathbf{q}}_a$ are the momentum operator and the coordinate of the $a$-th electron. 
The periodic Ewald pair potential is
\begin{eqnarray}
	\label{eq_ewaldpair}
	\Psi\left( \mathbf{r}, \mathbf{s} \right) = \frac{1}{V\pi}\sum_{\mathbf{G} \neq 0} G^{-2} \rm{e}^{ -\frac{\pi^2 G^2}{\kappa^2} + 2\pi\rm{i}\mathbf{G}(\mathbf{r}-\mathbf{s}) } - \frac{\pi}{\kappa^2 V} + \sum_{\mathbf{R}} \frac{\rm{erfc}\left( \kappa |\mathbf{r} - \mathbf{s} + \mathbf{R}| \right)}{|\mathbf{r} - \mathbf{s} + \mathbf{R}|},
\end{eqnarray}
and the Madelung constant is
\begin{eqnarray}
	\label{eq_madelung}
	\xi_{M}
	= \frac{1}{V\pi}\sum_{\mathbf{G} \neq 0} G^{-2} \rm{e}^{ -\frac{\pi^2 G^2}{\kappa^2} } - \frac{\pi}{\kappa^2 V} + \sum_{\mathbf{R} \neq 0} \frac{\rm{erfc}\left( \kappa |\mathbf{r} - \mathbf{s}| \right)}{|\mathbf{r} - \mathbf{s}|}.
\end{eqnarray}
Here $\mathbf{R} = \mathbf{n}_d L$ and $\mathbf{G} = \mathbf{n}_r / L$, with $\mathbf{n}_d$ and $\mathbf{n}_r$ having integer components.
Note that the Ewald parameter $\kappa$ does not affect on the values and should be chosen for efficiency reasons.

The momentum distribution functions (MDFs) of the UEG describes the probability densities of the electrons to have certain values of the momenta.
Due to the kinematic and dynamic identity of the electrons in the state of thermodynamic equilibrium, the single--particle MDF is of the most interest:
\begin{eqnarray}
\label{eq_defmdf}
F(\mathbf{p})=\idotsint {\rm{d}}\mathbf{p}_2 \dots {\rm{d}}\mathbf{p}_N F(\mathbf{p}_1, \dots, \mathbf{p}_N) \Bigr|_{\mathbf{p}=\mathbf{p}_1},
\end{eqnarray}
where $F(\mathbf{p}_1, \dots, \mathbf{p}_N)$ is the full MDF, and the integration is taken over all electrons except the first one (another electron also can be chosen).

The pair distribution functions (PDFs) $g_{uu}(r)$, $g_{dd}(r)$ and $g_{ud}(r)$ describes the distributions of the distances between the electrons with the corresponding spin projections:
\begin{eqnarray}
\label{eq_defpcf}
g_{uu}(r) \propto \idotsint {\rm{d}} \mathbf{q}_{u3} \dots {\rm{d}} \mathbf{q}_{uN} {\rm{d}} \mathbf{q}_{d1} \dots {\rm{d}} \mathbf{q}_{dN} g(\mathbf{q}_{1}, \dots, \mathbf{q}_{N}) \Bigr|_{r=|q_{u1}-q_{u2}|},
\nonumber \\
g_{dd}(r) \propto \idotsint {\rm{d}} \mathbf{q}_{u1} \dots {\rm{d}} \mathbf{q}_{uN} {\rm{d}} \mathbf{q}_{d3} \dots {\rm{d}} \mathbf{q}_{dN} g(\mathbf{q}_{1}, \dots, \mathbf{q}_{N}) \Bigr|_{r=|q_{d1}-q_{d2}|},
\nonumber \\
g_{ud}(r) \propto \idotsint {\rm{d}} \mathbf{q}_{u2} \dots {\rm{d}} \mathbf{q}_{uN} {\rm{d}} \mathbf{q}_{d2} \dots {\rm{d}} \mathbf{q}_{dN} g(\mathbf{q}_{1}, \dots, \mathbf{q}_{N}) \Bigr|_{r=|q_{u1}-q_{d1}|},
\end{eqnarray}
where $g(\mathbf{q}_{1}, \dots, \mathbf{q}_{N})$ is the full spatial distribution of the electrons, $\mathbf{q}_{ua}$ and $\mathbf{q}_{da}$ specifies the electrons with different spin projections, and the exact values of the proportional factors depending of $N_{u}$, $N_{d}$ and $V$ are not required for us.

\subsection{Single momentum approach}
To obtain the expressions for the single--particle MDF and PDFs required for the numerical calculations, we use the ``single momentum approach" presented in \cite{LarkinPoP2021}.
This approach is based on the pseudoprobability density $W(p,q)$ in the phase space known as the Wigner function:
\begin{eqnarray}
	\label{eq_wigfunc}
	W(p,q) = \int {\rm{d}}^{3N} \xi \, {\rm{e}}^{\frac{\rm{i}}{\hbar}(p,\xi)} \, \langle q-\xi/2 | \hat{\rho} | q+\xi/2 \rangle,
\end{eqnarray}
where $\hat{\rho} = \beta\hat{H}$, $\beta = 1/(k_B T)$, $(p,\xi) = \sum_{a=1}^{N} \mathbf{p}_a \boldsymbol{\xi}_a$. 
The quantum states $| q \rangle = | \mathbf{q}_1, \mathbf{q}_2, \dots, \mathbf{q}_N \rangle$ and $| p \rangle = | \mathbf{p}_1, \mathbf{p}_2, \dots, \mathbf{p}_N \rangle$ are the $N$--particle states with certain coordinates and momenta, antisymmetrized according to the Fermi-Dirac statistics:
\begin{eqnarray}
	\label{eq_statesqp}
	| q \rangle = \frac{1}{N!} \sum_{P_u,P_d} (-1)^{P_u+P_d} \, | \mathbf{q}_{p_1} \rangle | \mathbf{q}_{p_2} \rangle \dots | \mathbf{q}_{p_N} \rangle,
	\nonumber \\
	| p \rangle = \frac{1}{N!} \sum_{P_u,P_d} (-1)^{P_u+P_d} \, | \mathbf{p}_{a_1} \rangle | \mathbf{p}_{a_2} \rangle \dots | \mathbf{p}_{a_N} \rangle.
\end{eqnarray}
Here the sums are taken over all permutations $P_u$ and $P_d$ of electrons with the positive and negative spin projections with the factor $+1$ for each even and $-1$ for each  odd permutation.

The Weyl symbol can be associated with each quantum operator $\hat{A}$:
\begin{eqnarray}
	\label{eq_weylsymb}
	A(p,q) = (2\pi\hbar)^{-3N} \int {\rm{d}}^{3N}\xi \, {\rm{e}}^{\frac{\rm{i}}{\hbar}(p,\xi)} \, \langle q+\xi/2 | \hat{A} | q-\xi/2 \rangle,
\end{eqnarray}
so the average value of $\hat{A}$ over the canonical ensemble can be calculated via the classical--like expression:
\begin{eqnarray}
	\label{eq_averA}
	\langle \hat{A} \rangle = \iint {\rm{d}}^{3N}p \, {\rm{d}}^{3N}q \, A(p,q) \, W(p,q).
\end{eqnarray}

The full momentum and coordinate distribution functions can be obtained from the Wigner function via integration over coordinates and momenta respectively \cite{Tatarski1983}:
\begin{eqnarray}
	\label{eq_distrpq}
	F(\mathbf{p}_1, \dots, \mathbf{p}_N) = \int {\rm{d}}^{3N}q \, W(p,q),
	\quad
	g(\mathbf{q}_1, \dots, \mathbf{q}_N) = \int {\rm{d}}^{3N}p \, W(p,q).
\end{eqnarray}
The further integration over $\mathbf{p}_2, \dots, \mathbf{p}_N$ leads to the product of delta-functions $\prod_{a=2}^{N} \delta^{(3)}(\boldsymbol{\xi}_a)$, so the integrals over $\boldsymbol{\xi}_a$ ($a = 2,\dots,N$) disappear.
The density matrix with off-diagonal elements for particles with numbers $b \neq 1$ replaced with zeros is known as the ``single--momentum density matrix for particle $1$" \cite{LarkinPoP2021}
\begin{eqnarray}
	\label{eq_smdensmatr}
	\rho_{SM}(q,\boldsymbol{\xi}_1) = \langle q-\xi/2 | \hat{\rho} | q+\xi/2 \rangle_{\boldsymbol{\xi}_2 = \dots = \boldsymbol{\xi}_N = 0}.
\end{eqnarray}
Integrating it over coordinates $q$, one obtains the ``single--particle $\xi$-distribution function'':
\begin{eqnarray}
	\label{eq_smdistrx}
	f(\boldsymbol{\xi}) = \int {\rm{d}}q \,\rho_{SM}(q,\boldsymbol{\xi}_1) \Bigr|_{\boldsymbol{\xi}=\boldsymbol{\xi}_1}.
\end{eqnarray}
The Fourier transform of $\rho_{SM}$ results in the single--particle MDF:
\begin{eqnarray}
	\label{eq_smmdf}
	F(\mathbf{p}) = \int {\rm{d}}\boldsymbol{\xi} \, {\rm{e}}^{\frac{\rm{i}}{\hbar}\mathbf{p} \boldsymbol{\xi}} \, f(\boldsymbol{\xi}).
\end{eqnarray}
Because the macroscopic system is isotropic, the single--particle $\xi$-distribution function and MDF depend only on scalar lengths of the vectors $\boldsymbol{\xi}$ and $\mathbf{p}$, and
the $3$--dimensional Fourier transform can be reduced to the $1$--dimensional sine transform:
\begin{eqnarray}
	\label{eq_smmdf1}
	F({p}) = \int_{0}^{\infty} {\rm{d}}{\xi} \, 4\pi\xi^2 f({\xi}) \frac{\sin\left(p\xi/\hbar\right)}{p\xi/\hbar}.
\end{eqnarray}

The expressions for PDFs contains multiple integrals over $\mathbf{p}_1, \dots, \mathbf{p}_N$ and can be transformed in the similar way.
Due to the product of the delta--functions $\prod_{a=1}^{N} \delta^{(3)}(\boldsymbol{\xi}_a)$, only diagonal elements of the density matrix remain.
The density matrix with non-diagonal elements replaced with zeros is known as the ``diagonal density matrix" \cite{LarkinPoP2021}:
\begin{eqnarray}
	\label{eq_dgdensmatr}
	\rho_{DG}(q) = \langle q | \hat{\rho} | q \rangle.
\end{eqnarray}
Further integration over $N-2$ coordinates gives the corresponding PDF. 
Due to the identity of the electrons with the same spin projection, one can choose any pair of them with $\sigma=+1/2$ for $g_{uu}$ or $\sigma=-1/2$ for $g_{dd}$ and the one with $\sigma=+1/2$ and the other with $\sigma=-1/2$ for $g_{ud}$.
To improve the convergence of the numerical calculations, we consider each pair of electrons and average the PDF over all pairs:
\begin{eqnarray}
\label{eq_smpcf}
g_{uu}(r) \propto \frac{2}{N_u(N_u-1)} \sum_{ua=1}^{N_u}\sum_{ub=ua+1}^{N_u} \left[ \prod_{uc \neq ua,ub} {\rm{d}}\mathbf{q}_{uc} \right] \left[ \prod_{dc} {\rm{d}}\mathbf{q}_{dc} \right] \, \rho_{DG}(q) \Bigr|_{r=|q_{ua}-q_{ub}|},
\nonumber \\ 
g_{dd}(r) \propto \frac{2}{N_d(N_d-1)} \sum_{da=1}^{N_d}\sum_{db=da+1}^{N_d} \left[ \prod_{uc} {\rm{d}}\mathbf{q}_{uc} \right] \left[ \prod_{dc \neq da,db} {\rm{d}}\mathbf{q}_{dc} \right] \, \rho_{DG}(q) \Bigr|_{r=|q_{da}-q_{db}|},
\nonumber \\
g_{ud}(r) \propto \frac{1}{N_u N_d} \sum_{ua=1}^{N_u}\sum_{db=1}^{N_d} \left[ \prod_{uc \neq ua} {\rm{d}}\mathbf{q}_{uc} \right] \left[ \prod_{dc \neq db} {\rm{d}}\mathbf{q}_{dc} \right] \, \rho_{DG}(q) \Bigr|_{r=|q_{ua}-q_{db}|},
\end{eqnarray}
where we use the same notations as in (\ref{eq_defpcf}).

\subsection{Path integrals}
For calculation of the diagonal and the single--momentum density matrices (\ref{eq_smdensmatr}), (\ref{eq_dgdensmatr}), we use the method of path integrals \cite{FeynmanHibbs}.
Let us consider the density matrix $\rho(q^A,q^B) = \langle q^B | {\rm{e}}^{-\beta\hat H} | q^A \rangle$.
Decomposing the statistical operator into the product of $M$ high--temperature operators and using $M-1$ complete sets of q--states, one can represent the density matrix in form of a multiple integral:
\begin{eqnarray}
	\label{eq_pathint2}
	\rho(q^A,q^B) =	\idotsint {\rm{d}}^{3N}q^{1} \dots {\rm{d}}^{3N}q^{M-1} \prod_{k=0}^{M-1} \langle q^{k+1} | {\rm{e}}^{-\epsilon\hat{H}} | q^{k} \rangle_{\substack{q^{0}=q^A \\ q^{M}=q^B}},	
\end{eqnarray}
where $\epsilon = \beta/M$.
If we replace all permutations in (\ref{eq_pathint2}) for with the identical one $k=0,1,\dots,M-1$ leaving only $| q^M \rangle$ being antisymmetrized, the value of the integral does not be changed \cite{Filinov2003}.
Using the symbol $| \{\mathbf{q}_a\} \rangle$ for the non--antisymmetrized $q$-state $|\mathbf{q}_1\rangle |\mathbf{q}_2\rangle\dots|\mathbf{q}_N\rangle$ and the symbol $Pq$ for the permutation, one can rewrite the expression (\ref{eq_pathint2}) as follows:
\begin{eqnarray}
	\label{eq_pimc3}
	&&\rho(q^A,q^B) =		
	\idotsint {\rm{d}}^{3N}q^{1} \dots {\rm{d}}^{3N}q^{M-1} \sum_{P_u,P_d} (-1)^{P_u+P_d} 
	\nonumber \\
	&&\times\prod_{k=0}^{M-1} \langle \{\mathbf{q}_{a}^{k+1}\} | {\rm{e}}^{-\epsilon\hat{H}} | \{\mathbf{q}_{a}^{k}\} \rangle \Bigr|_{\substack{q^{0}=q^A \\ q^{M}=Pq^B}}.	
\end{eqnarray}
The non-antisymmetrized high--temperature matrix elements are well--known \cite{LarkinPoP2021}:
\begin{eqnarray}
	\label{eq_htdensmatr}
	&\langle \{\mathbf{q}_{a}^{k+1}\} | {\rm{e}}^{-\epsilon\hat{H}} | \{\mathbf{q}_{a}^{k}\} \rangle = \lambda_{\epsilon}^{-3N} & 
	\nonumber \\ 
	&\times \exp\Biggl\{ -\frac{\pi}{\lambda_{\epsilon}^2} \sum_{a=1}^{N} (\mathbf{q}_{a}^{k+1} - \mathbf{q}_{a}^{k})^2
	 - \frac{\epsilon}{2} \left( U(q^{k+1}) + U(q^k) \right)  \Biggr\} + O(M^{-2}),&
\end{eqnarray}
where $\lambda_{\epsilon}=\sqrt{2\pi\hbar^2\epsilon/m}$ .
Thus the approximation for the density matrix with accuracy $O(M^{-1})$ is
\begin{eqnarray}
	\label{eq_pimc4}
	\rho(q^A,q^B) \approx		
	\lambda^{-3N M}\idotsint {\rm{d}}^{3N}q^{1} \dots {\rm{d}}^{3N}q^{M-1} \sum_{P_u,P_d} (-1)^{P_u+P_d} 
	\nonumber \\
	\times \exp\left\{ -\sum_{k=0}^{M-1} \sum_{a=1}^{N} \frac{m}{2}\left(\frac{\mathbf{q}_{a}^{k+1} - \mathbf{q}_{a}^{k}}{\epsilon\hbar}\right)^2 - \sum_{k=0}^{M-1} \frac{\epsilon}{2}\left( U(q^{k+1}) + U(q^k) \right) \right\} \Biggr|_{\substack{q^{0}=q^A \\ q^{M}=Pq^B}},
\end{eqnarray}
where $\lambda = \sqrt{{2\pi\hbar\beta}/{m}}$ is the thermal wavelength of electrons and $C(M)$ is a constant depending on number of high--temperature terms.
At $M \to \infty$ the multiple integral turns into path integral over all $3N$-dimensional trajectories, and the expression becomes exact: 
\begin{eqnarray}
	\label{eq_pimc5}
	\rho(q^A,q^B) =		
	\lambda^{-3N} \sum_{P_u,P_d} (-1)^{P_u+P_d} \int_{\substack{ q(0)=q^A \\ q(\beta\hbar)=P q^B}} {\rm{D}}^{3N} q(t)  
	\nonumber \\
	\times \exp\left\{ -\frac{1}{\hbar} \int_{0}^{\beta\hbar} {\rm d}t \left[ \frac{m}{2} \sum_{a=1}^{N} \dot{\mathbf{q}_a^2}(t) +  U(q(t)) \right] \right\}.	
\end{eqnarray}
(Note that for some singular attractive potentials the continuous limit must be considered more carefully \cite{Kleinert}.) 

The formula (\ref{eq_pimc5}) contains the fermionic sign problem (FSP) due the sign--alternating permutations and, thus, cannot be used in Monte Carlo simulations directly.
To avoid this problem we substitute the variables:
\begin{eqnarray}
	\label{eq_changevars}
	\mathbf{q}_a(t) = \mathbf{z}_a(t) + \left( 1-\frac{t}{\beta\hbar} \right) \mathbf{q}_a^A + \frac{t}{\beta\hbar} \mathbf{q}_{Pa}^B \quad (a=1,\dots,N),
\end{eqnarray}
obtaining the path integral with zero boundary conditions:
\begin{eqnarray}
	\label{eq_pimc6}
	\rho(q^A,q^B) =		
	\lambda^{-3N} \sum_{P_u,P_d} (-1)^{P_u+P_d} \int_{\substack{ z(0)=0 \\ z(\beta\hbar)=0}} {\rm{D}}^{3N} z(t)  
	\nonumber \\
	\times \exp\left\{ -\frac{1}{\hbar} \int_{0}^{\beta\hbar} {\rm d}t \left[ \frac{m}{2}\sum_{a=1}^{N} \dot{\mathbf{z}}_a^2(t) +  U(q(t)) \right] \right\}
	\nonumber \\
	\times \exp\left\{ -\frac{\pi}{\lambda^2} \sum_{a=1}^{N} \left[ (\mathbf{q}_{Pa}^B)^2 - (\mathbf{q}_a^A)^2 \right] \right\}.
\end{eqnarray} 
Now we assume that all permutations in the potential function can be substituted with the identical one in the WDM regime \cite{LarkinJoP2017,LarkinPoP2021}.
This simplification allows one to move all permutations into the product of the exchange determinants:
\begin{eqnarray}
	\label{eq_exdet}
	D_{\lambda}^{\sigma}(q^A,q^B) = \det \left| \exp \left\{ -\frac{\pi}{\lambda^2} \left( \mathbf{q}_{a}^B - \mathbf{q}_{b}^A \right)^2 \right\} \right|,
\end{eqnarray}
where $\sigma = u,d$ is the index of the spin projection, $a,b$ are the indices of the electrons with the corresponding $\sigma$.
Finally, the path integral representation of the density matrix is
\begin{eqnarray}
	\label{eq_pimc7}
	\rho(q^A,q^B) \approx	\lambda^{-3N} 
	D_{\lambda}^{u}(q^A,q^B) \, D_{\lambda}^{d}(q^A,q^B)
	\nonumber \\
	\times \int_{\substack{ z(0)=0 \\ z(\beta\hbar)=0}} {\rm{D}}^{3N} z(t) 
	 \exp\left\{ -\frac{1}{\hbar} \int_{0}^{\beta\hbar} {\rm d}t \left[ \frac{m}{2}\sum_{a=1}^{N} \dot{\mathbf{z}}_a(t)^2 +  U(q(t)) \right] \right\}.
\end{eqnarray} 

For the numerical applications one have to use the discrete approximation of the path integral.
Replacing each continuous trajectory $z(t)$ with the poly-line $\{ z^0, z^1, \dots, z^{M-1} \}$ having the vertices called ``beads`` and the path integral with the multiple integral, we obtain the final expression for the density matrix: 
\begin{eqnarray}
	\label{eq_pimc8}
	\rho(q^A,q^B) \approx \lambda^{-3N} 
	D_{\lambda}^{u}(q^A,q^B) \, D_{\lambda}^{d}(q^A,q^B)
	\idotsint {\rm{d}}^{3N}z^{1} \dots {\rm{d}}^{3N}z^{M-1}
	\nonumber \\	
	\times \exp\left\{ -\epsilon \sum_{k=0}^{M-1} \left[ \frac{m}{2}\sum_{a=1}^{N} \left(\frac{\mathbf{z}_{a}^{k+1}-\mathbf{z}_{a}^{k}}{\hbar\epsilon}\right)^2 +  \frac{U(q^{k+1})+U(q^k)}{2} \right] \right\} \Biggr|_{\substack{z^{0}=0 \\ z^{M}=0}},
\end{eqnarray}
where $\mathbf{q}_a^k = \mathbf{z}_a^k + \left( 1-{k}/{M} \right) \mathbf{q}_a^A + {k}/{M} \mathbf{q}_{Pa}^B$.

In the case of the diagonal density matrix $\rho_{DG}$ one should set $\mathbf{q}_a^A = \mathbf{q}_a^B = \mathbf{q}_a$ for $a=1,2,\dots,N$,  
and in the case of the single--momentum density matrix $\rho_{SM}$ one should make the same choice for $a=2,\dots,N$ and $\mathbf{q}_1^{A,B} = \mathbf{q}_1 \pm \boldsymbol{\xi}_1/2$ for $a=1$.

\section{Numerical methods}

\subsection{Basic idea of path integrals Monte Carlo methods}
Let us consider the multiple integral defying average value of the function $a(\mathbf{x})$ with a distribution function $f(\mathbf{x})$ (sign-alternating in general case):
\begin{eqnarray}
	\label{eq_mcint}
	\langle a \rangle_{f} = \frac{1}{Z} \int a(\mathbf{x}) \, f(\mathbf{x}) \, {\rm{d}}^n\mathbf{x}, \quad Z = \int f(\mathbf{x}) \, {\rm{d}}^n\mathbf{x},
\end{eqnarray}
where $Z$ is the normalization factor.
Replacing the sign-alternating distribution function with the product of the normalized absolute value $w(\mathbf{x}) = |f(\mathbf{x})|/C$ and the weight function $g(\mathbf{x}) = {\rm{sign}}f(\mathbf{x})$, one can give the probabilistic interpretation of (\ref{eq_mcint}):
\begin{eqnarray}
	\label{eq_mcint1}
	\langle a \rangle_{f} = \frac{\langle a \, g \rangle_{w}}{\langle g \rangle_{w} }, 
	\quad \text{where} \quad
    \langle b \rangle_{w} = \int b(\mathbf{x}) \, w(\mathbf{x}) \, {\rm{d}}^n\mathbf{x}.
\end{eqnarray}
Now the function $w(\mathbf{x})$ can be interpreted as the probability density for the random vector $\mathbf{x}$.  
The basic idea of the Monte Carlo is to estimate the integral via averaging the integrands over the random sample with the probability density $w(\mathbf{x})$:
\begin{equation}
	\label{eq_sample}
	\langle a \rangle_{f} \approx \frac{\sum_{i=1}^{N} a(\mathbf{x}_i) \, g(\mathbf{x}_i)}{\sum_{i=1}^{N_{mc}} g(\mathbf{x}_i)}.
\end{equation}
If the random vectors $\mathbf{x}_i$ in the sample are not correlated, the statistical error is $O(N^{-1/2})$ and can be estimated via $3\sigma$-rule according to the law of large numbers.

The sample $\{\mathbf{x}_{1},\mathbf{x}_{2},\dots,\mathbf{x}_{N_{mc}}\}$ can be created via the Metropolis algorithm \cite{Hastings1970}.
This algorithm consists of sequential steps divided into two sub-steps: the proposal and the acceptance. 
If the value of the random vector on the $i$-th step is $\mathbf{x}_i$, and the new random value $\mathbf{x}'_i$ uniformly distributed in some $n$--dimensional cube is proposed, this new state has to be accepted with the probability
\begin{eqnarray}
	\label{eq_acceptprob}
	A(\mathbf{x}_i \to \mathbf{x}'_i) = \max\left( 1, \frac{w(\mathbf{x}_{i+1})}{w(\mathbf{x}_i)} \right).
\end{eqnarray}
In case of the acceptance the state on the $i+1$ step becomes $\mathbf{x}_{i+1} = \mathbf{x}'_i$, and in case of the rejection it saves the old value $\mathbf{x}_{i+1} = \mathbf{x}_i$. 
Performing the $N_{mc}$ steps of the Metropolis algorithm gives us the required sample.

\subsection{Periodic boundary conditions}
The available computer resources do not allow to simulate the macroscopic systems, so the number of electrons in the simulation box is limited.
Since the concentration $N/V$ is constant, the volume $V$ is relatively small and the finite--size effects can significant. 
To reduce them and reproduce the properties of the macroscopic system as accurately as possible, we consider the cubic main cell with side length $L=V^{1/3}$ and periodic boundary conditions (PBC) \cite{allen1988}. 
When any bead $\mathbf{q}_{a,k}$ leaves the simulation box, it should be replaced with the periodic image, entered into the main cell instead:
\begin{eqnarray}
	\label{eq_pbcq}
	q_{a,i}^k \to q_{a,i}^k - L\left[ \frac{q_{a,i}^k}{L} + \frac{1}{2} \right] 
	\quad (i=x,y,z).   
\end{eqnarray}
Here $[x]$ means the floor integer value of $x$.
In calculations of the distances between two beads $\Delta\mathbf{q}_{ab,k}$ the nearest images should be taken:
\begin{eqnarray}
	\label{eq_pbcqd}
	\Delta{q}_{ab,k,i} \to \Delta{q}_{ab,k,i} - L\left[ \frac{\Delta{q}_{ab,k,i}}{L} + \frac{1}{2} \right] \quad (i=x,y,z).   
\end{eqnarray}
In calculations of the $\xi$-distribution one has also to ``periodize" the coordinates $\boldsymbol{\xi}_a$ before adding them to the histogram:
\begin{eqnarray}
	\label{eq_pbcxi}
	\boldsymbol{\xi}_{a,i} \to \boldsymbol{\xi}_{a,i} - L\left[ \frac{\boldsymbol{\xi}_{a,i}}{L} + \frac{1}{2} \right] \quad (i=x,y,z).   
\end{eqnarray}

\subsection{SMPIMC algorithm for pair distribution functions}
Calculation of the PDFs with the SMPIMC method is based on the definition (\ref{eq_defpcf}) and the representation (\ref{eq_pimc8}) of the diagonal density matrix $\rho_{DG}$.
Integration over the variables $\mathbf{q}_a$ and $\mathbf{z}_a^k$ is performing via the Monte Carlo approach.
Each electron in the simulation box is represented as the closed poly--line with vertices $\mathbf{q}_a^k = \mathbf{q}_a + \mathbf{z}_a^k$, so the ``center"--coordinate $\mathbf{q}_a$ describes the position of the whole $a$-th electron, and ``bead"--coordinate $\mathbf{z}_a^k$ --- the relative position of the $k$-th bead.
Note that only coordinates $\mathbf{q}_a$ are used in calculations of the PDFs directly, while the beads are ``inner" coordinates and defy the weight and the probability density of the configuration.
The detailed SMPIMC algorithm for calculation of PDFs is presented below.
\begin{enumerate}
\item{
		Set the number of the run $l=0$ and the initial state $\mathbf{x}_0$: the coordinates $\mathbf{q}_a$ are uniformly distributed in the simulation box, the coordinates $\mathbf{z}_a^k$ are equal to zero ($a=1,\dots,N$, $k=0,\dots,M$). 
	 }
\item{
		Set the number of the step $i=1$ and the first state $\mathbf{x}_1=\mathbf{x}_0$.
	 }
\item{
		Select the number of particle $a=1,\dots,N$ randomly, then select the type of the step: $\delta{q}$--step with the probability $P_q$ or $\delta{z}$--step with the probability $P_z=1-P_q$.
		If $\delta{q}$--step has been chosen, modify $\mathbf{q}_a \to \mathbf{q}_a + \Delta\mathbf{q}$ with $\Delta\mathbf{q}$ uniformly distributed in the volume $\sim L^3/N$.
		If $\delta{z}$--step has been chosen, select the number of bead $k=1,\dots,M-1$ randomly and modify $\mathbf{z}_{a}^{k} \to \mathbf{z}_{a}^{k} + \Delta\mathbf{z}$ with $\Delta\mathbf{z}$ uniformly distributed in the volume $\sim \lambda^3/(4\pi K)^{3/2}$.
		In both cases take into account the PBC.
		The resulting state have to be set as the proposed state $\mathbf{x}'_{i}$.
	 }
\item{
		Accept the proposed state $\mathbf{x}'_{i}$ with the probability $A(\mathbf{x}_i \to \mathbf{x}'_i)$ (\ref{eq_acceptprob}) or reject it.
		In case of the acception --- set $\mathbf{x}_{i+1} = \mathbf{x}'_i$,
		in case of the rejection --- set $\mathbf{x}_{i+1} = \mathbf{x}_i$.
	 }	
\item{
		Calculate the distances between each pair of electrons $r_{ab} = |\mathbf{q}_a - \mathbf{q}_b|$ and build the related histograms: $h_{uu}^n(\mathbf{x}_i)$, $h_{dd}^n(\mathbf{x}_i)$ and $h_{ud}^n(\mathbf{x}_i)$, where $n=1,\dots,N_{q}$ is the number of the cell with the length $\Delta_r$, so $n = \left[ r_{ab}/\Delta_r \right]+1$.
	 }
\item{
		Repeat the steps (3.)---(5.) for $i = 1,\dots,N_{steps}$.
	 }
\item{		
		Calculate the average histograms for the obtained sample of $N_{steps}$ states via averaging of $h_{uu}^n$, $h_{dd}^n$ and $h_{ud}^n$ with $g(\mathbf{x})$ as the weight function:
		$$
			\langle h_{s_1 s_2}^n \rangle_l = \frac{ \sum_{i=1}^{N_{steps}} h_{s_1 s_2}^n(\mathbf{x}_i) \, g(\mathbf{x}_i) } { \sum_{i=1}^{N_{steps}} g(\mathbf{x}_i) }
			\quad (s_1,s_2 = \{u,d\}).
		$$
	 }	 
\item{	
		Repeat the steps (2.)---(7.) for $l=1,2,\dots,N_{runs}$, but instead of initialization $\mathbf{x}_1 = \mathbf{x}_0$ use the last state from the previous run: $\mathbf{x}_1\bigr|_{l} = \left(\mathbf{x}_{N_{steps}}\right)\bigr|_{l-1}$.			
	 }	
\item{		
		As a result, the sample of the average histograms $\langle h_{s_1 s_2}^n \rangle_{l}$, $l=0,1,\dots,N_{runs}$ is obtained.
		Considering the $0$-th run as idle and omitting it to eliminate the influence of the initial state, calculate the resulting average histograms over the sample and the statistical errors as follows:
		$$
			\langle h_{s_1 s_2}^n \rangle = \frac{\sum_{l=1}^{N_{runs}} \langle h_{s_1 s_2}^n \rangle_{l}}{N_{runs}},
			\quad
			\sigma( h_{s_1 s_2}^n ) = \sqrt{ \frac{\sum_{l=1}^{N_{runs}} (\langle h_{s_1 s_2}^n \rangle_{l} - \langle h_{s_1 s_2}^n \rangle)}{N_{runs}} }.
		$$
	}
\item{
		To obtain the final histograms of the PDFs with the statistical errors one take into account the angle distribution and the numbers of electron pairs for different spin projections:	
		$$
			g_{s_1 s_2}^n = \frac{ \langle h_{s_1 s_2}^n \rangle }{\Delta_n C_{s_1 s_2}},
			\quad
			\sigma( g_{s1 s2}^n ) = \frac{ \sigma( h_{s_1 s_2}^n ) }{\Delta_n C_{s_1 s_2}},
		$$
		where $\Delta_n = 4 \pi \Delta_r^3 \left( n^2 - n + 1/3 \right)$, $C_{s1 s2}$ is equal to $N_u(N_u-1)/2$, $N_d(N_d-1)/2$ and $N_u N_d$ for $s_1$, $s_2$ equal to $u$,$u$, $d$,$d$ and $u$,$d$ respectively. 
	 }	 	 
\end{enumerate}

\subsection{SMPIMC algorithm for momentum distribution functions}
Calculation of the single--particle MDF is based on the sine transform (\ref{eq_smmdf1}) of the $\xi$-distribution, which is obtained from the path integral representation (\ref{eq_pimc8}) via the Monte--Carlo procedure similar to the one for PDFs.
The difference is that the $1$-th electron is represented as the open poly--line with the vertices $
\mathbf{q}_1^k = \mathbf{z}_1^k + \left( 1-{k}/{M} \right) \left( \mathbf{q}_1 + \boldsymbol{\xi}/2 \right) + \left( {k}/{M} \right) \left( \mathbf{q}_1 - \boldsymbol{\xi}/2 \right)$.
Only the differential coordinate $\mathbf{\xi}$ is participating in calculation of the single--particle MDF, while the other coordinates defy only the weight and the probability density.
The detailed SMPIMC algorithm for calculation of the single--particle MDF is presented below.
\begin{enumerate}
\item{
		Set the number of the run $l=0$ and the initial state $\mathbf{x}_0$: the coordinates $\mathbf{q}_a$ are uniformly distributed in the simulation box, while the relative coordinates $\mathbf{z}_{a}^{k}$ and the differential coordinate $\boldsymbol{\xi}$ are equal to zero. 
	 }
\item{
		Set the number of the step $i=1$ and the first state $\mathbf{x}_1=\mathbf{x}_0$.
	 }	 
\item{
		Select the type of the step: $\delta{\xi}$--step with the probability $P_{\xi}$, $\delta{q}$--step with the probability $P_q$ or $\delta{z}$--step with the probability $P_z=1-P_{\xi}-P_{q}$.
		If $\delta{\xi}$--step has been chosen, modify $\boldsymbol{\xi} \to \boldsymbol{\xi} + \Delta\boldsymbol{\xi}$ with $\Delta\boldsymbol{\xi}$ uniformly distributed in the volume $\sim \lambda^3$.
		Else --- select the number of particle $a=1,\dots,N$ randomly;
		if $\delta{q}$--step has been chosen, modify $\mathbf{q}_a \to \mathbf{q}_a + \Delta\mathbf{q}$ with $\Delta\mathbf{q}$ uniformly distributed in the volume $\sim L^3/N$;
		if $\delta{z}$--step has been chosen, select the number of ``bead" $k=1,\dots,M-1$ randomly and modify $\mathbf{z}_{a}^{k} \to \mathbf{z}_{a}^{k} + \Delta\mathbf{z}$ with $\Delta\mathbf{z}$ uniformly distributed in the volume $\sim \lambda^3/(4\pi K)^{3/2}$.
		Take into account the PBC.
		The resulting state has to be set as the proposed state $\mathbf{x}'_{i}$.
	 }
\item{
		Accept the proposed state $\mathbf{x}'_{i}$ with the probability $A(\mathbf{x}_i \to \mathbf{x}'_i)$ (\ref{eq_acceptprob}) or reject it.
		In case of acception --- set $\mathbf{x}_{i+1} = \mathbf{x}'_i$,
		in case of rejection --- set $\mathbf{x}_{i+1} = \mathbf{x}_i$.
	 }	 
\item{
		Calculate the absolute value $\xi = |\boldsymbol{\xi}|$ and build the related histogram $F_{\Omega}^n(\mathbf{x}_i)$, where $n=1,\dots,N_{\xi}$ is the number of the cell with the length $\Delta_{\xi}$, so $n = \left[ \xi/\Delta_{\xi} \right]+1$.
	 }
\item{
		Repeat the steps (3.)---(5.) for $i = 1,\dots,N_{steps}$.
	 }
\item{		
		Calculate the average histogram for the obtained sample of $N_{steps}$ states via averaging of $f_{\Omega}^n$ with $g(\mathbf{x})$ as the weight function:
		$$
			\langle f_{\Omega}^n \rangle_l = \frac{ \sum_{i=1}^{N_{steps}} f_{\Omega}^n(\mathbf{x}_i) \, g(\mathbf{x}_i) } { \sum_{i=1}^{N_{steps}} g(\mathbf{x}_i) }.
		$$
	 }	 	 
\item{
		To obtain the histogram of the $\xi$-distribution on the $l$-th run take into account the angle distribution:	
		$$
			\langle f^n \rangle_l = \frac{ \langle f_{\Omega}^n \rangle_l }{\Delta_n},
			\quad
			\Delta_n = 4 \pi \Delta_{\xi}^3 \left( n^2 - n + 1/3 \right).
		$$
	 }
\item{	
		Repeat the steps (2.)---(8.) for $l=1,2,\dots,N_{runs}$, but instead of the initialization $\mathbf{x}_1 = \mathbf{x}_0$ use the last state from the previous run: $\mathbf{x}_1\bigr|_{l} = \left(\mathbf{x}_{N_{steps}}\right)\bigr|_{l-1}$.			
	 }
\item{		
		As a result, the sample of average histograms of $\xi$-distribution $\langle f^n \rangle_{l}$, $l=0,1,\dots,N_{runs}$ is obtained.
		Considering the $0$-th run as idle and omitting it to eliminate the influence of the initial state, calculate the resulting average histogram over the sample with the statistical error as follows: 
		$$
			\langle f^n \rangle = \frac{\sum_{l=1}^{N_{runs}} \langle f^n \rangle_{l}}{N_{runs}},
			\quad
			\sigma( f^n ) = \sqrt{ \frac{\sum_{l=1}^{N_{runs}} (\langle f^n \rangle_{l} - \langle f^n \rangle)}{N_{runs}} }.
		$$
	}
\item{
		To obtain the histogram of the MDF perform the discrete sine transform of the lattice function $f(\xi_n)=\langle f^n \rangle$:
		$$
			F({p}) = \sum_{n=1}^{N_{\xi}} \Delta_{\xi} \, 4\pi\xi_n^2 f({\xi_n}) \frac{\sin\left(p\xi_n/\hbar\right)}{p\xi_n/\hbar},
		$$		 
		for the lattice values of $p$. The statistical error can be easily calculated via the similar procedure.
	 }	 				 	 	 	 	 		 
\end{enumerate}

\section{Results of the simulations}

We studied the unpolarized UEG in the states with $0.2 \le r_s \le 36$, $0.5 \le \theta \le 4$, so the coupling parameter $0.03 \lesssim \Gamma \lesssim 20$ and the degeneracy parameter $0.2 \lesssim \chi \lesssim 4$.
Therefore a wide range of states have been covered: from almost ideal to strongly non-ideal and, in parallel, from almost classical to degenerate system.

We carried out our simulations for $N=66$ electrons in the cubic cell with PBCs.
Such number of electrons has been chosen for conformity with the papers \cite{LarkinPoP2021,Dornheim2018,Dornheim2016} and is quite enough for $\theta \gtrsim 0.5$.
Also we simulated the unpolarized ideal Fermi gas (IFG) under the same conditions, i.e the UEG with ``turned--off" Coulomb interactions with the same $\theta$ and $r_s$.
Comparison of these results with the analytical Fermi distributions allows us to take into account the finite--size effects and control the related systematic errors in the normalization factors in the MDFs.
The influence of the finite--size effects on the MDFs manifests in deviation of the calculated MDF for the IFG from the analytical Fermi distribution beginning at high value of the momentum $p_{FS}$,  so the interval $p \lesssim p_{FS}$ contains the reliable data.
We have been estimated the value of $p_{FS}$ from the graphic of $F(p)$ built in the logarithmic scale  visually.
The values of $p_{FS}$ for each graphic one can find in Table~\ref{tab_1}.
This table also contains the approximate positions $r_{max}$ and $r_{min}$ of the first maxima and minima of the PDFs $g_{uu}$, $g_dd$ and $g_{ud}$.

\begin{figure}[ht]	
		\includegraphics[width=0.49\linewidth]{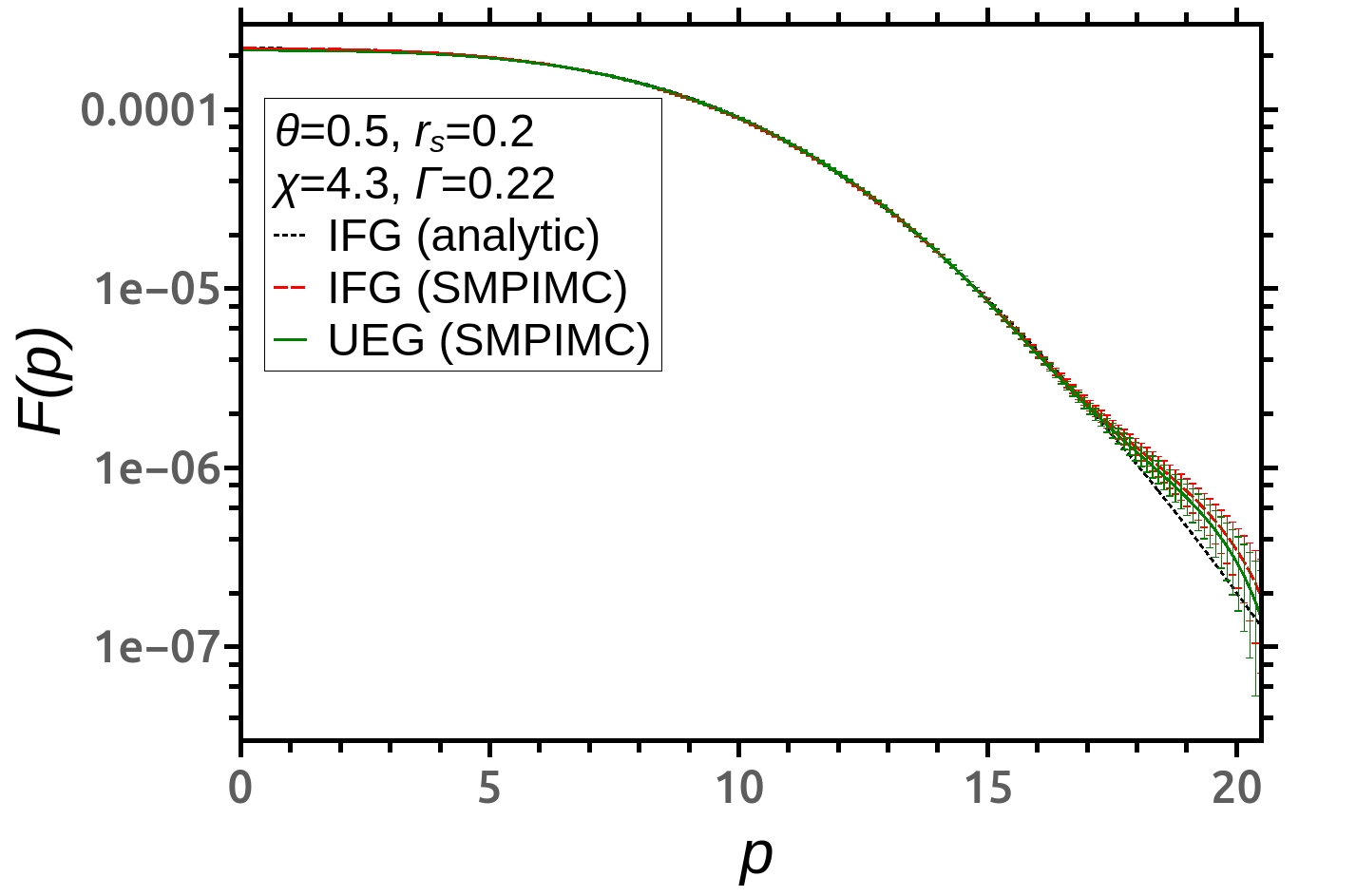}
    	\includegraphics[width=0.49\linewidth]{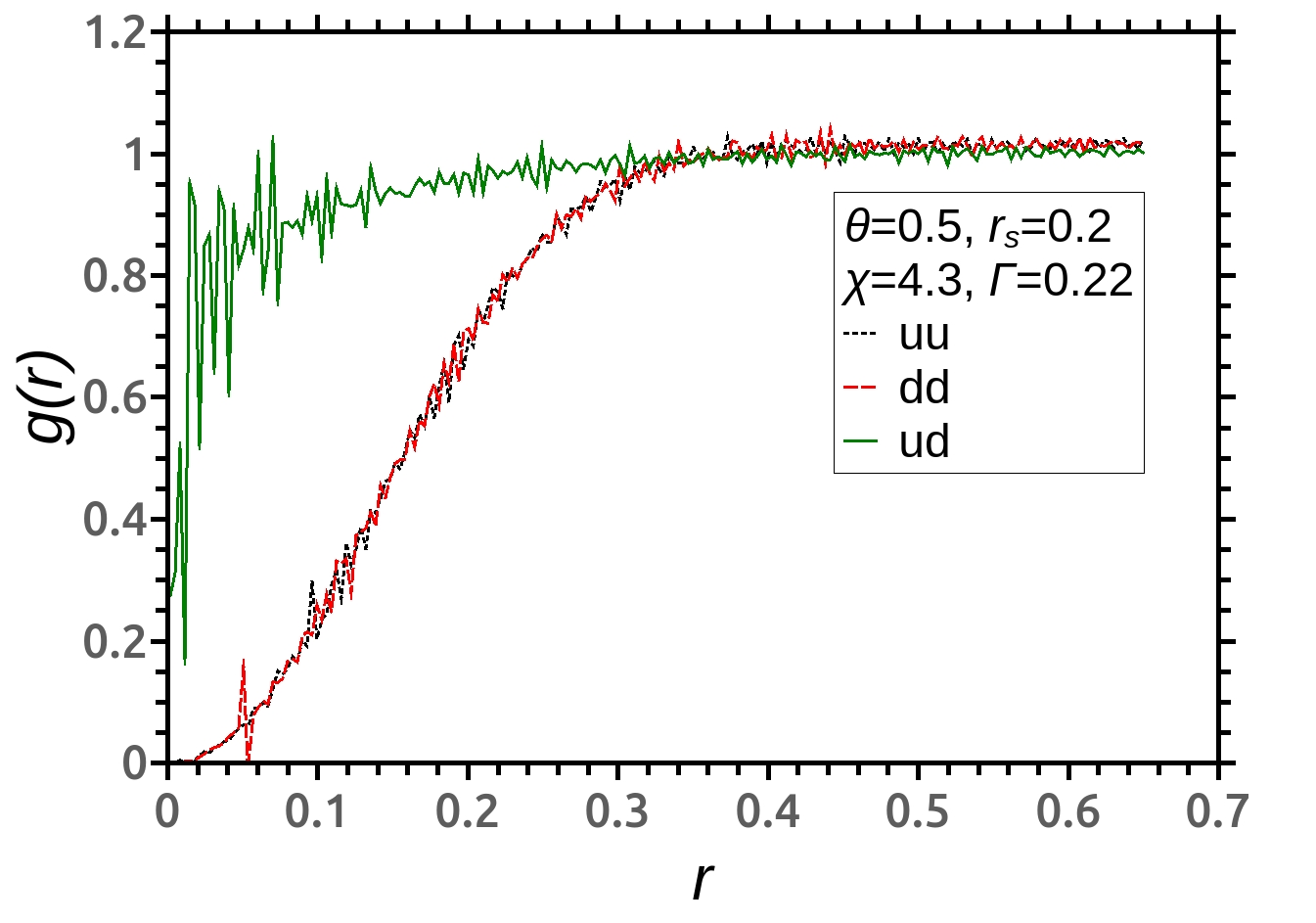}
    	\includegraphics[width=0.49\linewidth]{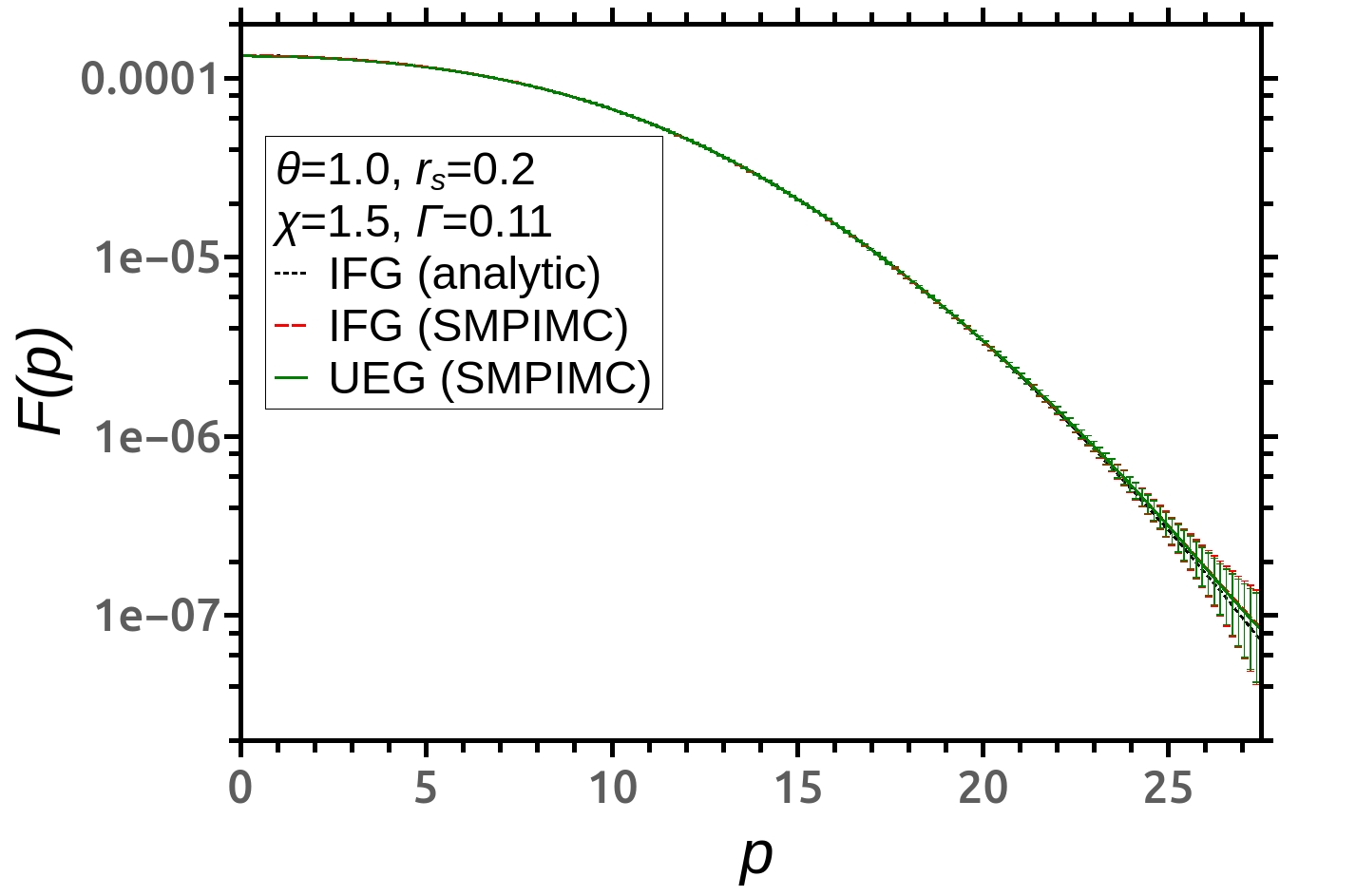}
    	\includegraphics[width=0.49\linewidth]{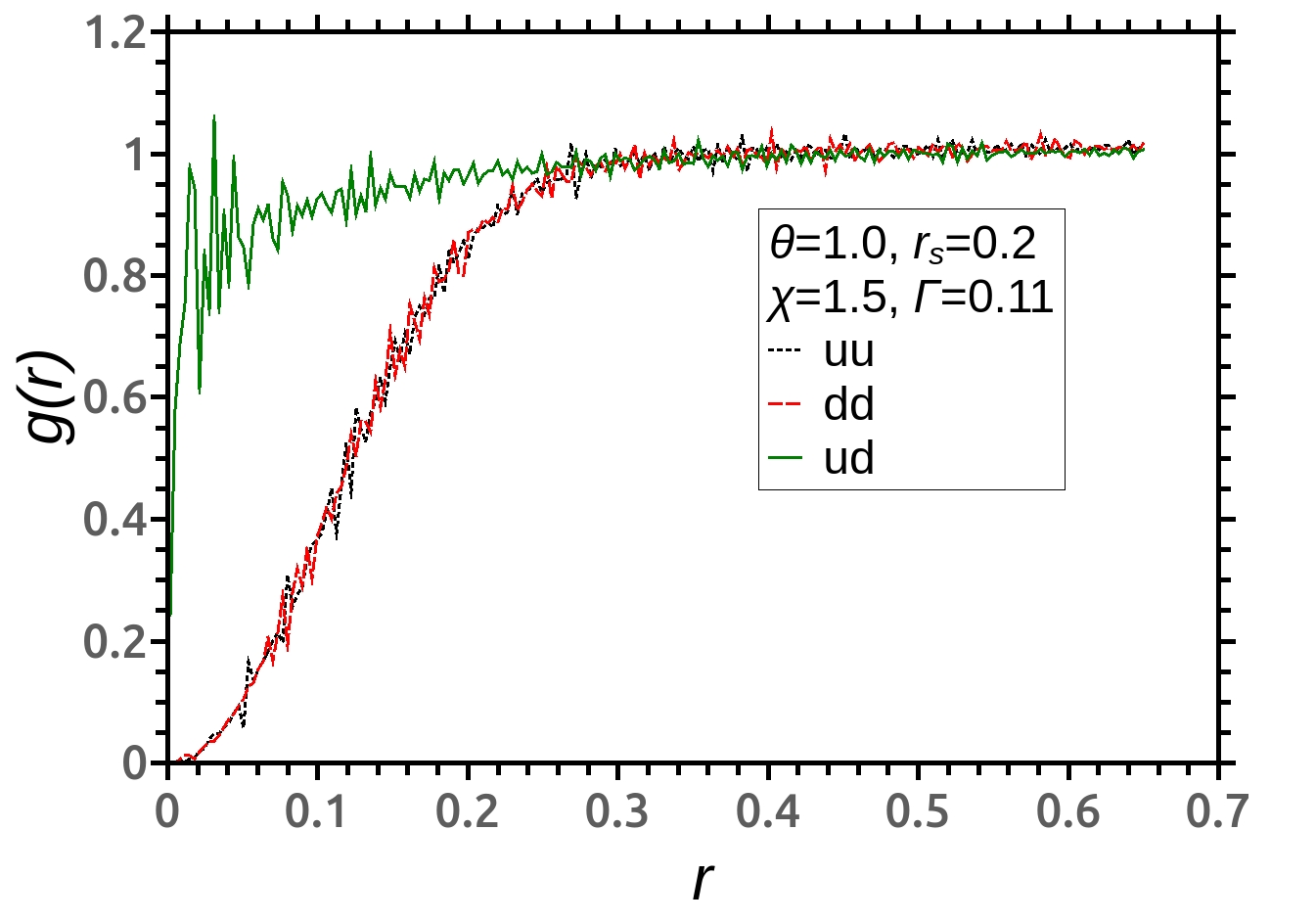}
    	\includegraphics[width=0.49\linewidth]{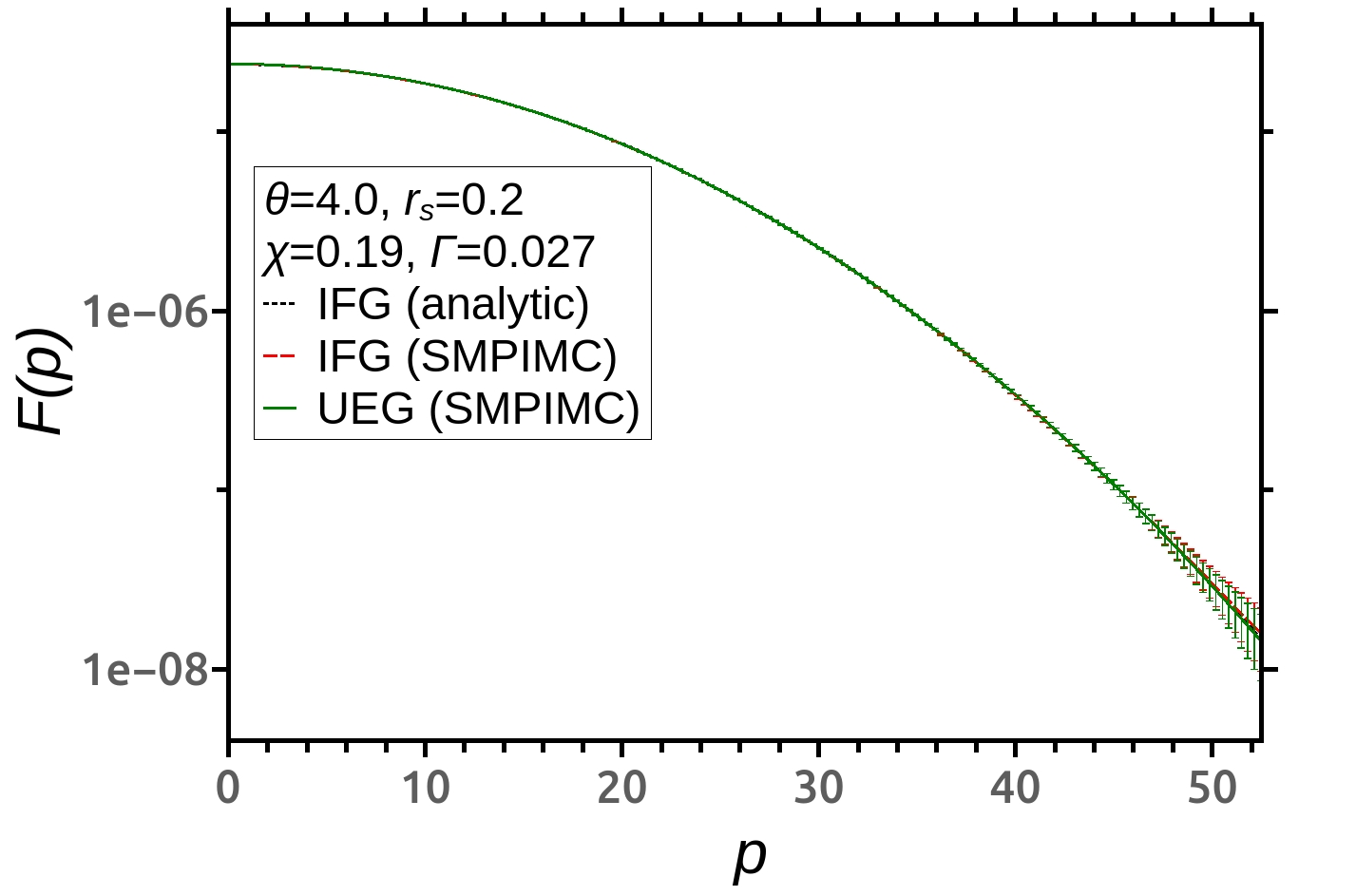}
    	\includegraphics[width=0.49\linewidth]{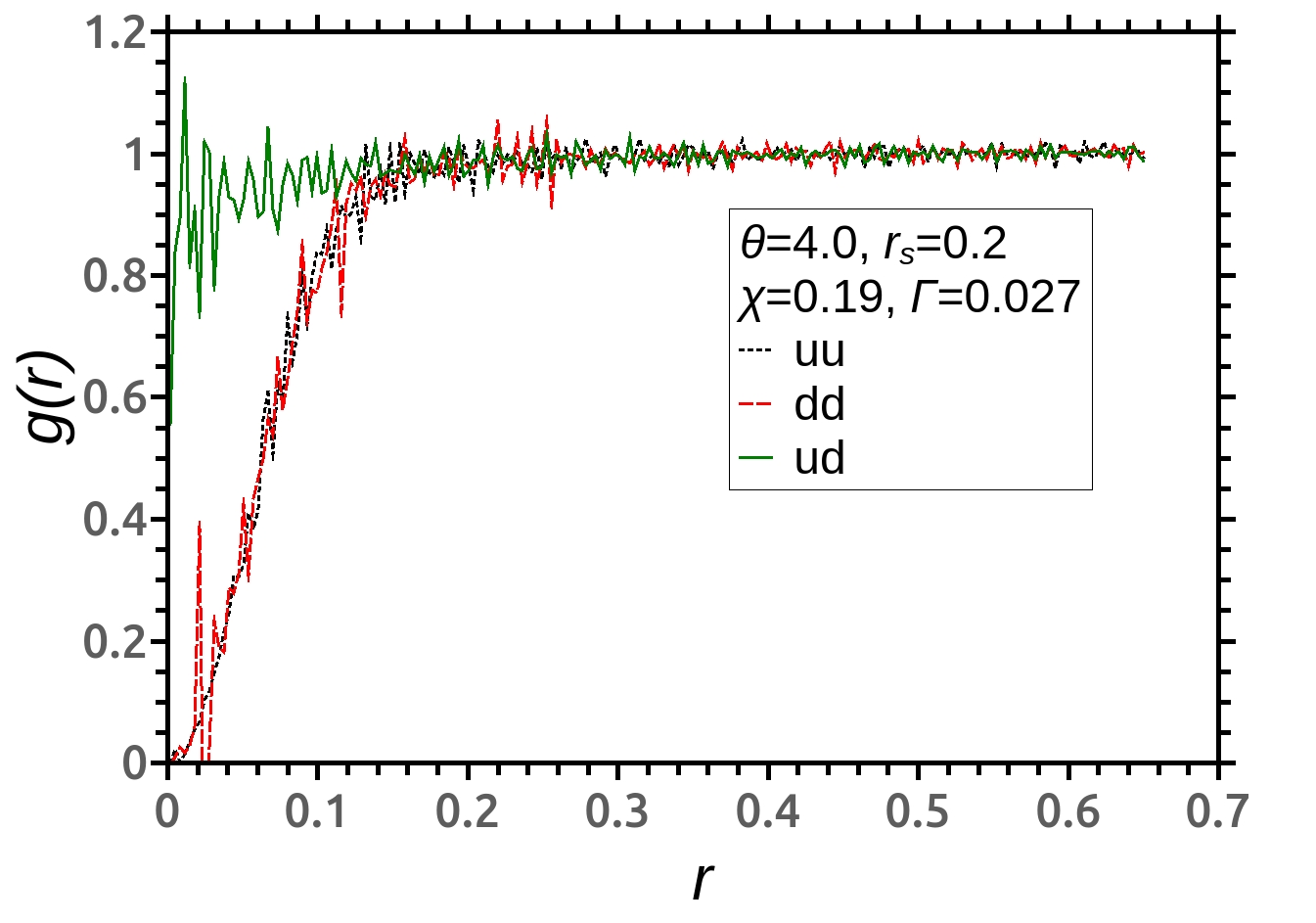}
	\caption{Left plots: single--particle momentum distribution functions of the unpolarized UEG at $r_s=0.2$ and $\theta=0.5,1,4$ calculated with the SMPIMC method.
	The results are compared with the SMPIMC results for the IFG in the same conditions and with the analytical Fermi distributions. 
	Right plots: the pair distribution functions $g_{uu}$, $g_{dd}$ and $g_{ud}$ of the UEG. }
	\label{fig_r02}
\end{figure}

The results for $r_s=0.2$ are shown in Fig.~\ref{fig_r02}.
In this case the UEG is almost ideal, and the coupling parameter $\Gamma$ varies from $0.22$ at $\theta=0.5$ to $0.027$ at $\theta=0.5$.
The finite--size effects become more significant with increasing degeneracy due to the growing relation $\lambda/L$.
For the values of the momentum $p \lesssim p_{FS}$ the MDF of the UEG coincides with the Fermi distribution. 
The PDFs of the UEG also behaves almost similarly to the PDFs of the IFG: $g_{uu}$ and $g_{dd}$ (the same spin projections) slowly rises from $0$ to $1$ with growing $r$ due to the exchange repulsion, and $g_{ud}$ (the opposite spin projections) becomes equal to $1$ rapidly due to very weak Coulomb repulsion.
The exchange repulsion rapidly decays with the decreasing degeneracy.

\begin{figure}[ht]	
		\includegraphics[width=0.49\linewidth]{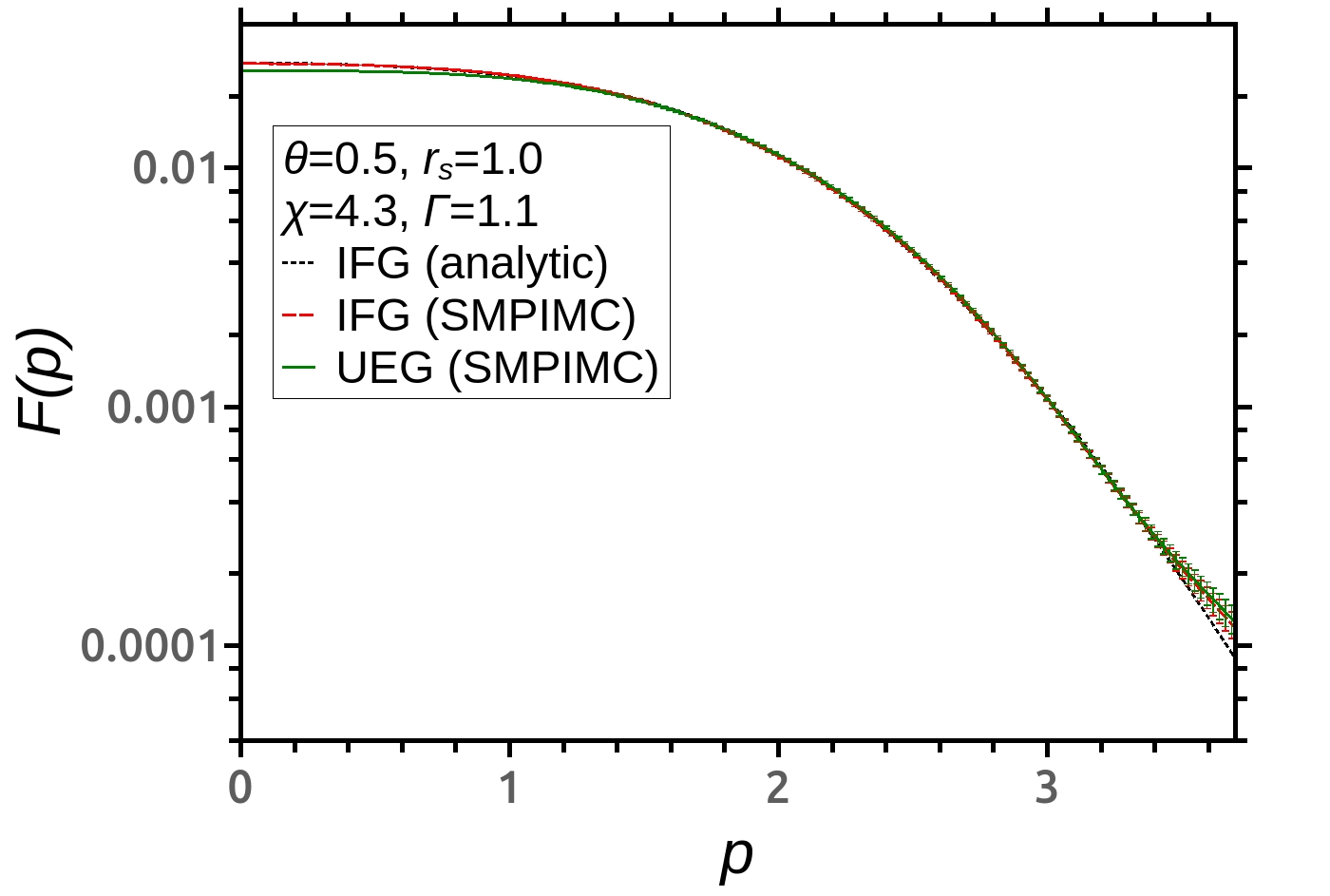}
    	\includegraphics[width=0.49\linewidth]{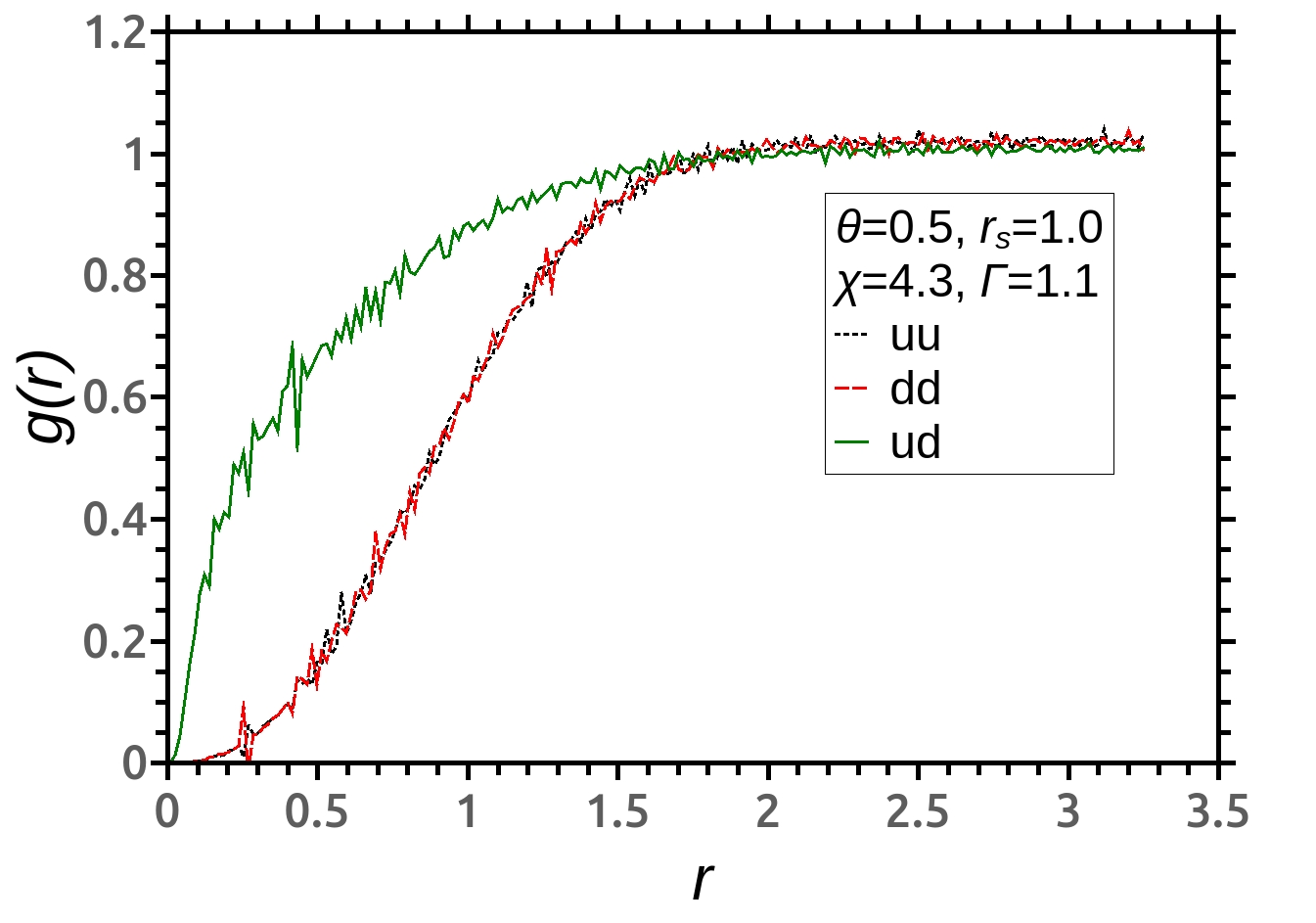}
    	\includegraphics[width=0.49\linewidth]{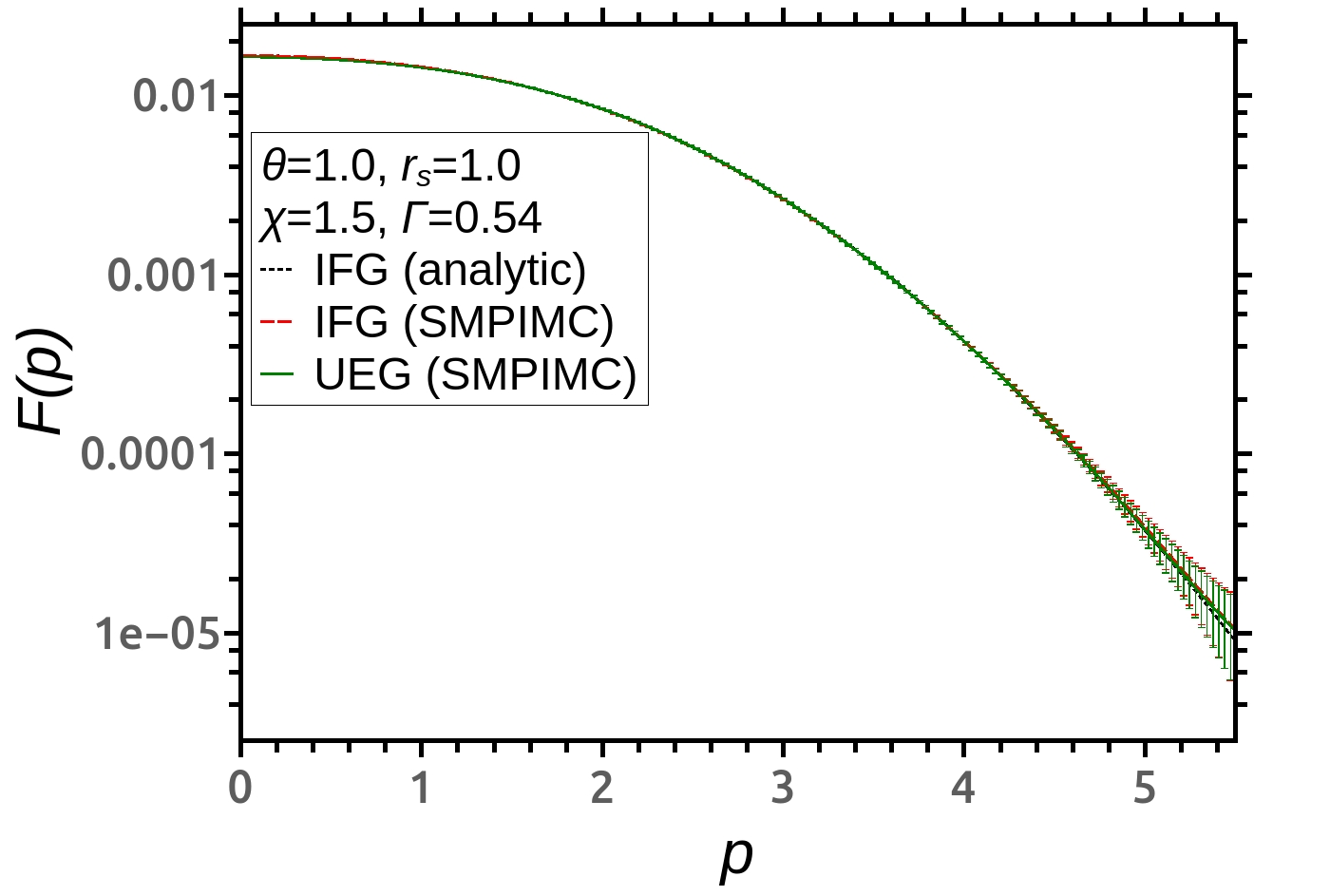}
    	\includegraphics[width=0.49\linewidth]{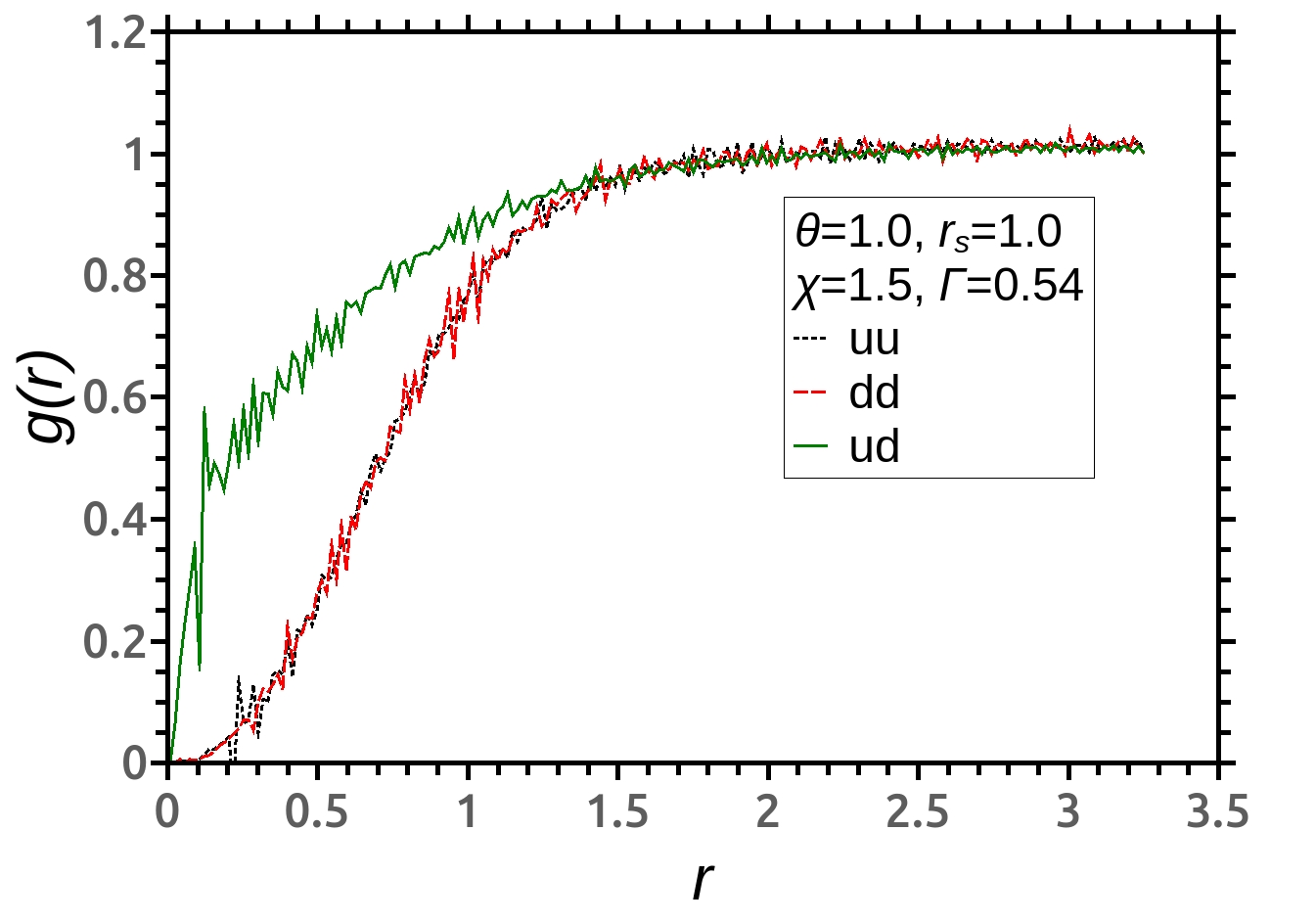}
    	\includegraphics[width=0.49\linewidth]{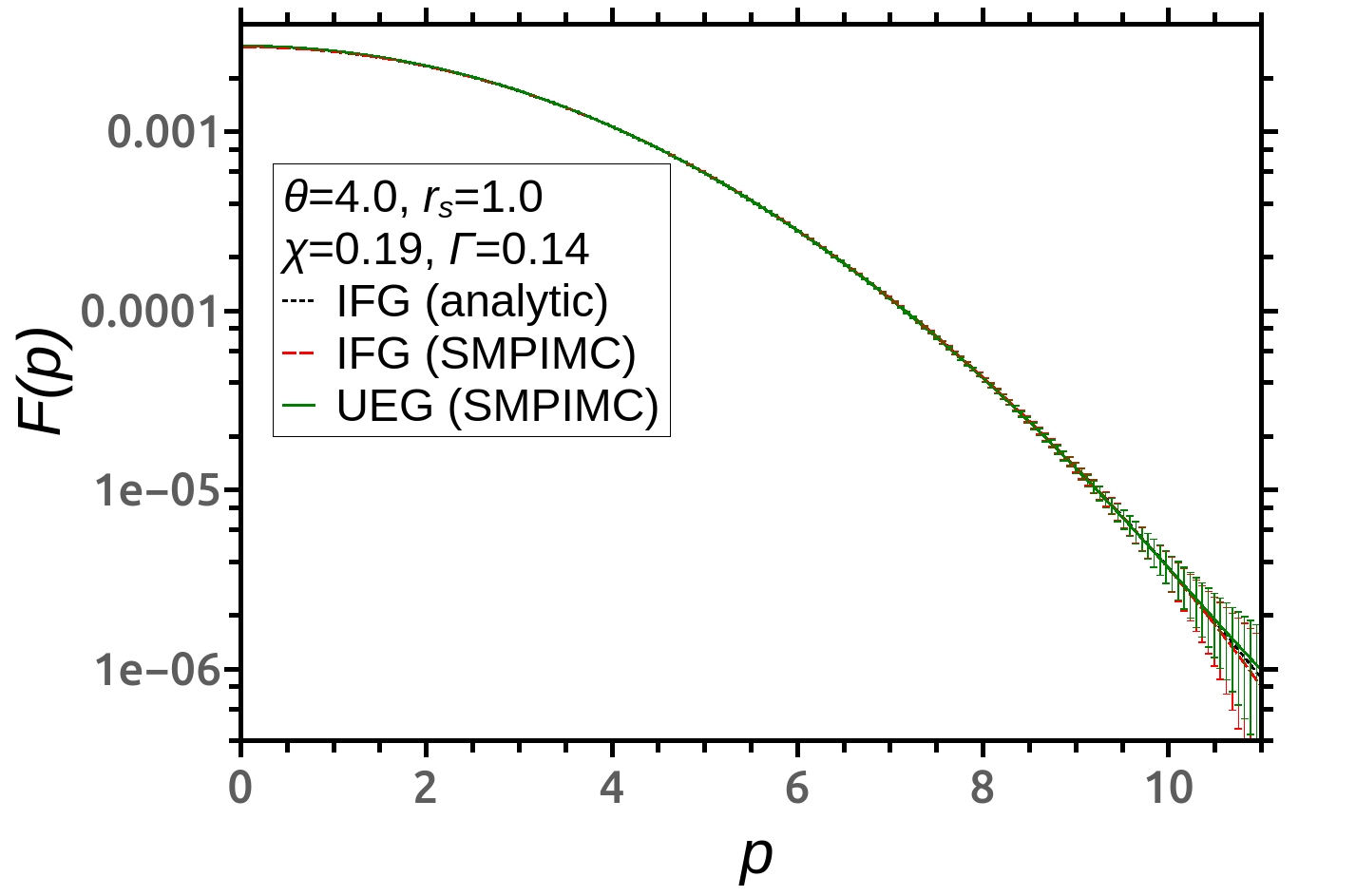}
    	\includegraphics[width=0.49\linewidth]{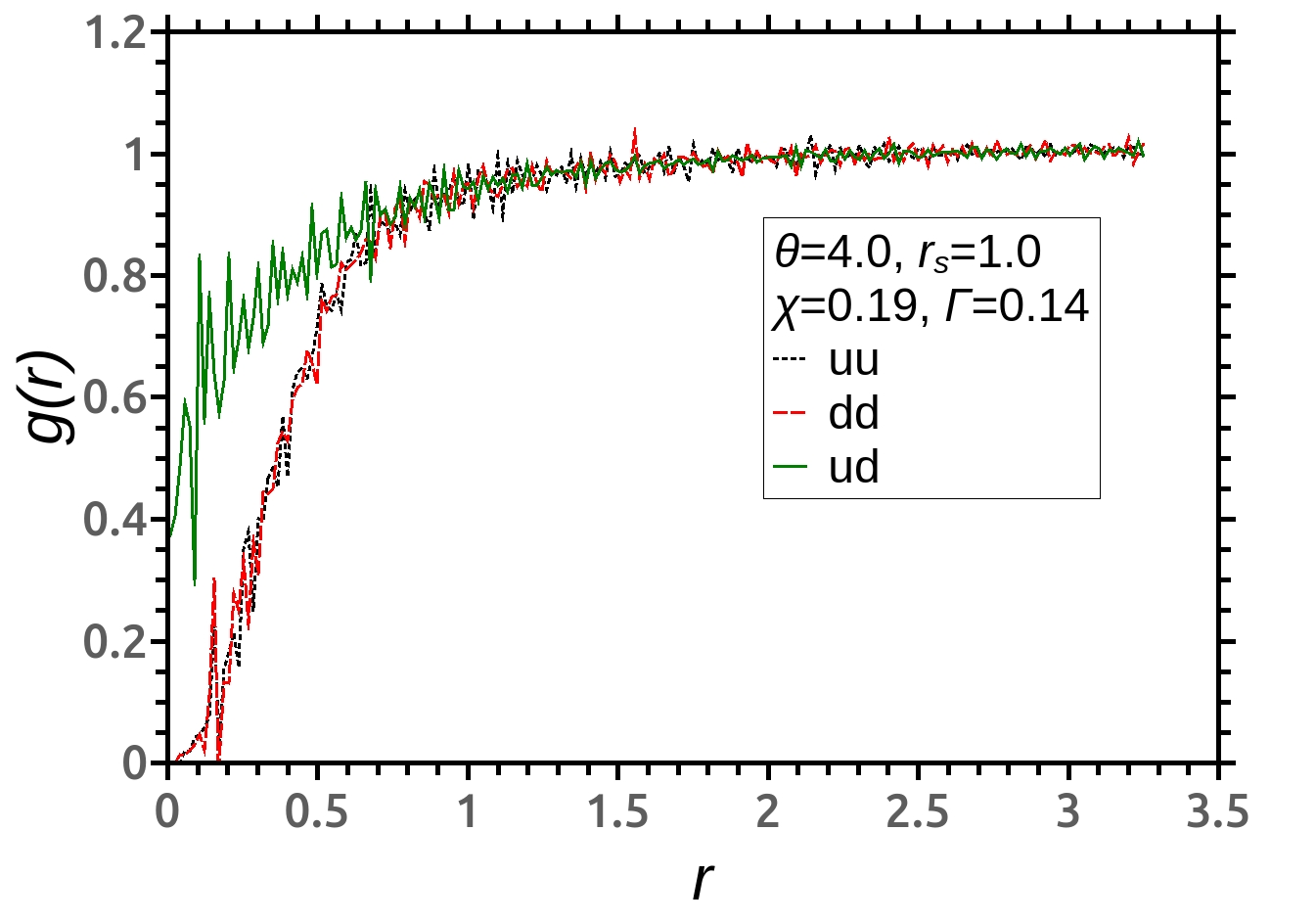}
	\caption{Left plots: single--particle momentum distribution functions of the unpolarized UEG at $r_s=1.0$ and $\theta=0.5,1,4$ calculated with the SMPIMC method.
	The results are compared with the SMPIMC results for the IFG in the same conditions and with the analytical Fermi distributions.
	Right plots: the pair distribution functions $g_{uu}$, $g_{dd}$ and $g_{ud}$ of the UEG.}
	\label{fig_r10}
\end{figure}

The results for $r_s=1.0$ are shown in Fig.~\ref{fig_r10}.
In case of $\theta=0.5$ the UEG is significantly non-ideal and the average kinetic and potential energy are of the same order ($\Gamma \approx 1.1$).
The MDF coincides with the Fermi distribution for $p \lesssim p_{FS}$.
The PDF $g_{ud}$ rises from $0$ to $1$ much slower than in the case of $r_s=0.2$, but faster than the PDFs $g_{uu}$ and $g_{dd}$,
because the exchange repulsion is significantly stronger than the Coulomb repulsion yet. 
In cases of $\theta=1.0$ and $\theta=4.0$ the UEG is weakly non-ideal ($\Gamma \approx 0.54$ and $\Gamma \approx 0.14$ respectively), the MDFs also do not differ from the Fermi distribution.
However the Coulomb interaction starts to affect on $g_{ud}$, as one can find out from the PDFs behavior.

\begin{figure}[ht]	
		\includegraphics[width=0.49\linewidth]{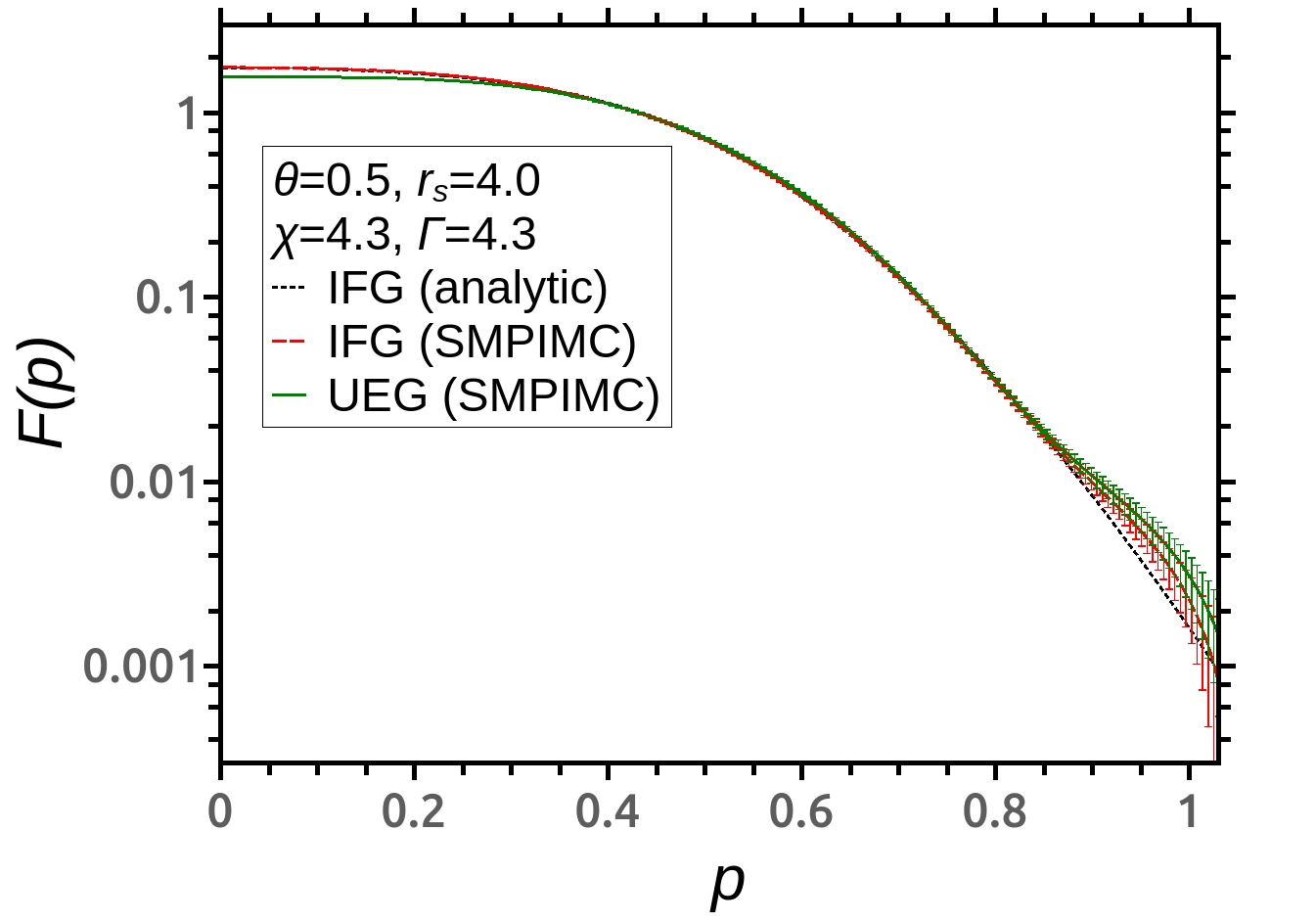}
    	\includegraphics[width=0.49\linewidth]{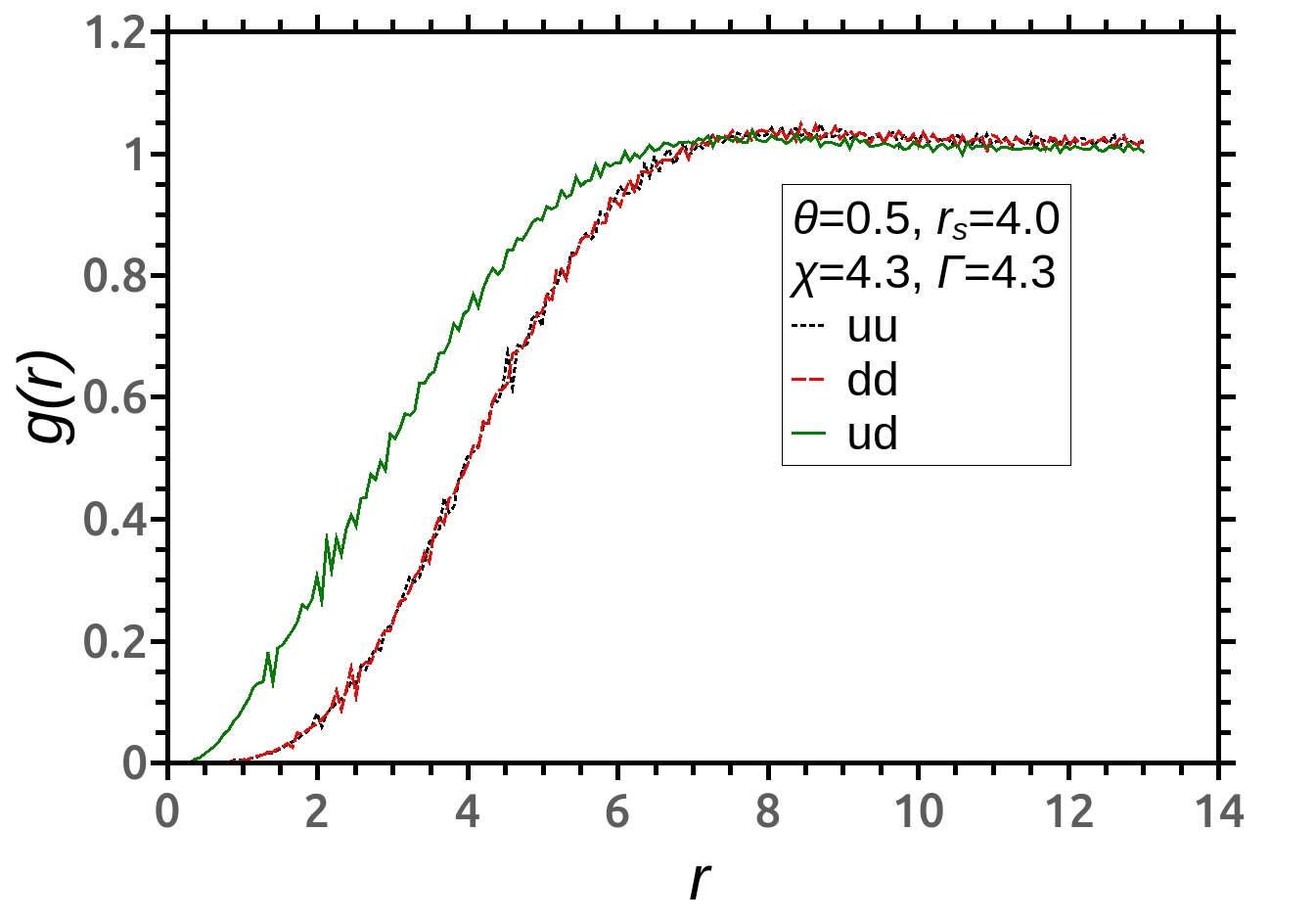}
    	\includegraphics[width=0.49\linewidth]{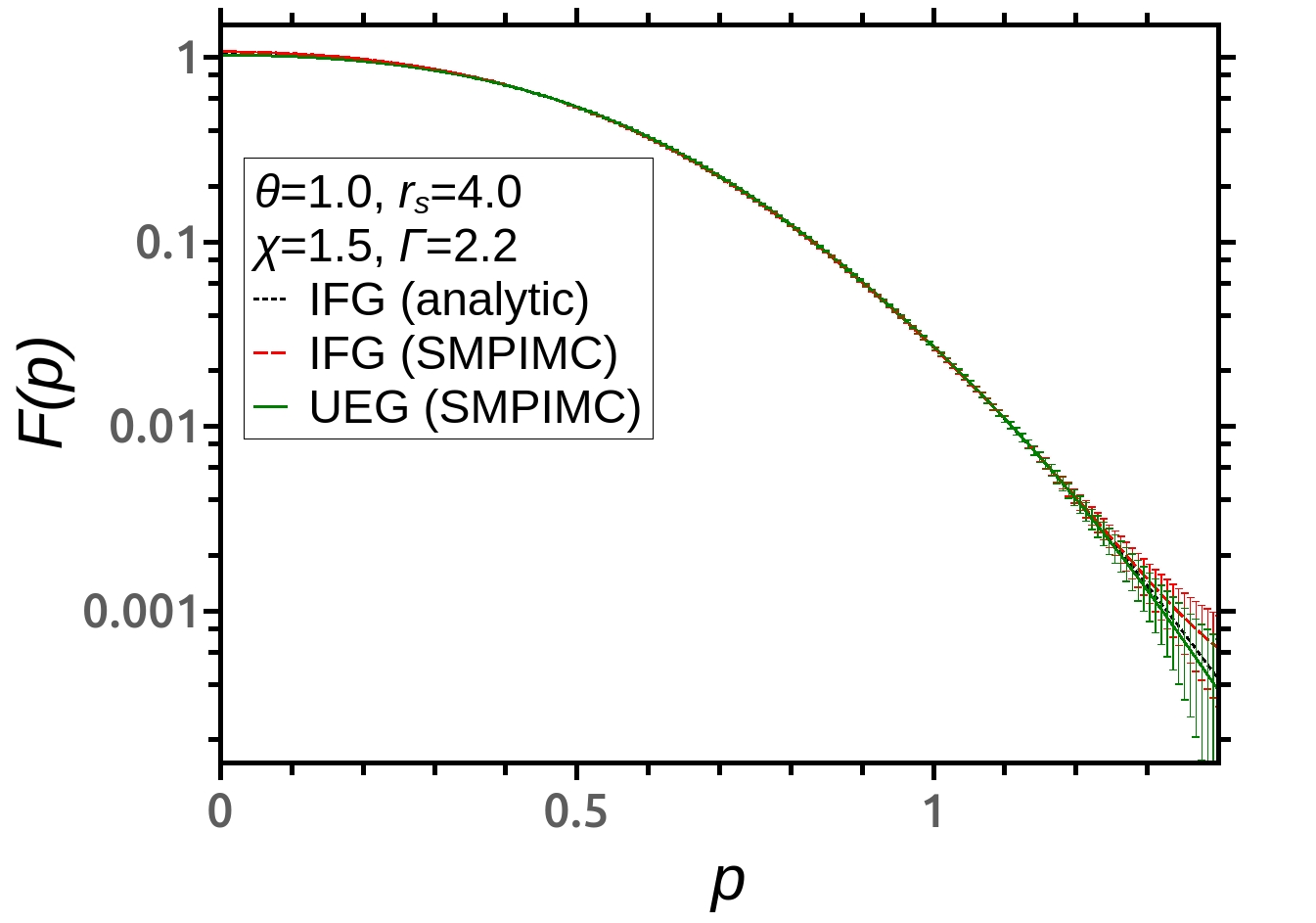}
    	\includegraphics[width=0.49\linewidth]{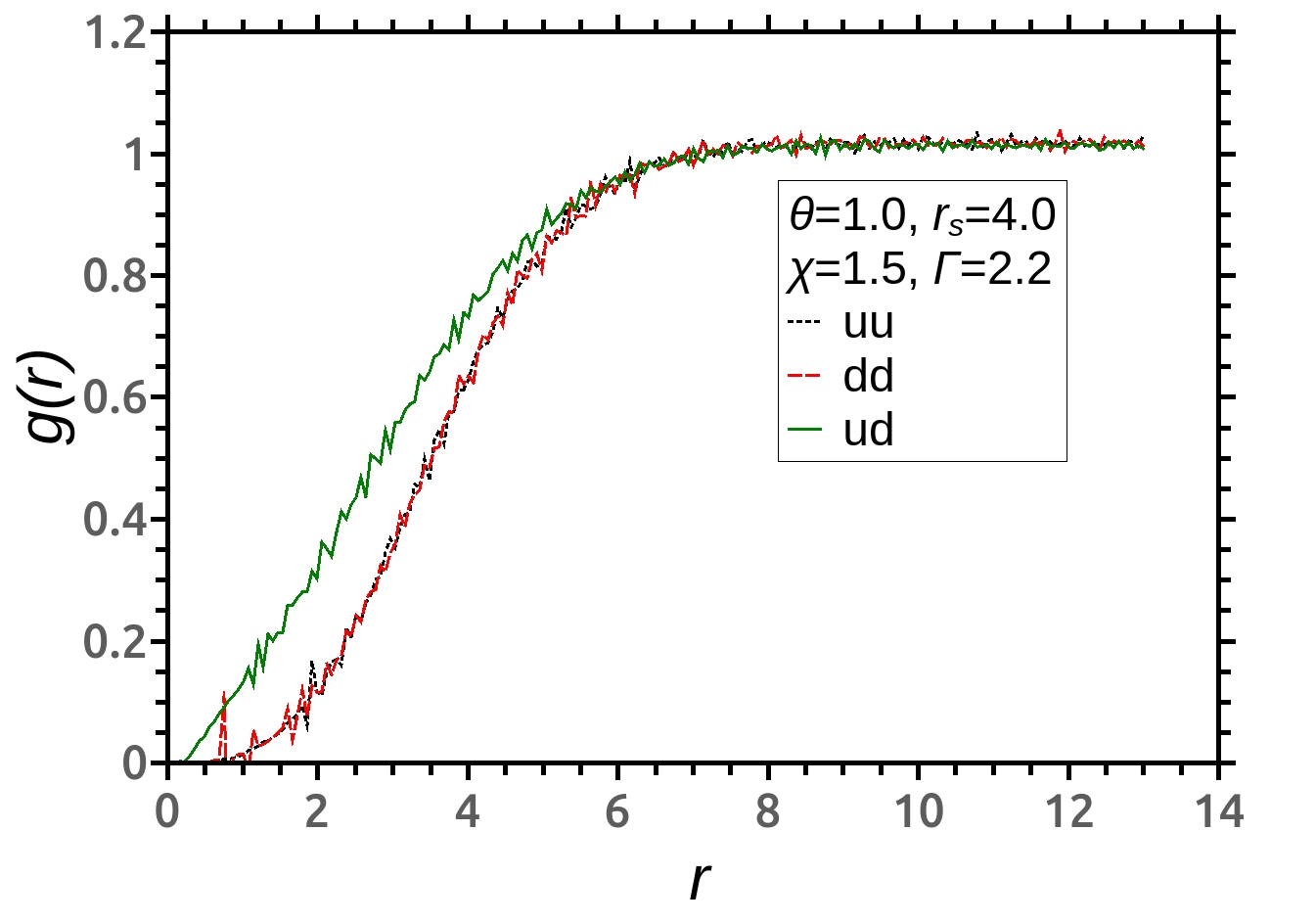}
    	\includegraphics[width=0.49\linewidth]{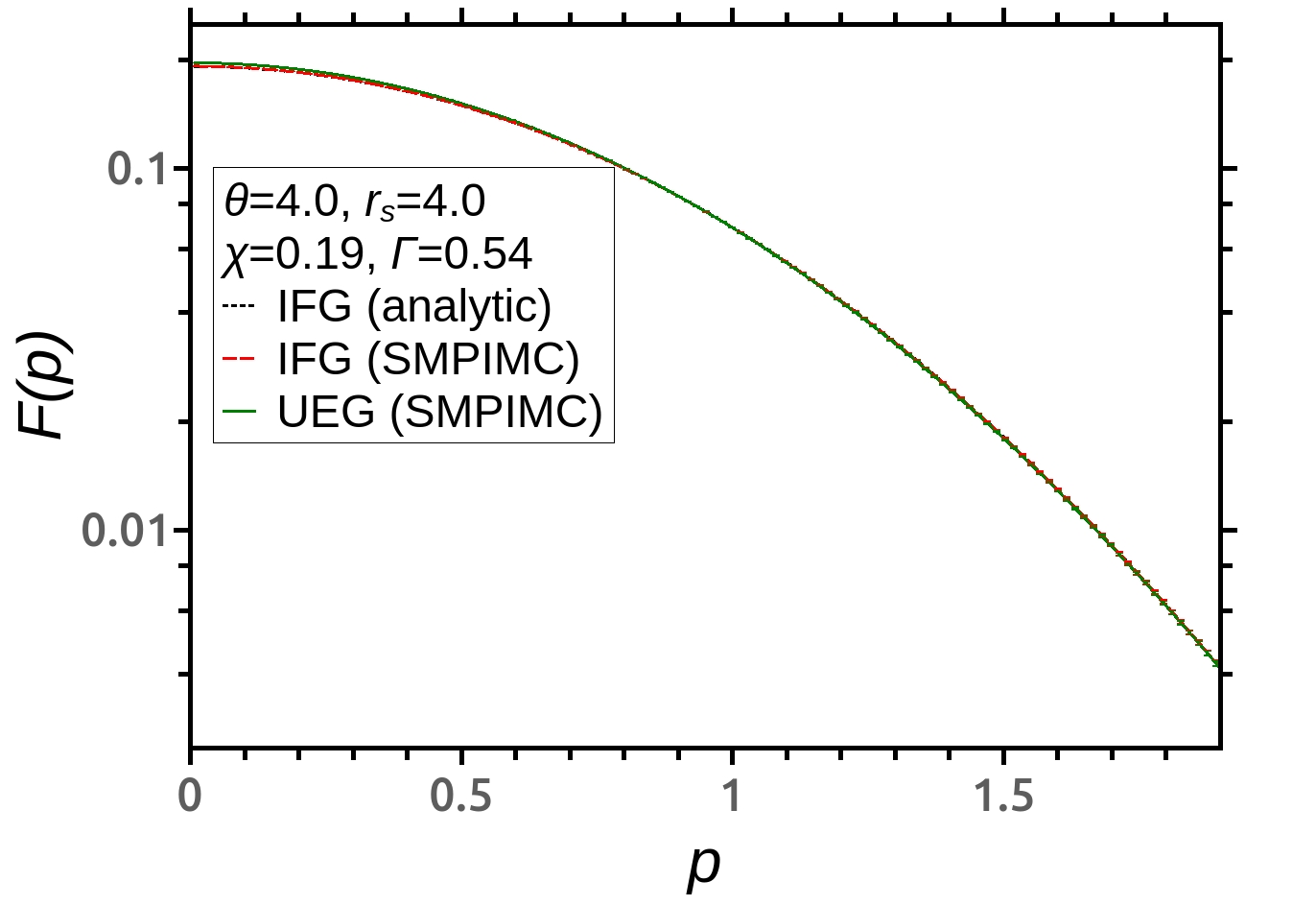}
    	\includegraphics[width=0.49\linewidth]{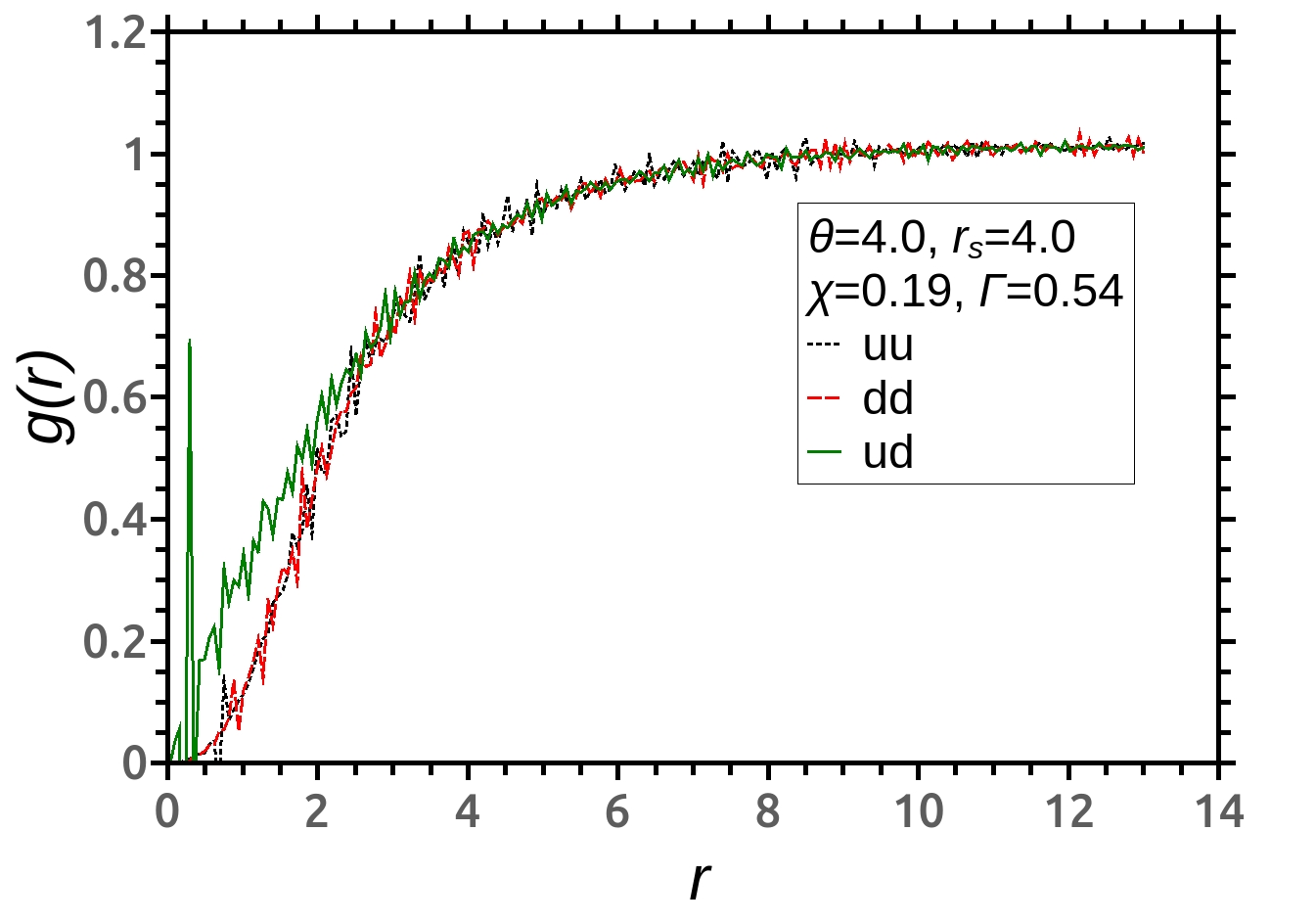}
	\caption{Left plots: single--particle momentum distribution functions of the unpolarized UEG at $r_s=4.0$ and $\theta=0.5,1,4$ calculated with the SMPIMC method.
	The results are compared with the SMPIMC results for the IFG in the same conditions and with the analytical Fermi distributions. 
	Right plots: the pair distribution functions $g_{uu}$, $g_{dd}$ and $g_{ud}$ of the UEG. }
	\label{fig_r40}
\end{figure}

The results for $r_s=4.0$ are shown in Fig.~\ref{fig_r40}.
In cases of $\theta=0.5$ and $\theta=1$ the UEG becomes strongly non-ideal ($\Gamma \approx 4.3$ and $\Gamma \approx 2.2$), while at $\theta =4$ it is weakly non-ideal ($\Gamma \approx 0.54$).
The difference between MDFs and the related Fermi distribution does not exceed the statistical error.
The PDFs $g_{uu}$, $g_{dd}$ and $g_{ud}$ differ from each other only slightly (especially at $\theta = 4.0$), so the Coulomb and the exchange interactions are almost of the same order.

\begin{figure}[ht]	
		\includegraphics[width=0.49\linewidth]{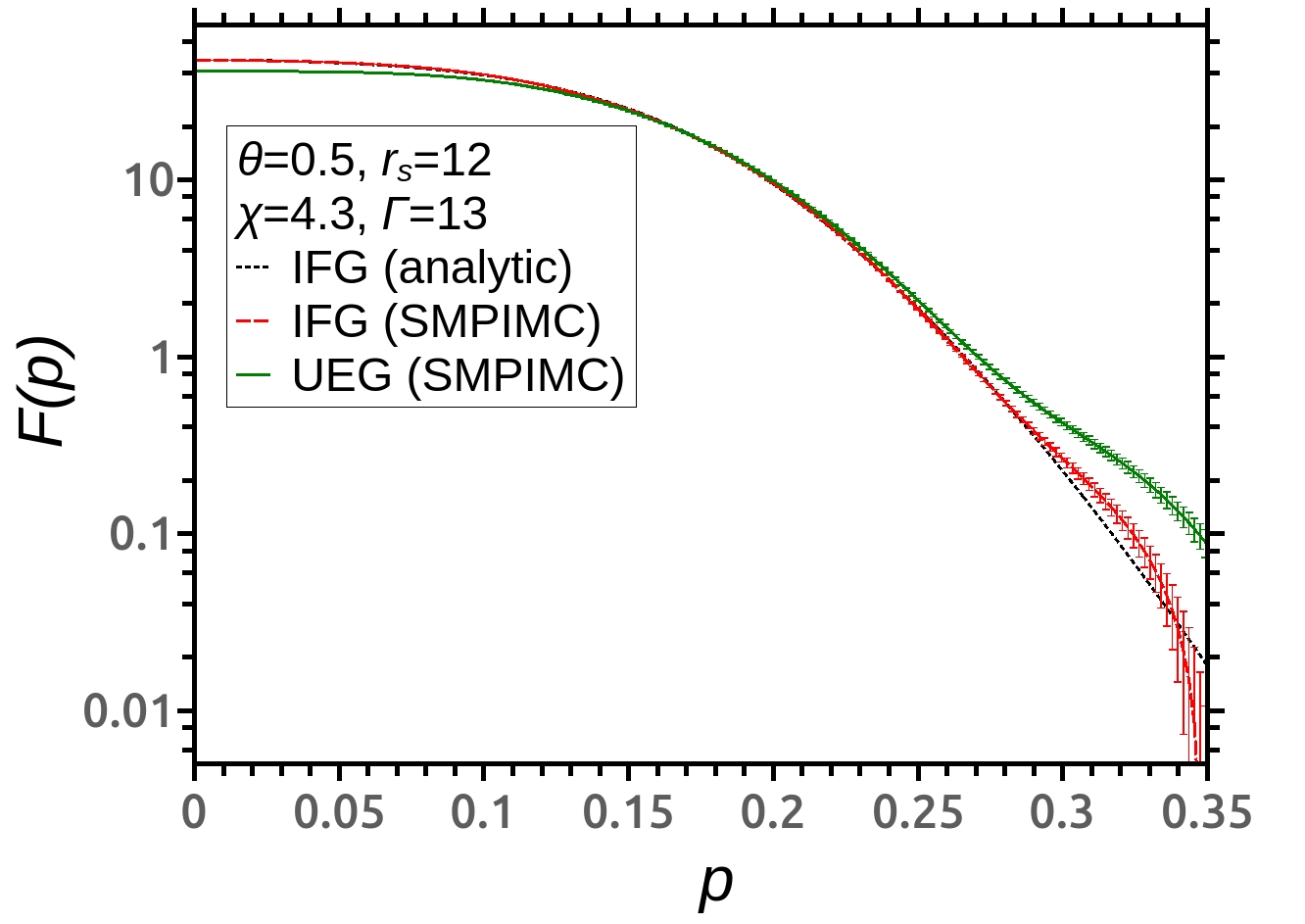}
    	\includegraphics[width=0.49\linewidth]{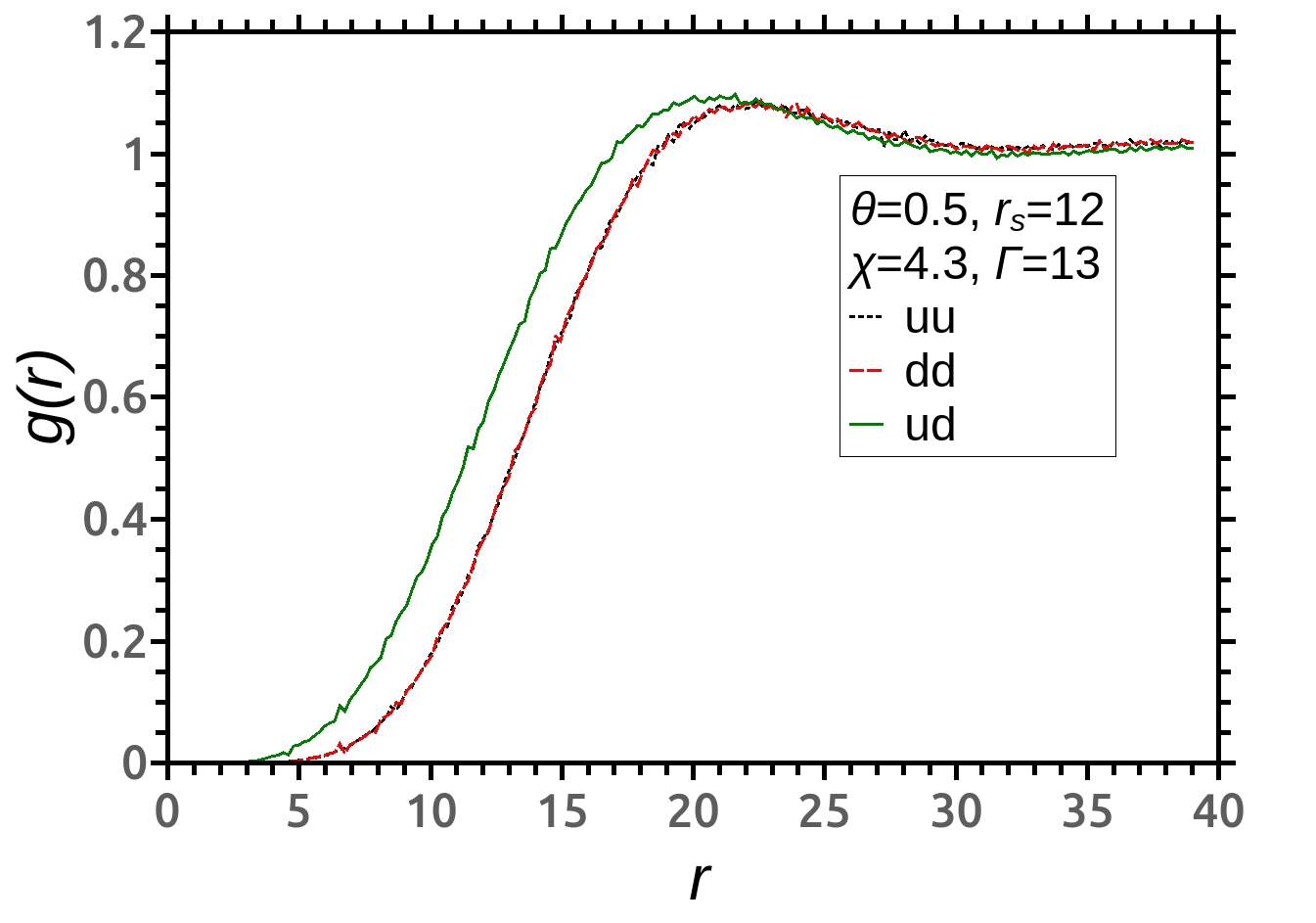}
    	\includegraphics[width=0.49\linewidth]{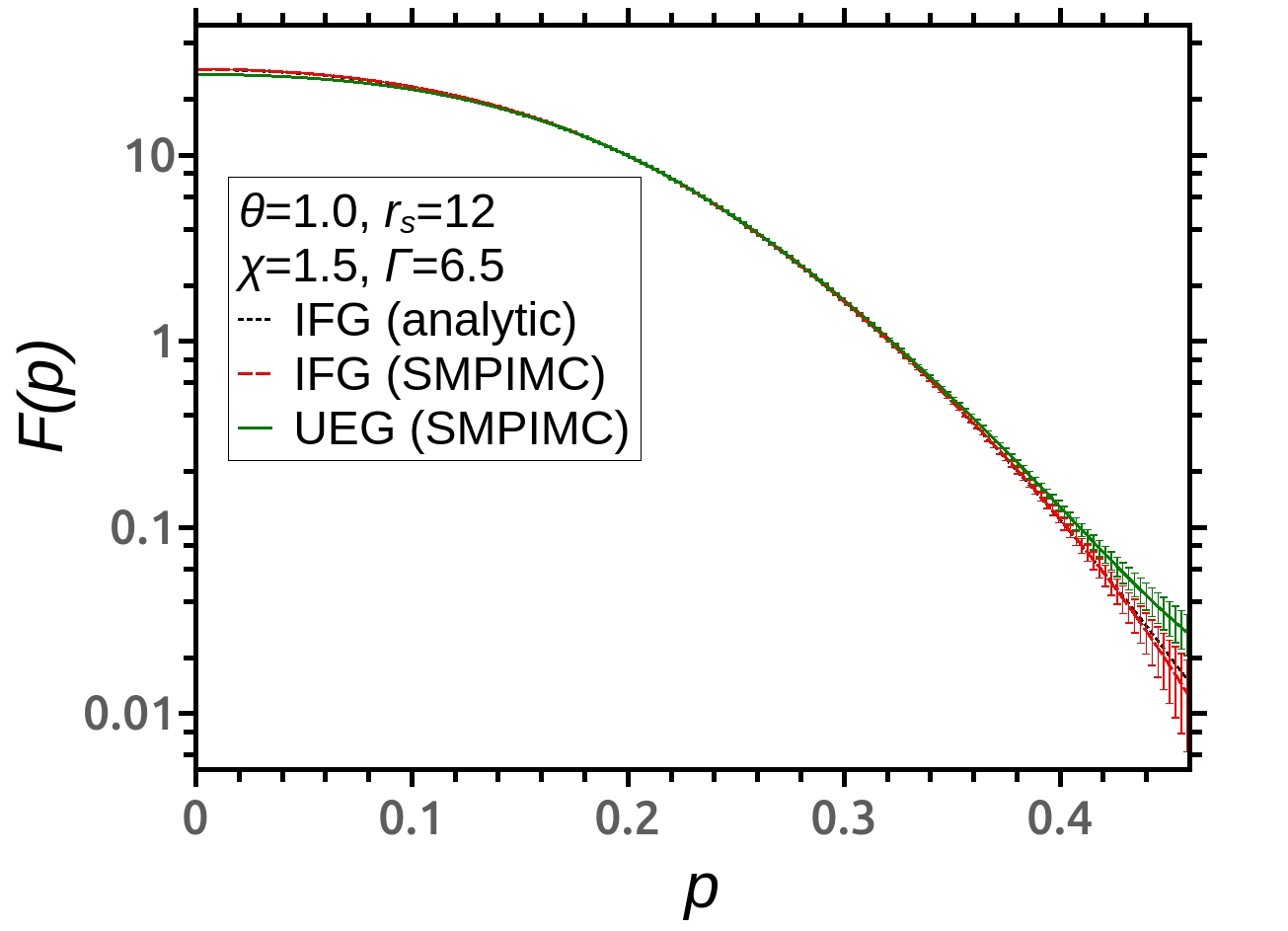}
    	\includegraphics[width=0.49\linewidth]{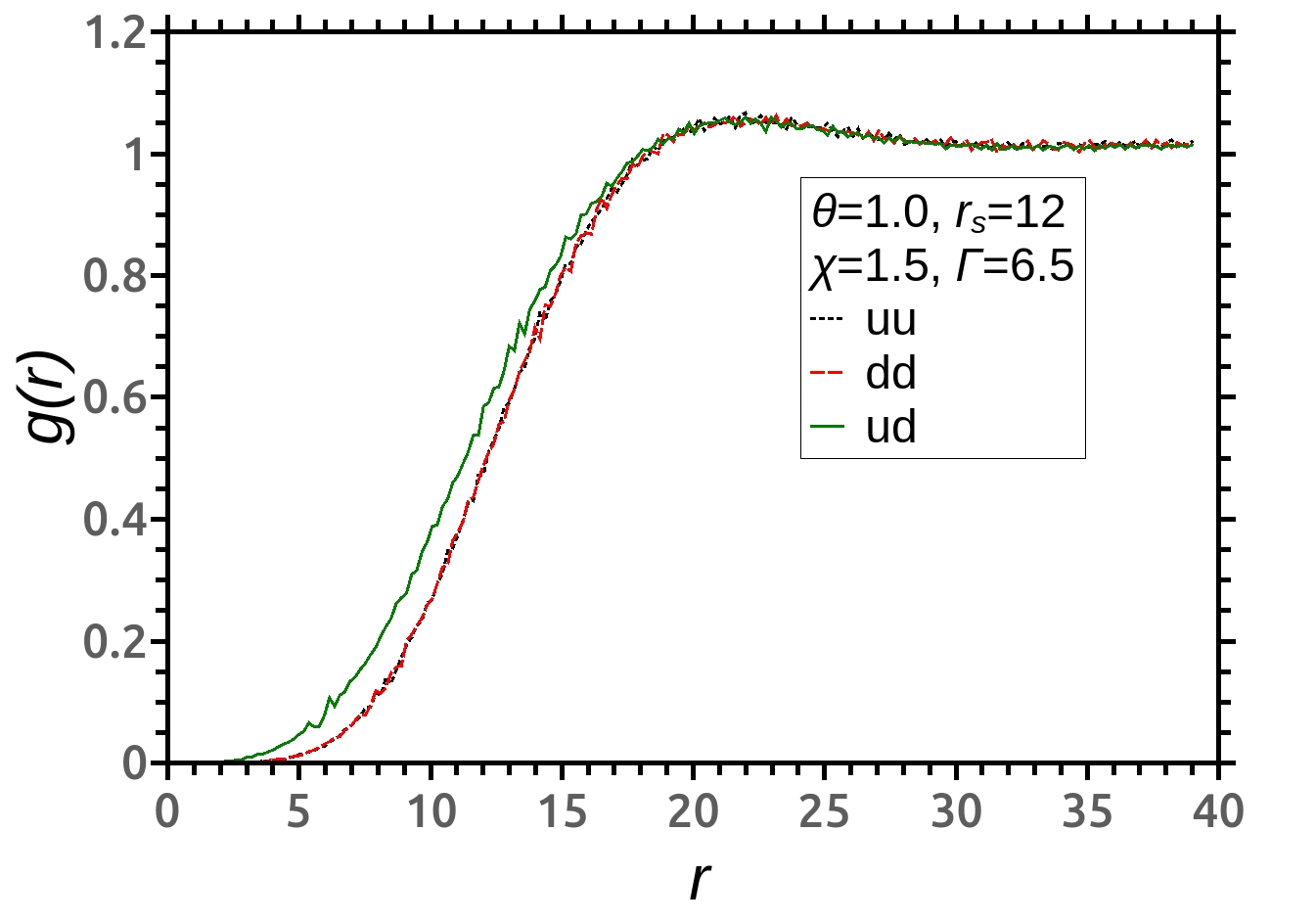}
    	\includegraphics[width=0.49\linewidth]{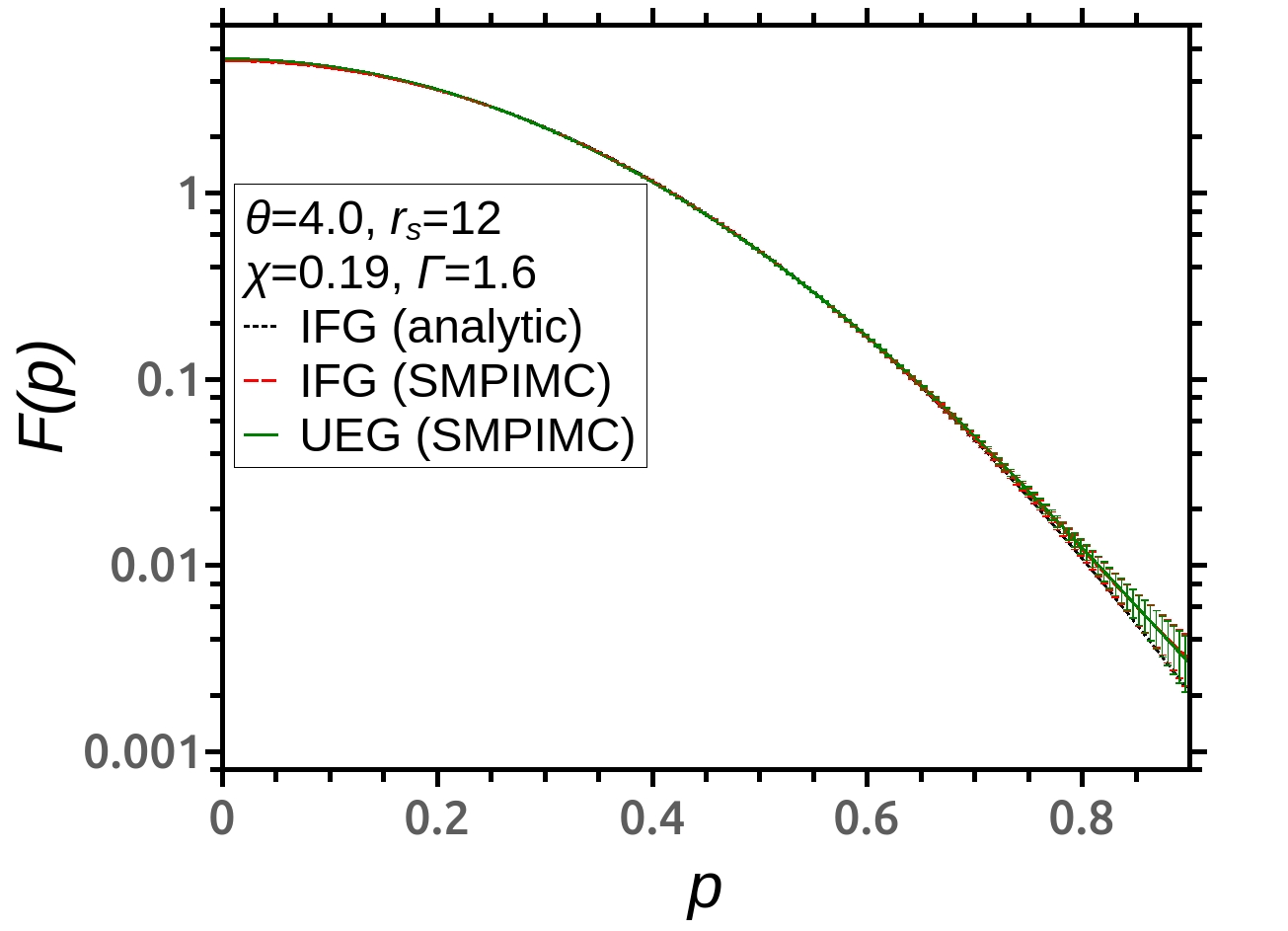}
    	\includegraphics[width=0.49\linewidth]{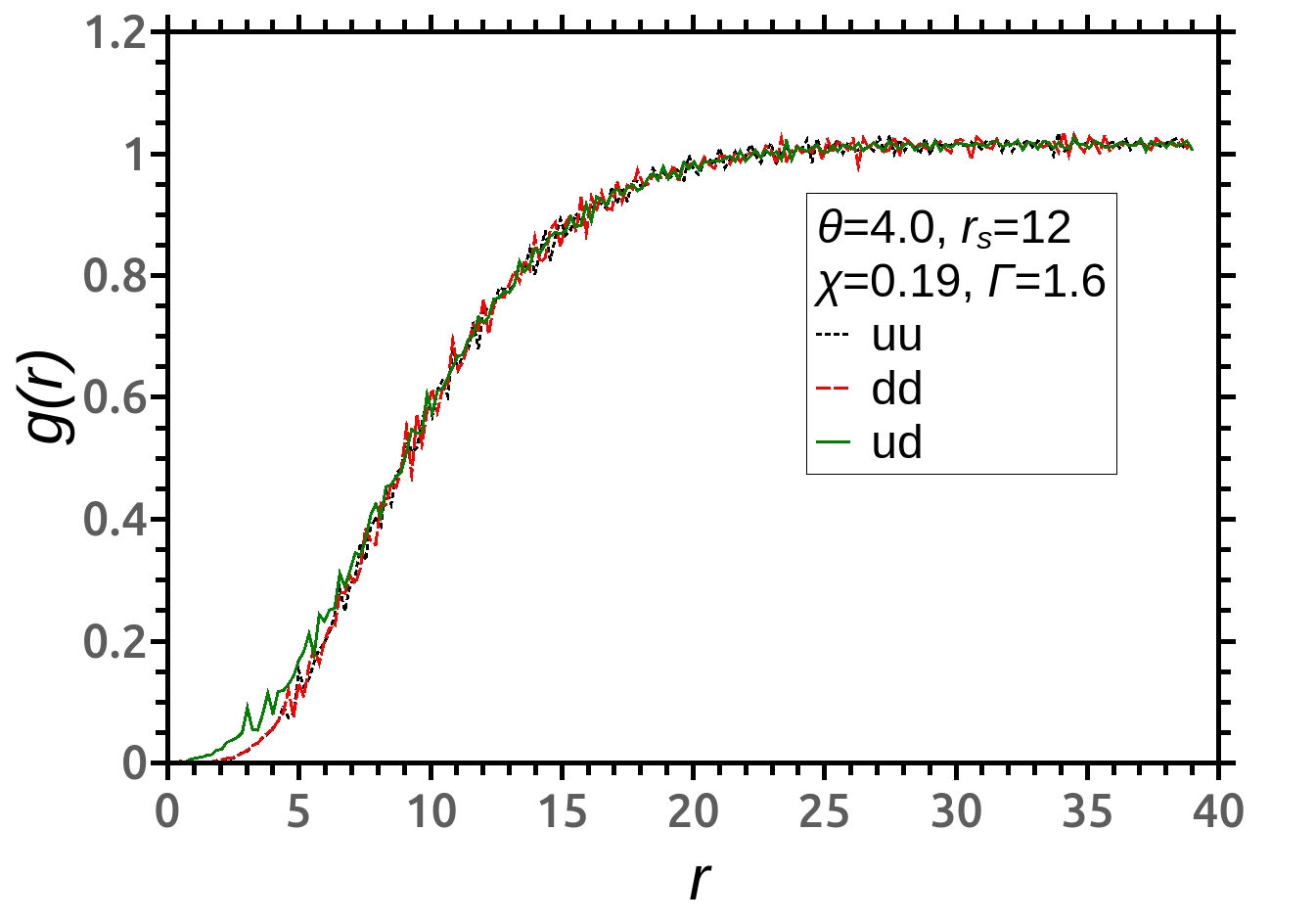}
	\caption{Left plots: single--particle momentum distribution functions of the unpolarized UEG at $r_s=12$ and $\theta=0.5,1,4$ calculated with the SMPIMC method.
	The results are compared with the SMPIMC results for the IFD in the same conditions and with the analytical Fermi distributions.
	Right plots: the pair distribution functions $g_{uu}$, $g_{dd}$ and $g_{ud}$ of the UEG. }
	\label{fig_r120}
\end{figure}

The results for $r_s=12.0$ are shown in Fig.~\ref{fig_r120}.
In case of $\theta = 0.5$ the coupling is very strong ($\Gamma \approx 13$), and the MDF significantly differs from the Fermi distribution via forming an exceeding ``tail" at high momenta, which is not camouflaged by the finite--size effects.
All PDFs have maxima exceeding $1$ at $r_{max}$, and $g_{ud}$ grows almost in the same rate as $g_{uu}$ and $g_{dd}$, because the Coulomb repulsion plays the major role.
In case of $\theta=1.0$ the coupling is quite strong ($\Gamma \approx 6.5$). 
The difference between MDF and the Fermi distribution does not exceed the statistical error.
However there are weak maxima of PDFs, and $g_{ud}$ almost coincides with $g_{uu}$ and $g_{dd}$.
At $\theta = 4.0$ the UEG is significantly non-ideal ($\Gamma \approx 1.6$), the MDF does not differ from the Fermi distribution and all PDFs are the same and do not have any maxima.

\begin{figure}[ht]	
		\includegraphics[width=0.49\linewidth]{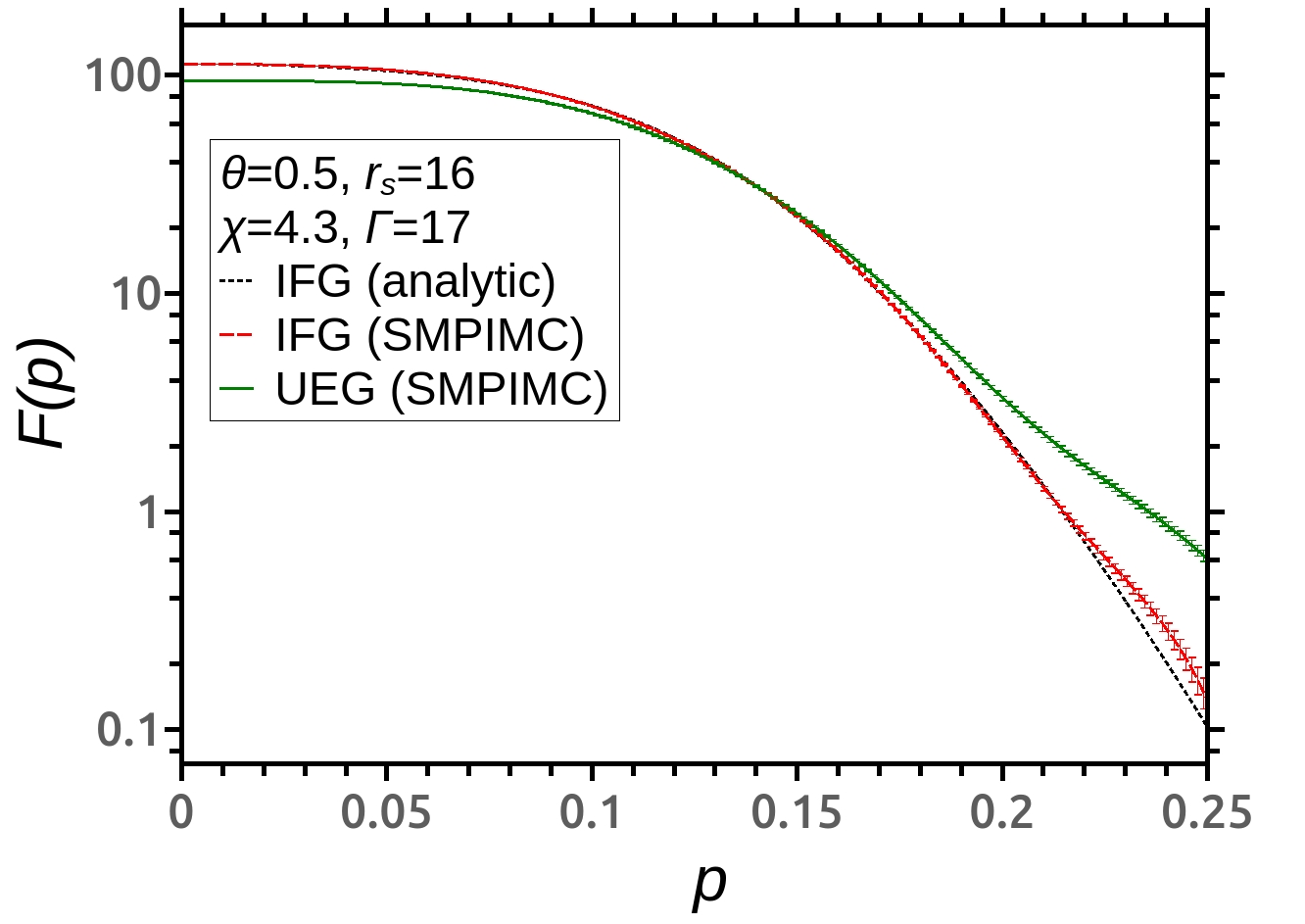}
    	\includegraphics[width=0.49\linewidth]{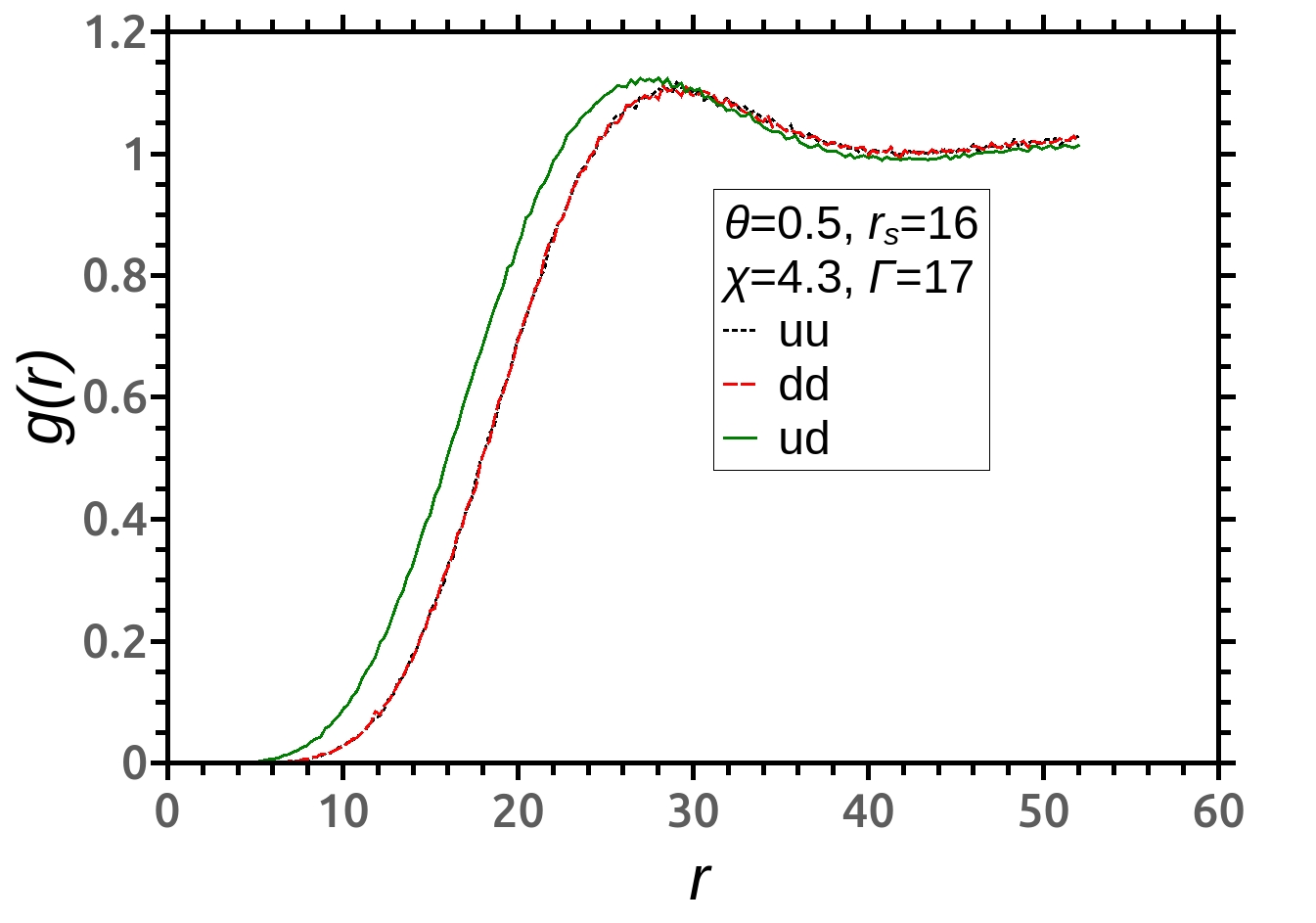}
    	\includegraphics[width=0.49\linewidth]{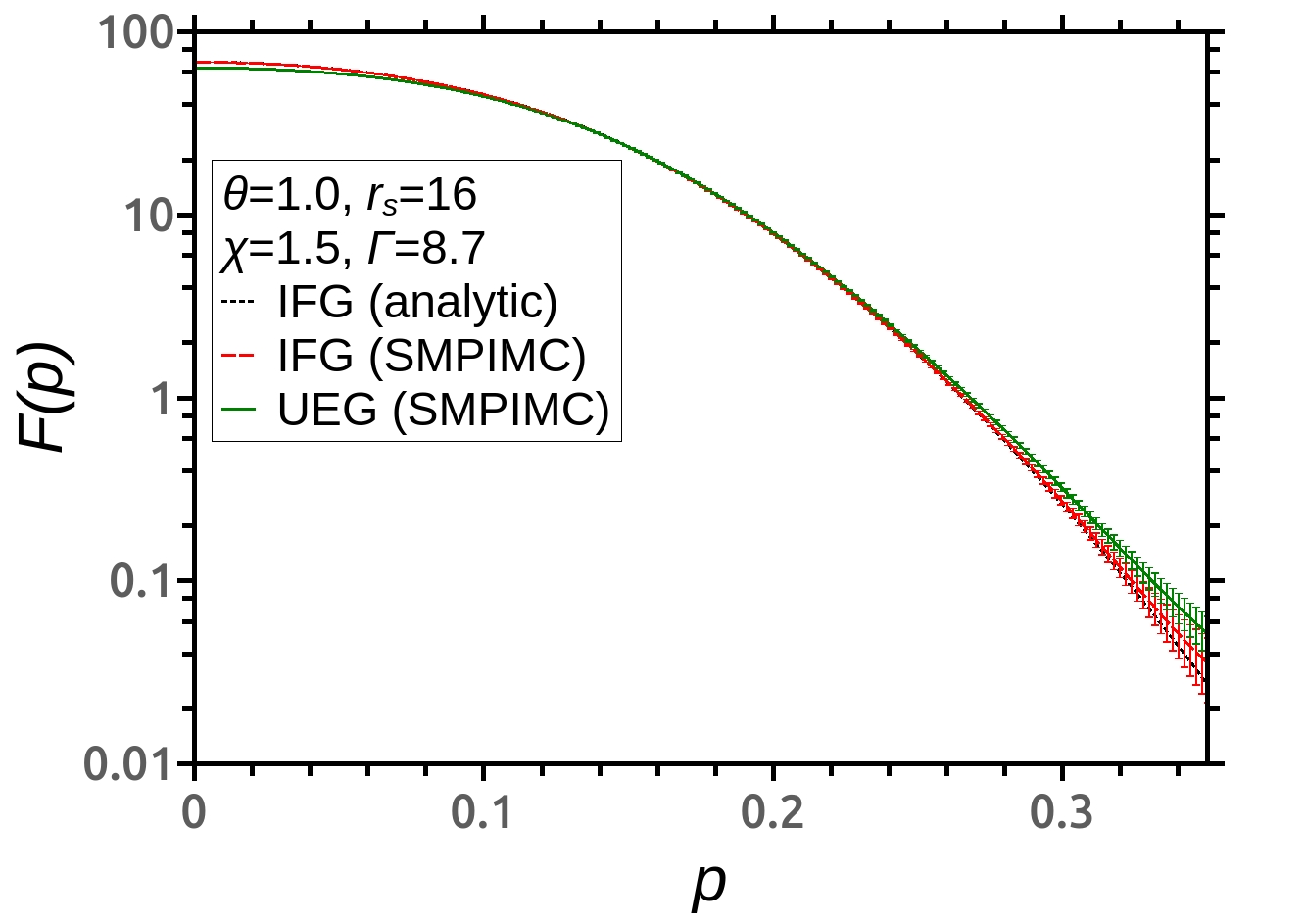}
    	\includegraphics[width=0.49\linewidth]{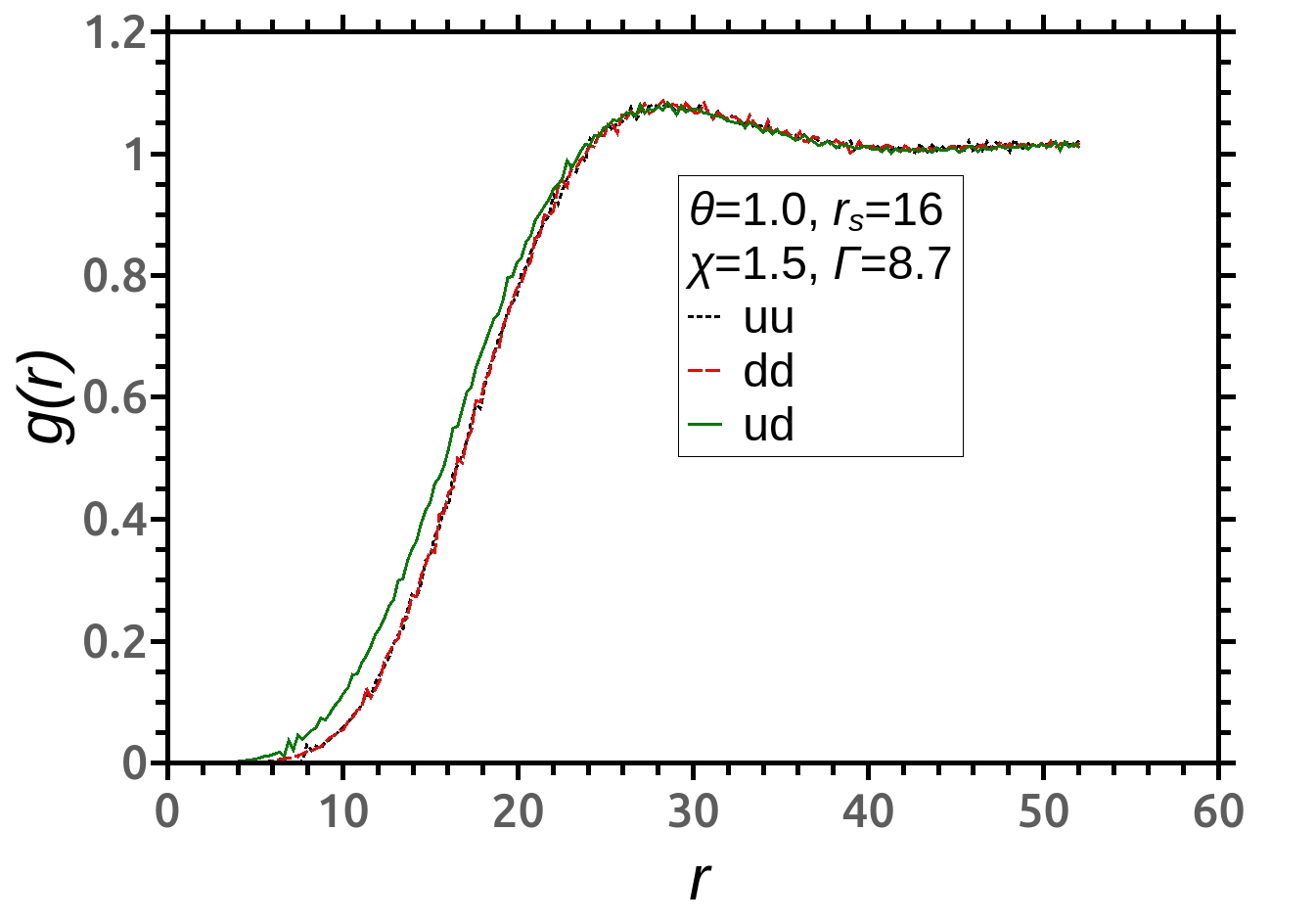}
    	\includegraphics[width=0.49\linewidth]{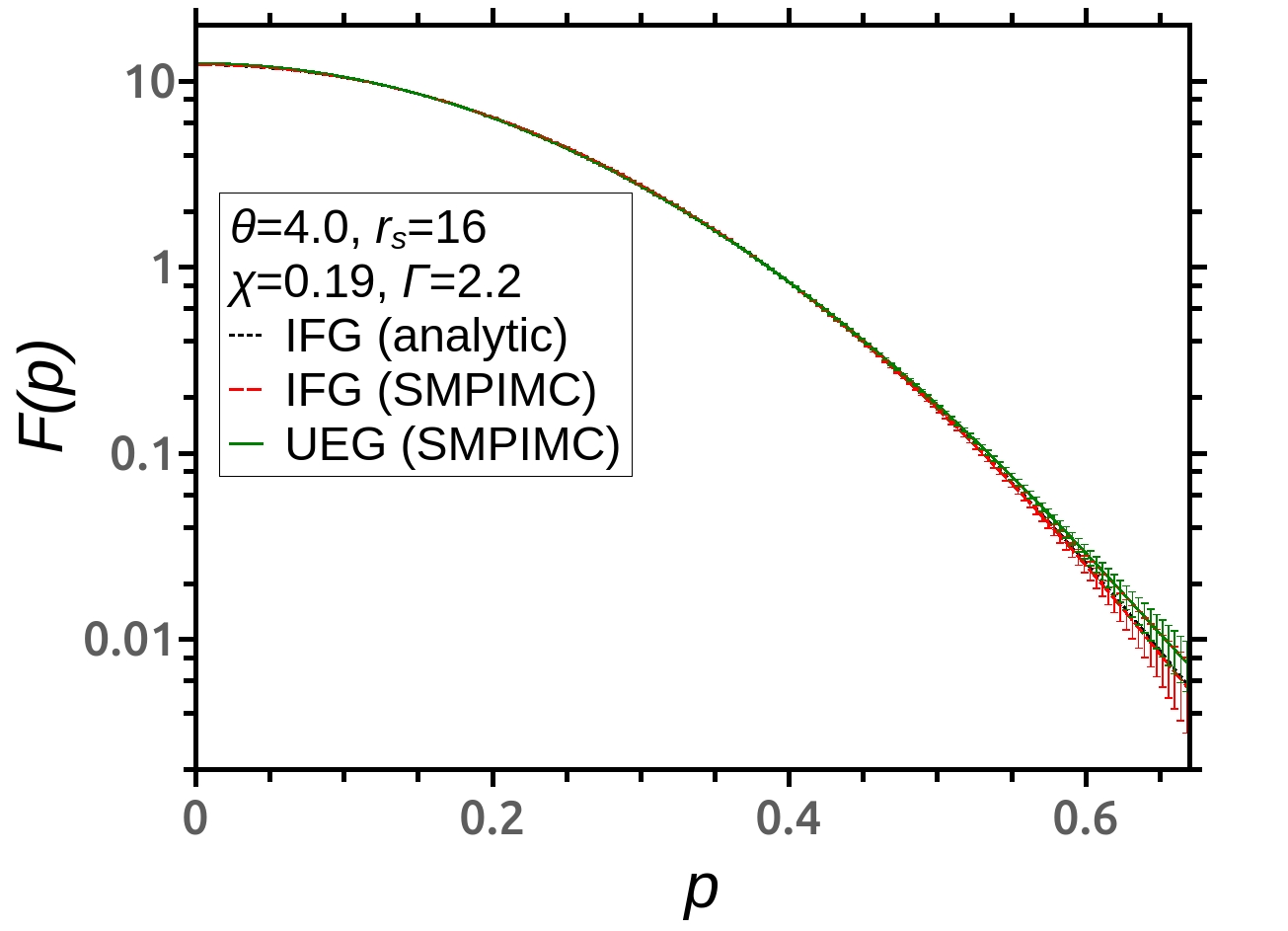}
    	\includegraphics[width=0.49\linewidth]{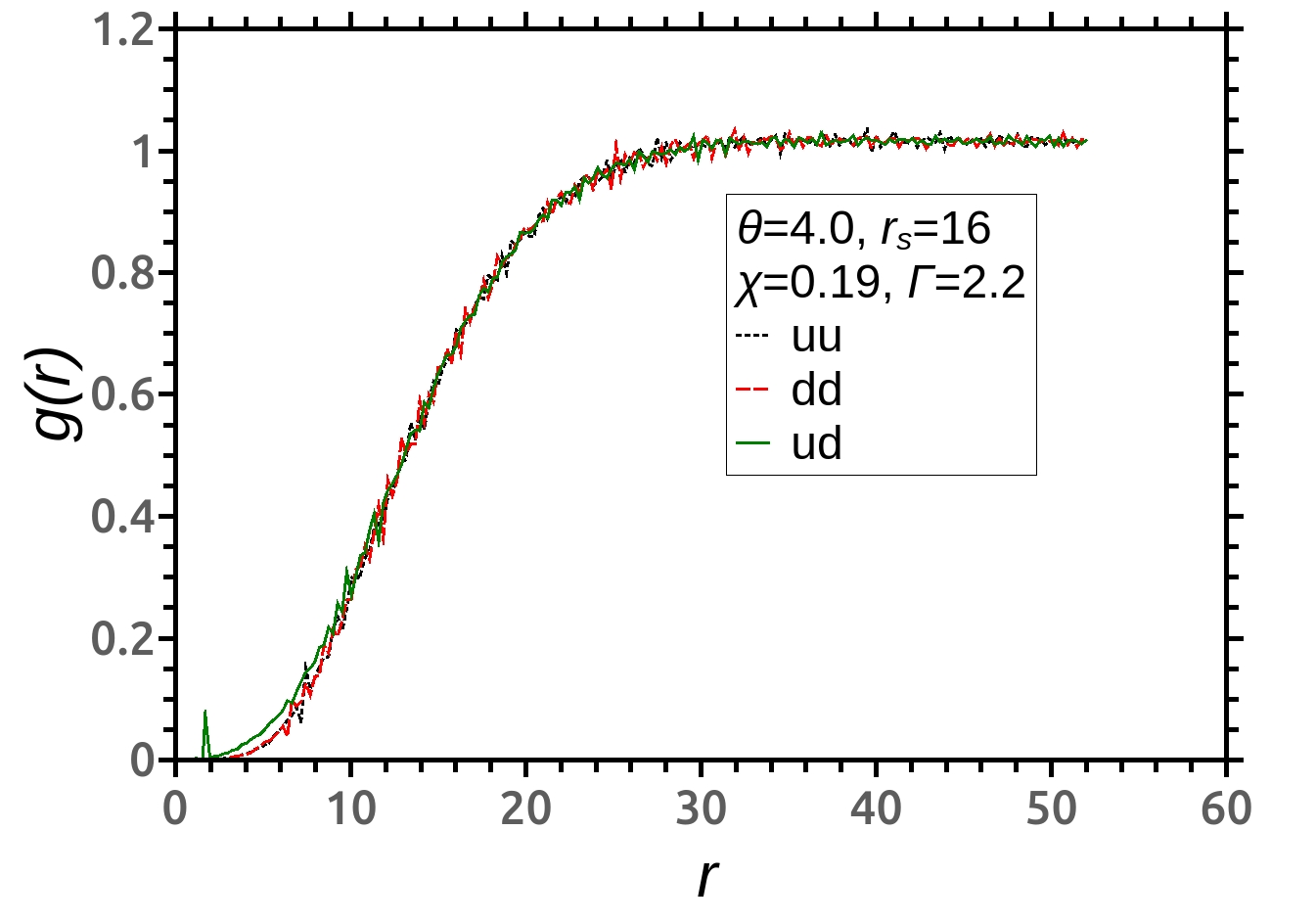}
	\caption{Left plots: single--particle momentum distribution functions of the unpolarized UEG at $r_s=16$ and $\theta=0.5,1,4$ calculated with the SMPIMC method.
	The results are compared with the SMPIMC results for the IFG in the same conditions, and with the analytical Fermi distributions.
	Right plots: the pair distribution functions $g_{uu}$, $g_{dd}$ and $g_{ud}$ of the UEG. }
	\label{fig_r160}
\end{figure}

The results for $r_s=16.0$ are shown in Fig.~\ref{fig_r160}.
In case of $\theta=0.5$  ($\Gamma \approx 17$) the MDF is quite similar to the previously considered one with $r_s=12$, but the difference between it and the Fermi distribution is more significant.
The PDFs are also analogous to that case, but the maxima are more pronounced and the minima (at $r_{min}$) begin to appear.
In case of $\theta = 1.0$ the coupling is also strong ($\Gamma \approx 8.7$), and the MDF begins to differ from the Fermi distribution. 
The PDFs have distinct maxima at $r \approx 28 a_0$.
In case of $\theta = 4.0$ the coupling strength $\Gamma \approx 2.2$, and the MDF coincides with the Fermi distribution with difference less than the statistical error.
The PDFs do not have any maxima yet, and there are no difference between $g_{uu}$, $g_{dd}$ and $g_{ud}$, so the exchange interaction is negligible in comparison with the Coulomb one.

\begin{figure}[ht]		
    	\includegraphics[width=0.49\linewidth]{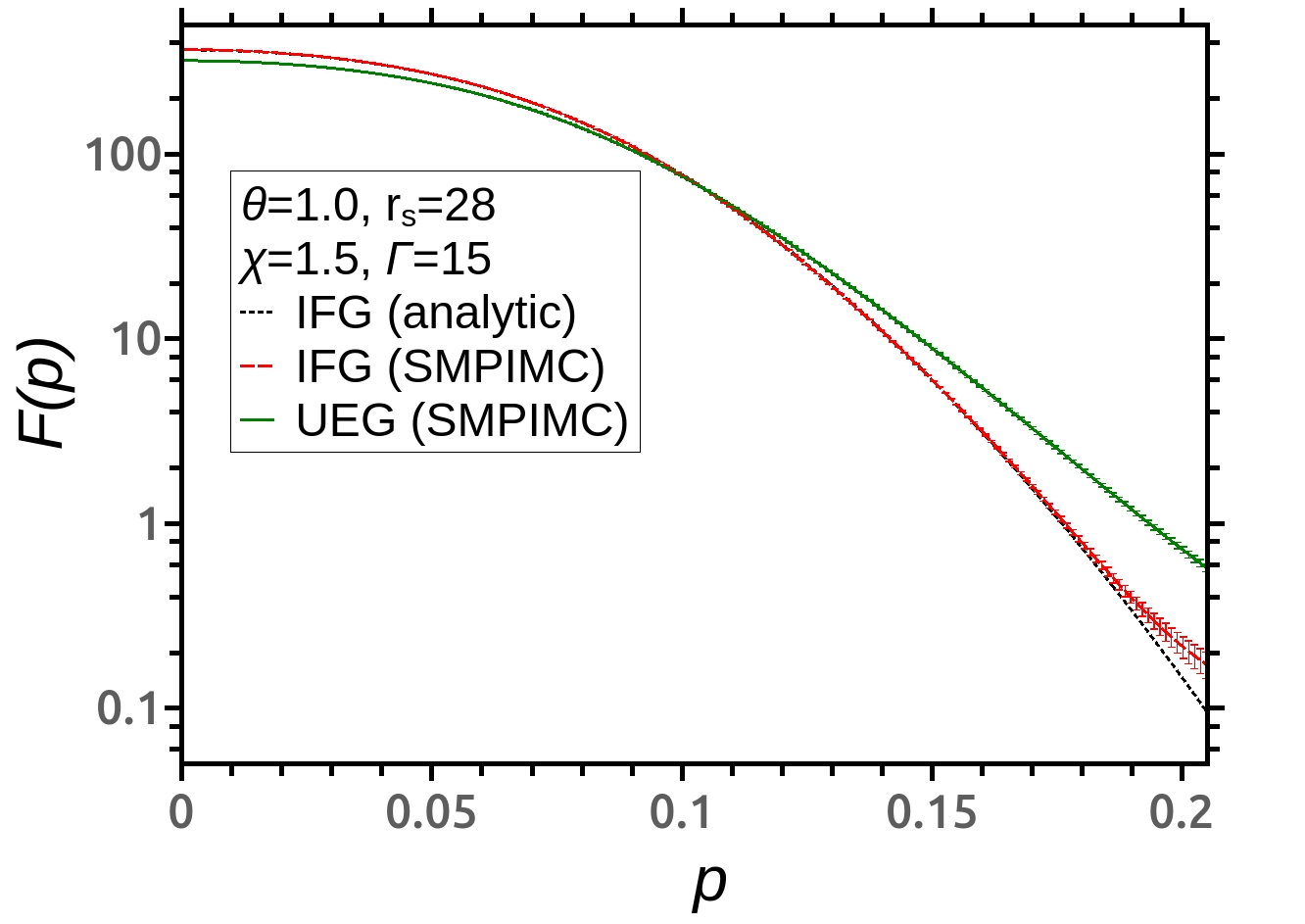}
    	\includegraphics[width=0.49\linewidth]{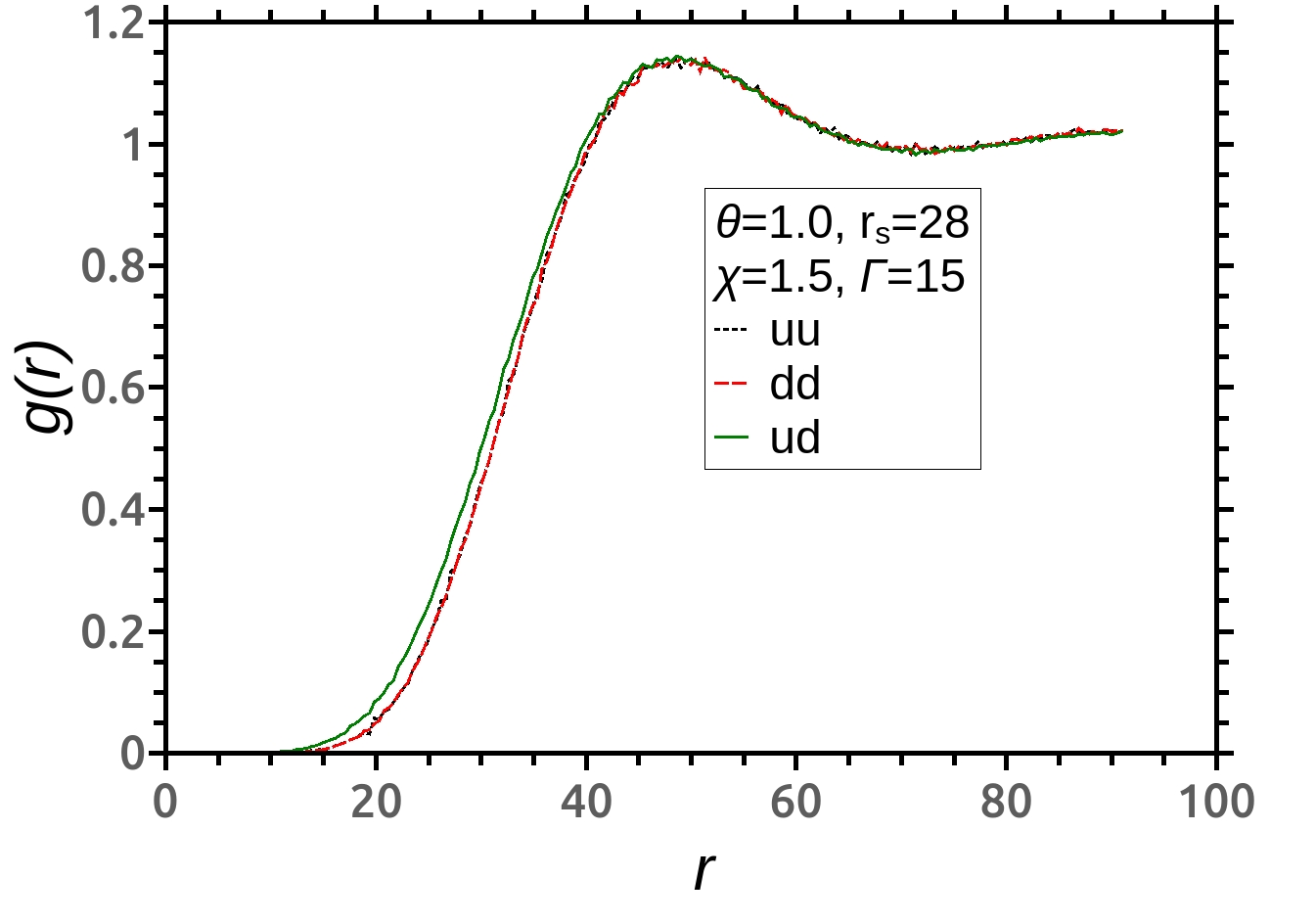}
    	\includegraphics[width=0.49\linewidth]{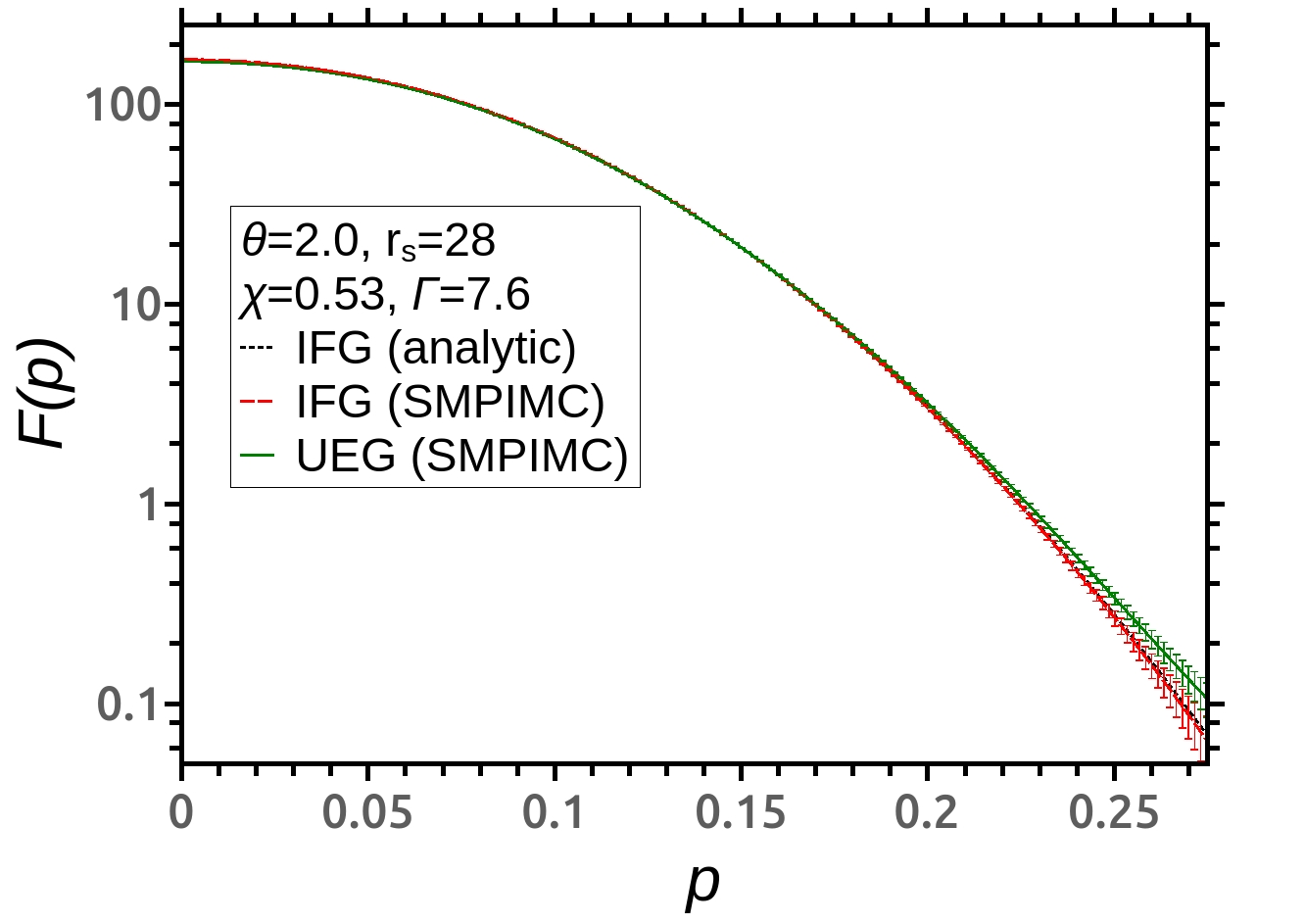}
    	\includegraphics[width=0.49\linewidth]{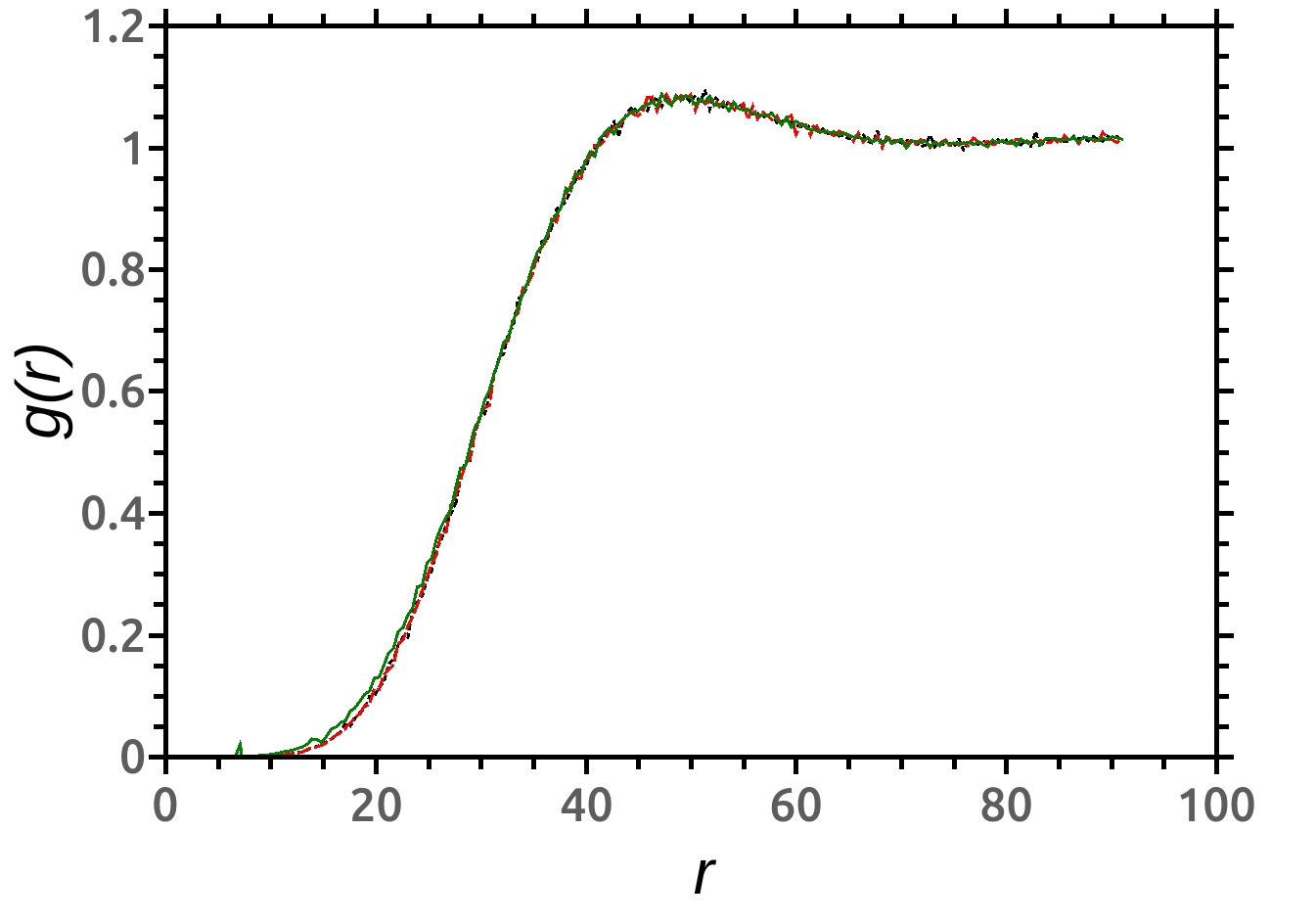}
    	\includegraphics[width=0.49\linewidth]{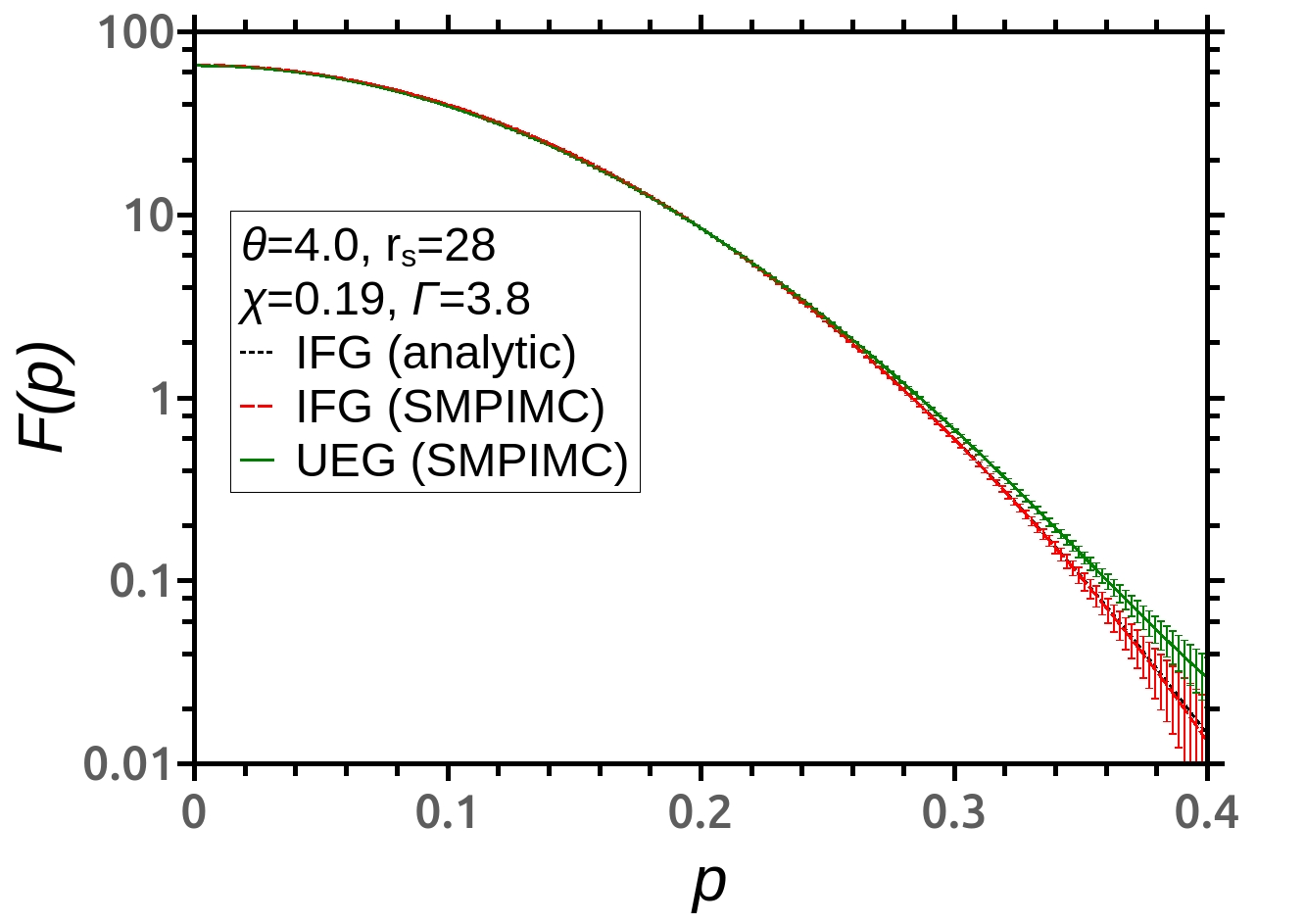}
    	\includegraphics[width=0.49\linewidth]{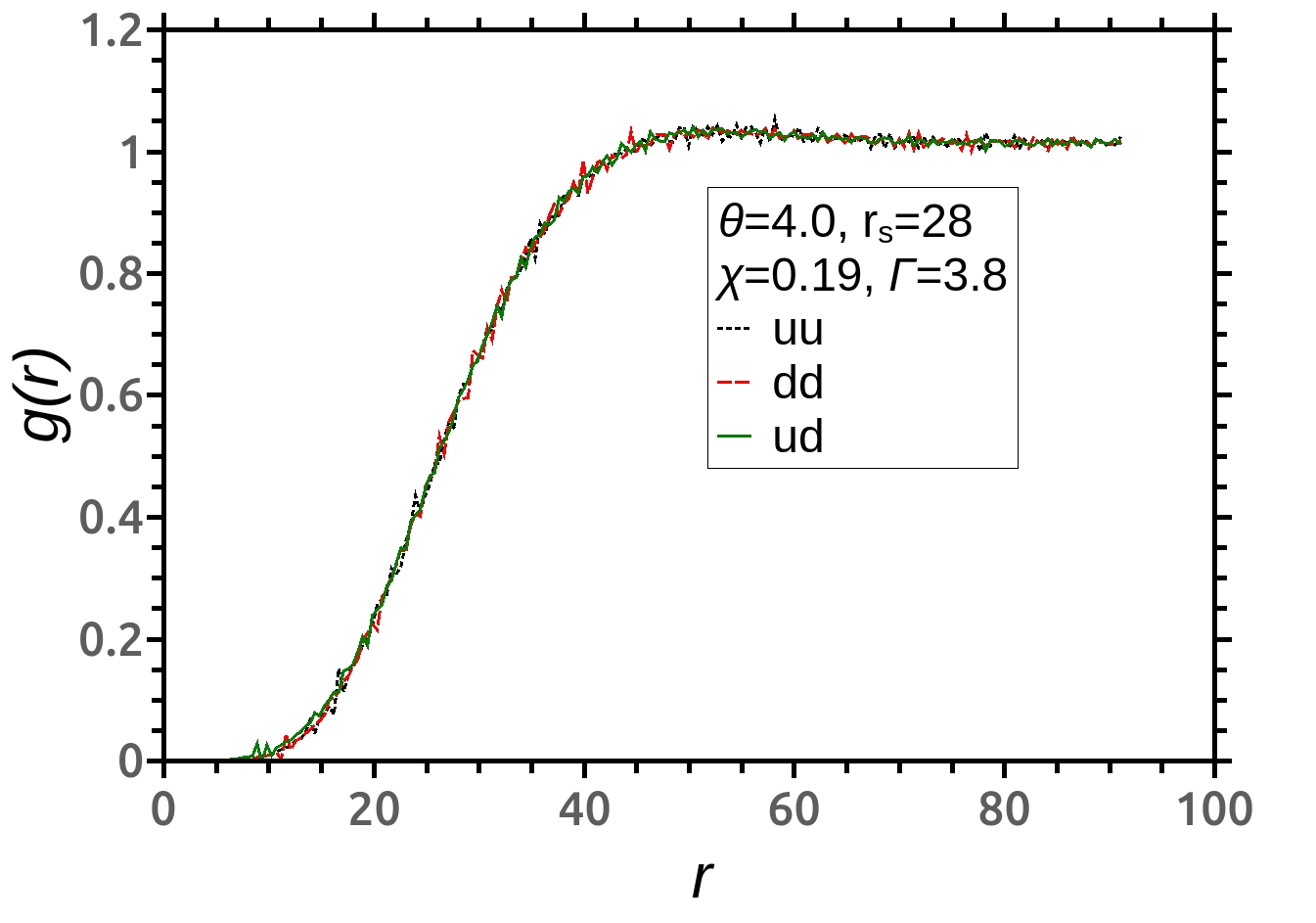}
	\caption{Left plots: single--particle momentum distribution functions of the unpolarized UEG at $r_s=28$ and $\theta=0.5,1,4$ calculated with the SMPIMC method.
	The results are compared with the SMPIMC results for the IFG in the same conditions and with the analytical Fermi distributions.
	Right plots: the pair distribution functions $g_{uu}$, $g_{dd}$ and $g_{ud}$ of the UEG. }
	\label{fig_r280}
\end{figure}

The results for $r_s=28.0$ are shown in Fig.~\ref{fig_r280}.
In case of $\theta=1$ the UEG is strongly non-ideal ($\Gamma \approx 15$), and the MDF differs from the Fermi distribution very distinctly.
All PDFs are almost the same and have maxima and minima.
In case of $\theta \approx 2$ the coupling strength $\Gamma \approx 7.6$ and non-ideality is quite strong.
The MDF begins to deviate from the Fermi distribution, while
the PDFs $g_{uu}$, $g_{dd}$ and $g_{ud}$ are exactly the same and have distinct maxima.
In case of $\theta \approx 4.0$ the coupling strength $\Gamma \approx 3.8$, and the MDF slightly differs from the Fermi distribution.
The maxima of the PDFs is just forming.

\begin{figure}[ht]		
    	\includegraphics[width=0.49\linewidth]{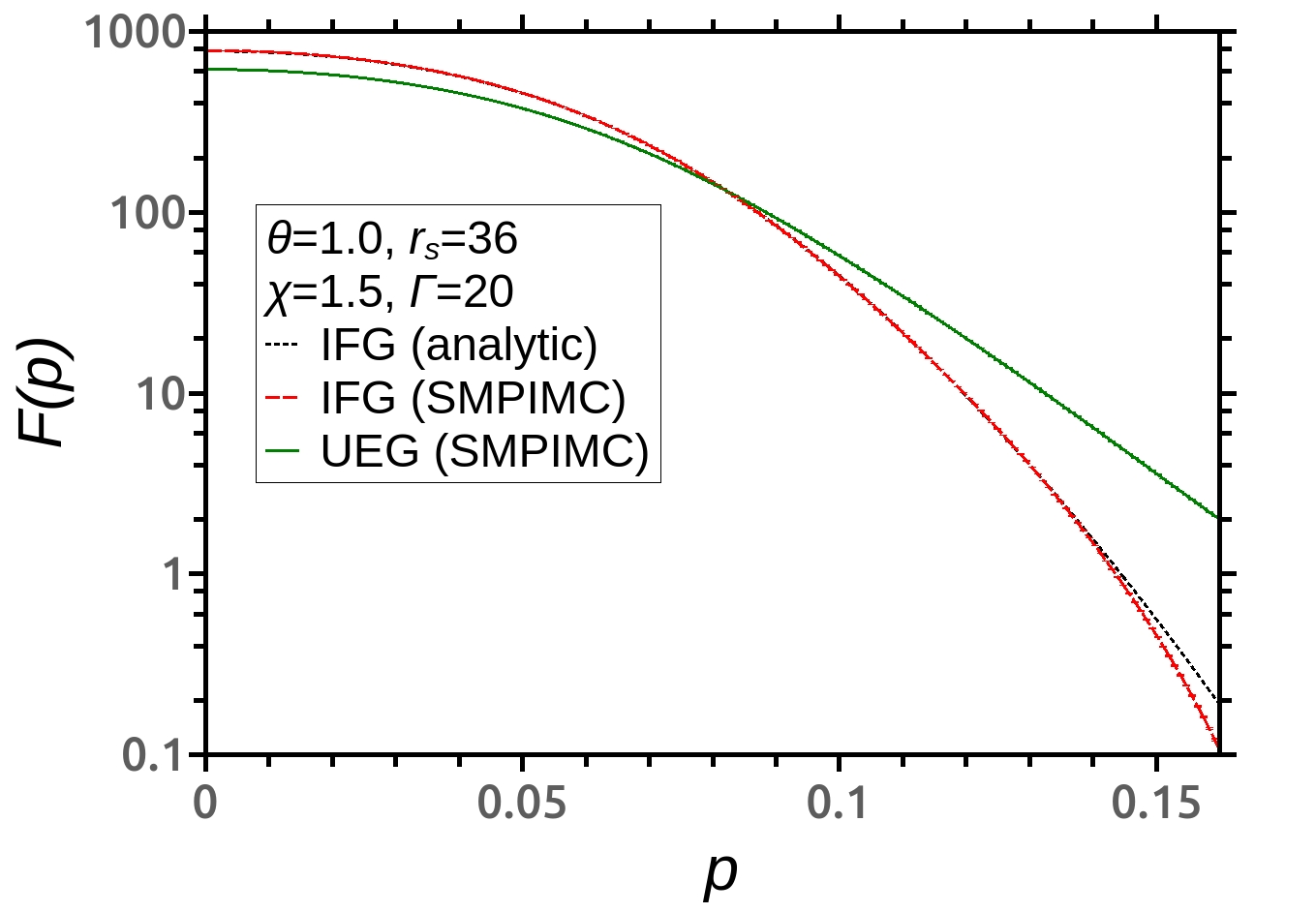}
    	\includegraphics[width=0.49\linewidth]{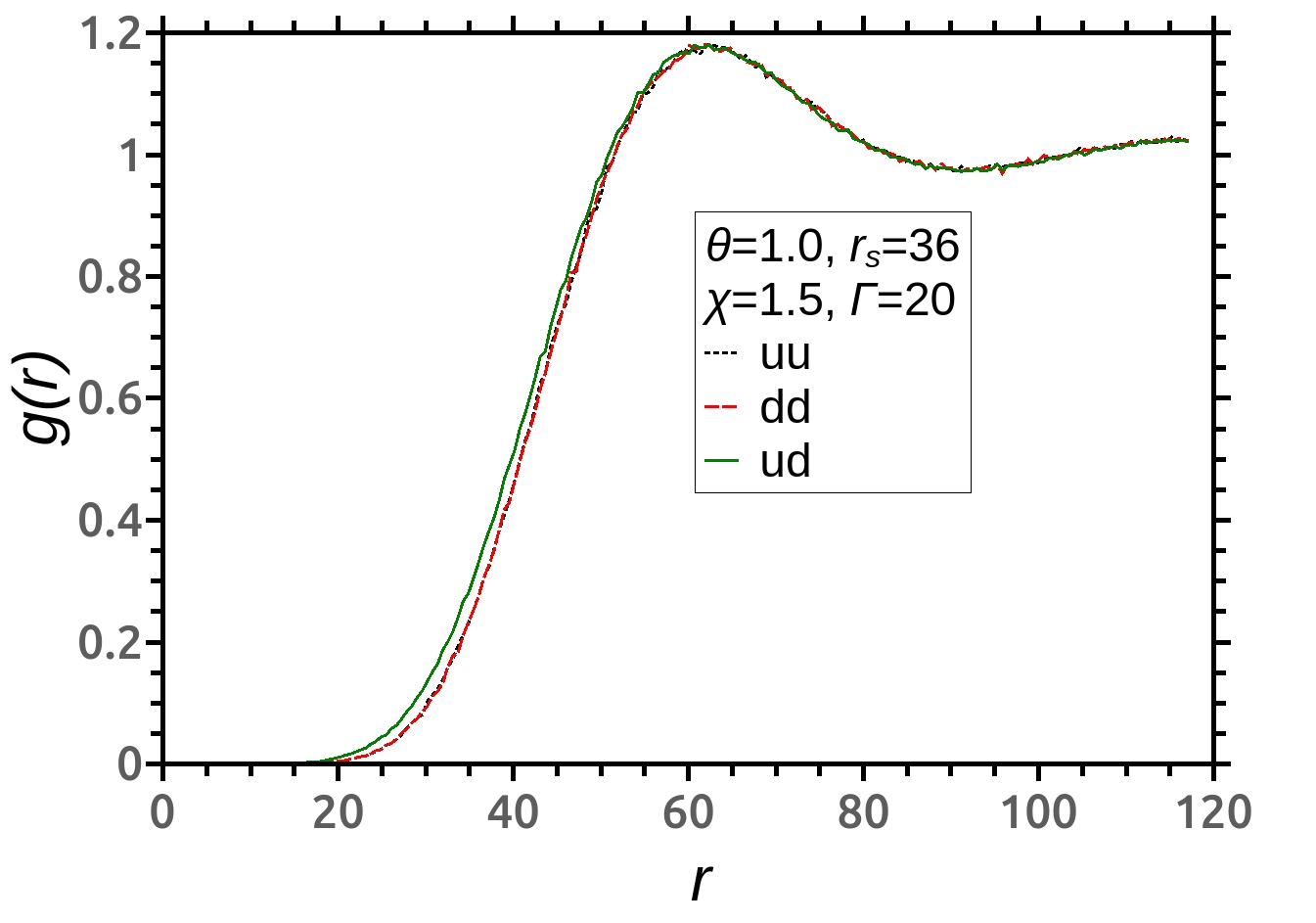}
    	\includegraphics[width=0.49\linewidth]{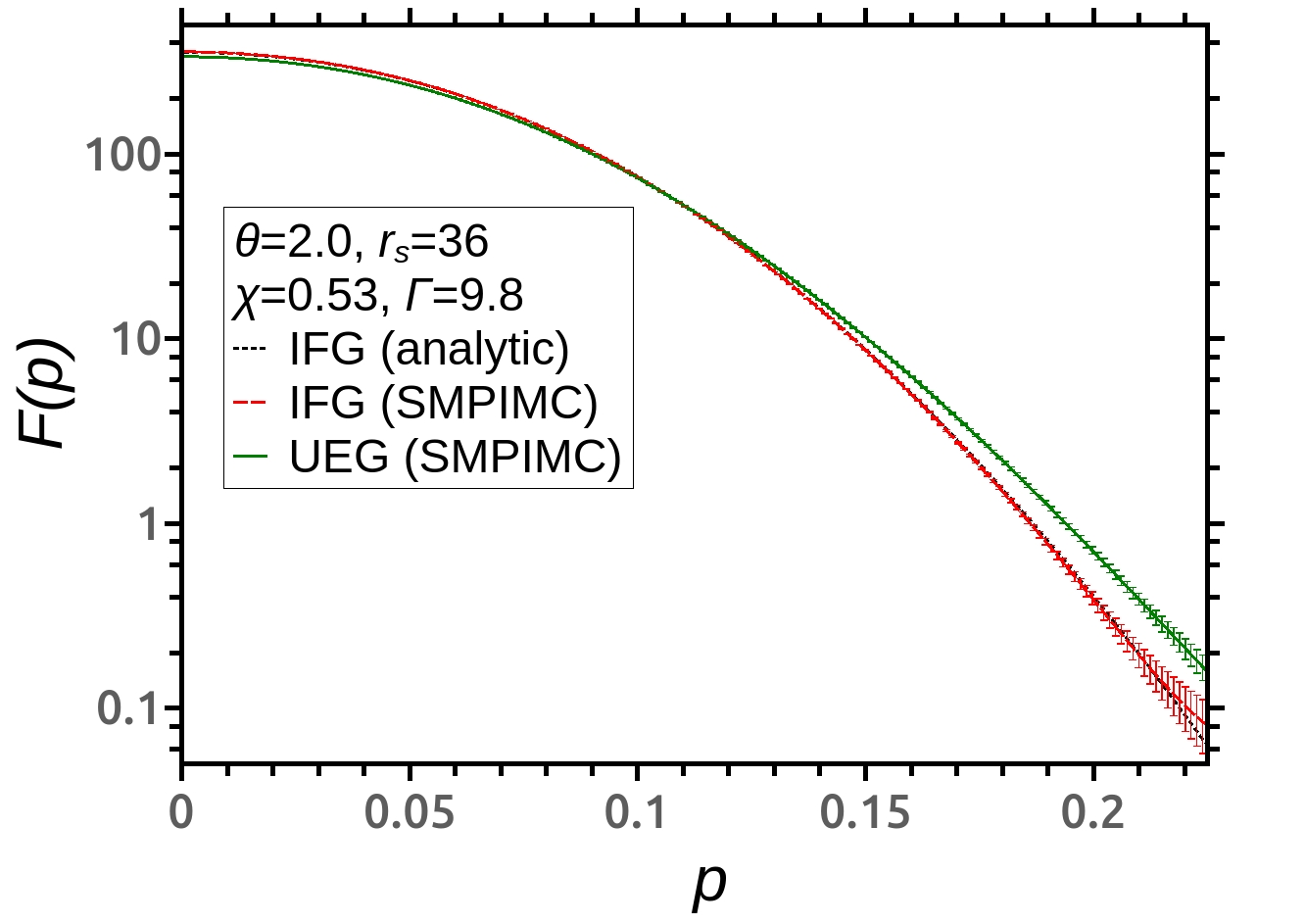}
    	\includegraphics[width=0.49\linewidth]{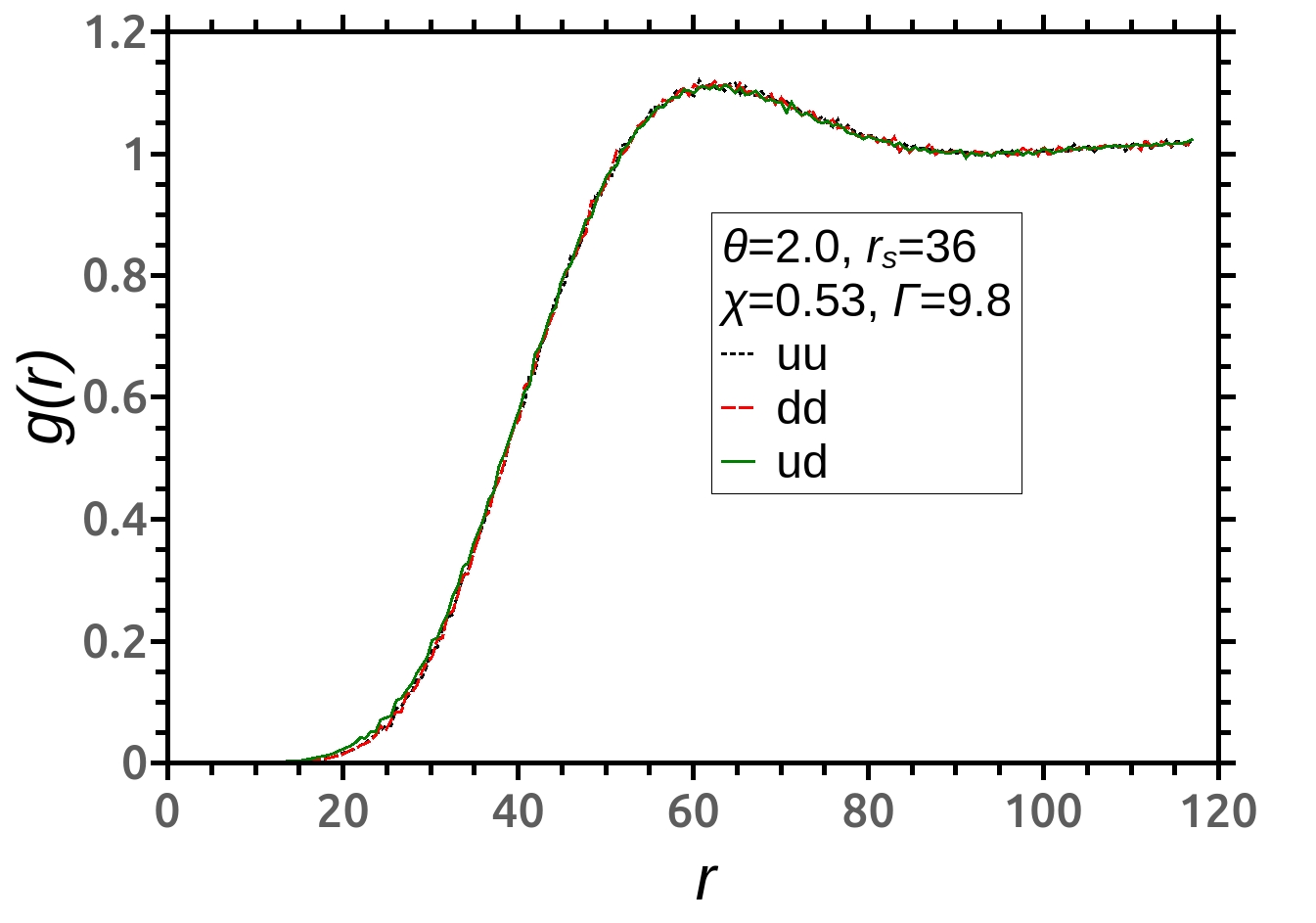}
    	\includegraphics[width=0.49\linewidth]{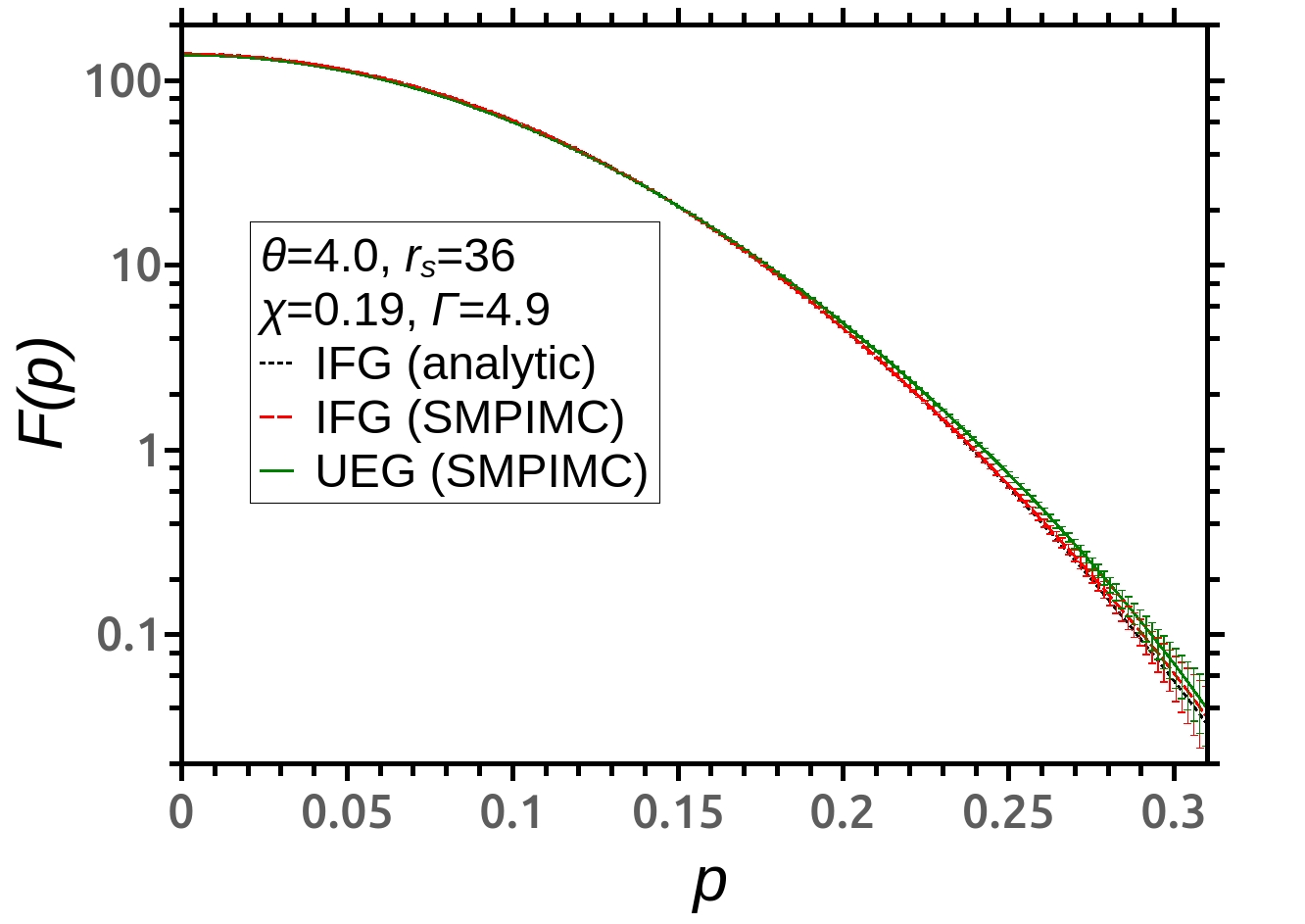}
    	\includegraphics[width=0.49\linewidth]{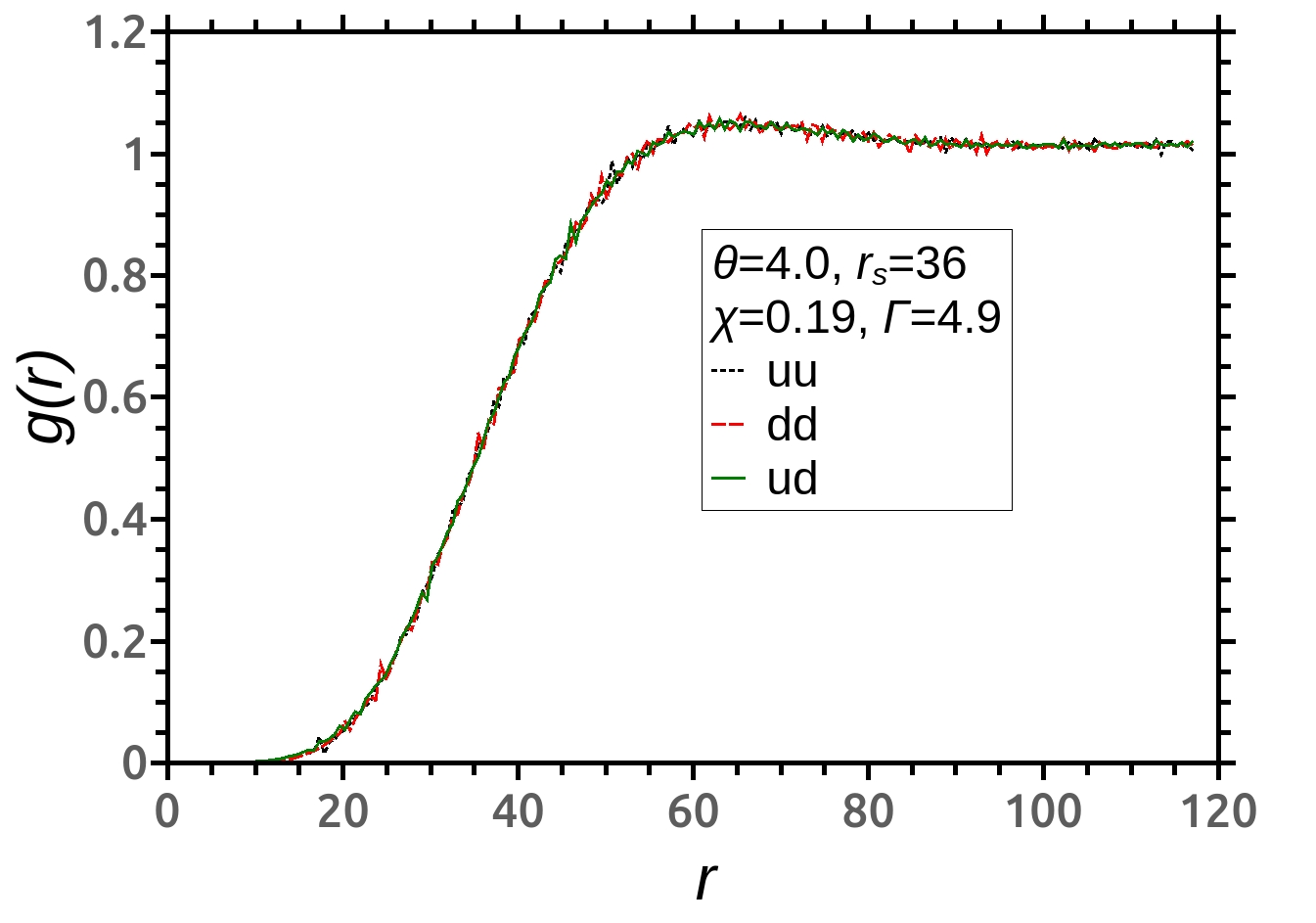}
	\caption{Left plots: single--particle momentum distribution functions of the unpolarized UEG at $r_s=36$ and $\theta=0.5,1,4$ calculated with the SMPIMC method.
	The results are compared with the SMPIMC results for the IFG in the same conditions and with the analytical Fermi distributions.
	Right plots: the pair distribution functions $g_{uu}$, $g_{dd}$ and $g_{ud}$ of the UEG. }
	\label{fig_r360}
\end{figure}

The results for $r_s=36.0$ are shown in Fig.~\ref{fig_r360}.
In case of $\theta=0.5$ the coupling strength is very strong ($\Gamma \approx 20$). 
The difference between the MDF and the Fermi distribution becomes more significant.
The first maxima and minima of the PDFs increases, also the second maxima starts to appear at the $r \approx 120 a_0$, however the simulation box with number of electrons has to be increased for the further resolution.
In case of $\theta \approx 2$ ($\Gamma \approx 9.8$) the MDF have quite distinct ``tail".
The PDFs $g_{uu}$, $g_{dd}$ and $g_{ud}$ are completely the same and have maxima and very slight minima.
In case of $\theta \approx 4.0$ the coupling strength $\Gamma \approx 4.9$, the MDF slightly differs from the Fermi distribution and there are maxima at the PDFs.

\begin{table}
\caption{ Some parameters and data describing the MDFs and PDFs from the Figs.~\ref{fig_r02}--\ref{fig_r360}.
The values are shown via two significant digits.}
\begin{tabular}{|r|r|r|r|r|r|r|r|r|}
\hline
	$r_s$ & $\theta$ & $\Gamma$ & $\chi$ & $\lambda, a_0$ & $L, a_0$ & $p_{FS}, \hbar/a_0$ & $r_{max}, a_0$ & $r_{min}, a_0$  \\
\hline
	$0.2$ & $0.5$ & $0.22$ & $4.3$ & $0.52$ & $1.3$ & $17$ & --- & --- \\
\hline
	$0.2$ & $1.0$ & $0.11$ & $1.5$ & $0.37$ & $1.3$ & $23$ & --- & --- \\
\hline
	$0.2$ & $4.0$ & $0.027$ & $0.19$ & $0.18$ & $1.3$ & $48$ & --- & --- \\
\hline
	$1.0$ & $0.5$ & $1.1$ & $4.3$ & $2.6$ & $6.5$ & $3.4$ & --- & --- \\
\hline
	$1.0$ & $1.0$ & $0.54$ & $1.5$ & $1.9$ & $6.5$ & $5.0$ & --- & --- \\
\hline
	$1.0$ & $4.0$ & $0.14$ & $0.19$ & $0.92$ & $6.5$ & $10$ & --- & --- \\
\hline
	$4.0$ & $0.5$ & $4.3$ & $4.3$ & $10$ & $26$ & $0.85$ & --- & --- \\
\hline
	$4.0$ & $1.0$ & $2.2$ & $1.5$ & $7.4$ & $26$ & $1.2$ & --- & --- \\
\hline
	$4.0$ & $4.0$ & $0.54$ & $0.19$ & $3.7$ & $26$ & $>1.9$ & --- & --- \\
\hline
	$12$ & $0.5$ & $13$ & $4.3$ & $31$ & $78$ & $0.28$ & $20$--$22$ & $30$--$32$ \\
\hline
	$12$ & $1.0$ & $6.5$ & $1.5$ & $22$ & $78$ & $0.42$ & $20$--$22$ & --- \\
\hline
	$12$ & $4.0$ & $1.6$ & $0.19$ & $11$ & $78$ & $0.65$ & --- & --- \\
\hline
	$16$ & $0.5$ & $17$ & $4.3$ & $42$ & $100$ & $0.21$ & $26$--$28$ & $40$--$42$ \\
\hline
	$16$ & $1.0$ & $8.7$ & $1.5$ & $30$ & $100$ & $0.30$ & $26$--$28$ & --- \\
\hline
	$16$ & $4.0$ & $2.2$ & $0.19$ & $15$ & $100$ & $0.60$ & --- & --- \\
\hline
	$28$ & $1.0$ & $15$ & $1.5$ & $52$ & $180$ & $0.17$ & $45$--$50$ & $70$ \\
\hline
	$28$ & $2.0$ & $7.6$ & $0.53$ & $37$ & $180$ & $0.26$ & $45$--$50$ & $70$--$75$ \\
\hline
	$28$ & $4.0$ & $3.8$ & $0.19$ & $26$ & $180$ & $0.38$ & $50$ & --- \\
\hline
	$36$ & $1.0$ & $20$ & $1.5$ & $66$ & $230$ & $0.14$ & $60$--$65$ & $90$ \\
\hline
	$36$ & $2.0$ & $9.8$ & $0.53$ & $47$ & $230$ & $0.21$ & $60$--$65$ & $90$--$95$ \\
\hline
	$36$ & $4.0$ & $4.9$ & $0.19$ & $33$ & $230$ & $0.28$ & $60$--$65$ & --- \\	
\hline
\end{tabular}
\label{tab_1}
\end{table}

In addition we calculated the average kinetic energy $E_{kin}$ and the potential energy $E_{pot}$ of the UEG which correspond to the Figs.~\ref{fig_r02}-\ref{fig_r360}.
These results are presented in Table~\ref{tab_2} and compared with the internal (kinetic) energy of the IFG, calculated with the SMPIMC method ($E_{IFG}$) and obtained from the analytical Fermi distribution directly ($E_{IFG0}$).
While $E_{IFG}$ coincides with $E_{IFG0}$ within the ranges of the statistical error, $E_{kin}$ start to exceed it in agreement with the behavior of the MDFs described above.
For example, $E_{kin}$ becomes almost $30\%$ higher than $E_{IFG}$ at $r_s=36$, $\theta=0.5$.
Also the dependence of $E_{pot}$ on $\theta$ at fixed $r_s$ is more sharp at low values of $r_s$ and becomes weaker at the high values.
\begin{table}
\caption{ $E_{IFG0}$ --- internal (kinetic) energy of the macroscopic IFG obtained from the analytical Fermi distribution;  $E_{IFG}$  --- internal (kinetic) energy of the finite--size IFG obtained from the SMPIMC; $E_{kin}$ and $E_{pot}$  --- kinetic and potential energy of the UEG. Two significant digits of the statistical error $3\sigma$ are written in the brackets. All energies are given per one electron. }
\begin{tabular}{|r|r|r|r|r|r|}
\hline
	$r_s$ & $\theta$ & $E_{IFG0}$, Ha & $E_{IFG}$, Ha & $E_{kin}$, Ha & $E_{pot}$, Ha \\
\hline
	$0.2$ & $0.5$ & $47.04$ & $47.29(95)$ & $47.23(90)$ & $-3.3896(34)$  \\
\hline
	$0.2$ & $1.0$ & $78.1$ & $78.5(22)$ & $78.64(20)$ & $-3.0596(45)$  \\
\hline
	$0.2$ & $4.0$ & $280.8$ & $282.2(82)$ & $280.6(75)$ & $-2.5034(68)$  \\	
\hline
	$1.0$ & $0.5$ & $1.881$ & $1.889(35)$ & $1.913(40)$ & $-0.78157(49)$  \\
\hline
	$1.0$ & $1.0$ & $3.125$ & $3.140(81)$ & $3.143(82)$ & $-0.72225(60)$  \\
\hline
	$1.0$ & $4.0$ & $11.23$ & $11.21(32)$ & $11.23(31)$ & $-0.57291(98)$  \\	
\hline
	$4.0$ & $0.5$ & $0.1176$ & $0.1177(18)$ & $0.1206(23)$ & $-0.22386(12)$  \\
\hline
	$4.0$ & $1.0$ & $0.1953$ & $0.1966(48)$ & $0.1953(47)$ & $-0.21480(10)$  \\
\hline
	$4.0$ & $4.0$ & $0.702$ & $0.702(18)$ & $0.6989(18)$ & $-0.17822(14)$  \\
\hline
	$12$ & $0.5$ & $ 0.01307$ & $0.01309(13)$ & $0.01412(25)$ & $-0.081073(40)$  \\
\hline
	$12$ & $1.0$ & $0.02170$ & $0.02161(32)$ & $0.02239(42)$ & $-0.079157(35)$  \\
\hline
	$16$ & $4.0$ & $0.07801$ & $0.07930(18)$ & $0.07892(18)$ & $-0.070567(24)$  \\
\hline
	$16$ & $0.5$ & $0.007349$ & $0.007346(54)$ & $0.00841(12)$ & $-0.061467(25)$  \\
\hline
	$16$ & $1.0$ & $0.01221$ & $0.01230(18)$ & $0.01264(21)$ & $-0.060535(18)$  \\
\hline
	$16$ & $4.0$ & $0.04388$ & $0.0438(10)$ & $0.04457(92)$ & $-0.054942(17)$  \\
\hline
	$28$ & $0.5$ & $0.003986$ & $0.004019(24)$ & $0.004640(34)$ & $-0.035732(12)$  \\
\hline
	$28$ & $1.0$ & $0.007377$ & $0.007346(82)$ & $0.007548(93)$ & $-0.034815(10)$  \\
\hline
	$28$ & $4.0$ & $0.01433$ & $0.01428(21)$ & $0.01489(22)$ & $-0.033396(10)$  \\
\hline
	$36$ & $0.5$ & $0.0024110$ & $0.0023974(04)$ & $0.0031391(07)$ & $-0.028246(08)$  \\
\hline
	$36$ & $1.0$ & $0.004462$ & $0.004463(36)$ & $0.004801(40)$ & $-0.027628(10)$  \\
\hline
	$36$ & $4.0$ & $0.00867$ & $0.00869(11)$ & $0.00894(12)$ & $-0.026520(05)$  \\	
\hline
\end{tabular}
\label{tab_2}
\end{table}

\begin{figure}[ht]		
    	\includegraphics[width=1.0\linewidth]{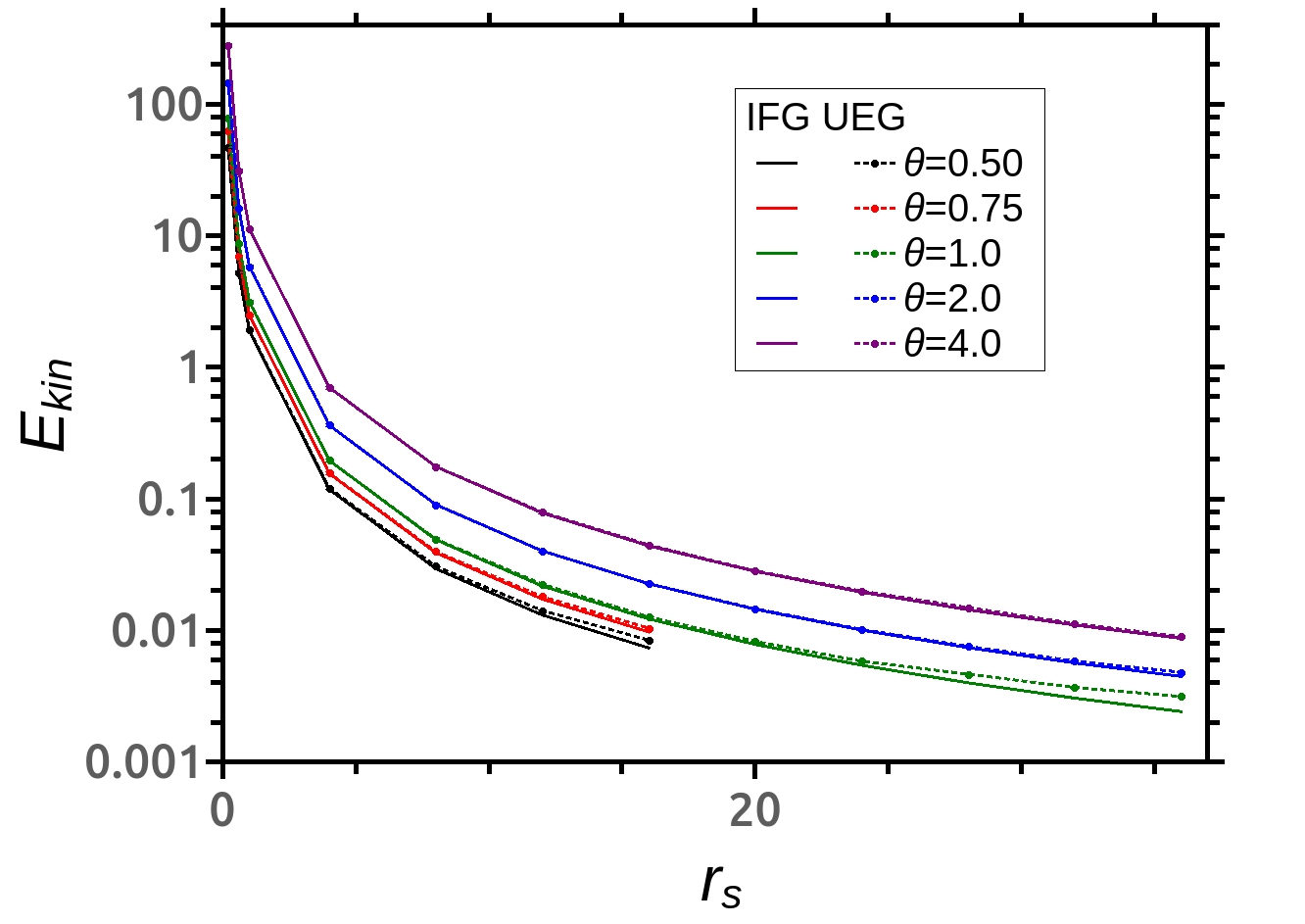}
	\caption{ Average kinetic energy of the UEG at different values of $r_s$ and $\theta$ compared with the internal (kinetic) energy of the IFG. }
	\label{fig_ekin}
\end{figure}

The dependence of the kinetic energies of the UEG and the IFG on $r_s$ at different values of $\theta$ is also shown in Fig.~\ref{fig_ekin} with additional intermediate points.

\section{Discussion}

First of all, let us discuss the phenomena of the ``quantum tails" of the MDFs and make some suggestions about it's physical nature.
As it follows from Figs.~\ref{fig_r02}-\ref{fig_r360}, the single--particle MDF of the UEG start to differ from the Fermi distribution when the system becomes strongly non-ideal.
The value of the coupling parameter $\Gamma$ at which the ``tail" appears depends on the degeneracy rate: in case of $\theta = 0.5$ (significant degeneracy) they appears at $\Gamma$ from $8$ to $9$ ($r_s \approx 8$), but in case of $\theta = 4.0$ the $\Gamma \approx 5$ is already enough ($r_s \approx 36$).

In the same time, the deviation of the MDFs from the Fermi distribution is always followed by the appearance of the visible maxima on the PDFs, and with increasing of the maxima the deviation also grows.
This observation points at the possible physical reason of the ``quantum tails": with increasing non-ideality the UEG becomes more liquid--like and the short--range order appears.
This results in the potential wells surrounding the electrons at the average distance $r_{max}$, so the spatial localization of the electrons increases.
According to the basic principles of quantum mechanics, the localization of the electron in the momentum space weakens and the MDFs become wider.
When the coupling increases, the potential wells becomes deeper and the effect grows.

Also this hypothesis can explain the fact that the ``quantum tail" of the MDF appears at less value of $\Gamma$ when the degeneracy is low and at higher values in the opposite case.
In fact, at low $\theta$ the thermal wavelength $\lambda$ becomes of order of the interparticle distances, so the electron can escape from the potential well more easily via quantum tunneling and becomes less localized.
Besides that, the definition of the coupling strength $\Gamma$ uses the classical kinetic energy $kT$ and is gives overestimated value of non-ideality in the degenerate case.

Secondly, the difference between PDFs $g_{uu}$, $g_{dd}$ and $g_{ud}$ becomes lower with increasing $r_s$ even at low values of $\theta$ (degenerate case).
Because the exchange repulsion is applied only to the electrons with the same spin projections, while the Coulomb interaction is universal,  we can conclude that the PDFs of the degenerate UEG are basically conditioned by the last one.
This reasoning is approved by the fact that the value of the $r_{max}$ is almost does not depend on $\theta$ when $r_s$ is fixed.
However the value of the maxima depends strongly on $\theta$, following the coupling strength $\Gamma$.

\section{Conclusion}

In this paper we have continued our research of the UEG and study the MDFs and the PDFs using the SMPIMC method in a wide range of the parameters $0.5 \le \theta \le 4$, $0.2 \le r_s \le 36$.
Thus, the different thermodynamic states from almost ideal gas (Boltzmann of Fermi) to strongly non-ideal system with weak and significant degeneracy has been considered.

Firstly, we have discovered that the single--particle MDFs start to exceed the Fermi distribution at high momentum values, when the UEG is strongly non--ideal.
In the degenerate case $\theta = 0.5$ this occurs at $r_s \gtrsim 8$ ($\Gamma \gtrsim 9$), in the semi--degenerate case $\theta = 1$ --- at $r_s \gtrsim 16$ ($\Gamma \gtrsim 8$), in the almost classical case $\theta = 4$ --- at $r_s = 36$ ($\Gamma \gtrsim 5$).
The deviation has form of the ``quantum tail'' at the high momenta and is reliable for the high momenta $p \lesssim p_{FS}$, where the value of $p_{FS}$ is conditioned by the finite--size effects. 

We explain the phenomena of the ``quantum tails'' as follows.
The deviation of the MDFs from the Fermi distribution is always followed by the appearance of the short--range order on the PDFs,
so with increasing coupling strength the UEG becomes more liquid--like. 
and the potential wells surrounding the electrons appear.
As a result, the spatial localization of the electrons increases, so the localization in the momentum space becomes weaker and the MDFs become wider.
Besides that, at high degeneracy the electrons are tunneling from the potential wells more easily, so the ``quantum tail'' appears at higher values of $\Gamma$ in comparison with the non-degenerate case.

Secondly, we have obtained that the difference between the PDFs of the electrons with the same and the opposite spine projections becomes negligible at high values of $\Gamma$ in the degenerate case.
Thus, the spatial structure of the strongly non-ideal UEG is basically conditioned by the Coulomb interaction rather than the exchange repulsion.

Finally, we have calculated the average kinetic and potential energies for $r_s$ up to $36$, expanding our results from \cite{LarkinPoP2021} significantly.

\section{\label{sec:ack}Acknowledgments}
We acknowledge stimulating discussions with Prof. M. Bonitz, T. Schoof, S. Groth and T. Dornheim (Kiel). 
This research was supported by the by the Russian Science Foundation, Grant No.~20-42-04421.


\begin{thebibliography}{10}

\bibitem{Knudson2012}
M.~D. Knudson, M.~P. Desjarlais, R.~W. Lemke, T.~R. Mattsson, M.~French,
  N.~Nettelmann, and R.~Redmer, ``Probing the interiors of the ice giants:
  Shock compression of water to 700 gpa and 3.8 $\rm g/cm^3$,'' {\em Phys. Rev.
  Lett.}, vol.~108, p.~091102, Feb 2012.

\bibitem{Nettelmann2012}
N.~Nettelmann, A.~Becker, B.~Holst, and R.~Redmer, ``{JUPITER} {MODELS} {WITH}
  {IMPROVED} {AB} {INITIO} {HYDROGEN} {EQUATION} {OF} {STATE} (h-{REOS}.2),''
  {\em The Astrophysical Journal}, vol.~750, p.~52, apr 2012.

\bibitem{Mazevet2019}
{Mazevet, S.}, {Licari, A.}, {Chabrier, G.}, and {Potekhin, A. Y.}, ``Ab initio
  based equation of state of dense water for planetary and exoplanetary
  modeling,'' {\em A\&A}, vol.~621, p.~A128, 2019.

\bibitem{Hubbard1997}
W.~B. Hubbard, T.~Guillot, J.~I. Lunine, A.~Burrows, D.~Saumon, M.~S. Marley,
  and R.~S. Freedman, ``Liquid metallic hydrogen and the structure of brown
  dwarfs and giant planets,'' {\em Physics of Plasmas}, vol.~4, no.~5,
  pp.~2011--2015, 1997.

\bibitem{Chabrier2000}
G.~Chabrier, P.~Brassard, G.~Fontaine, and D.~Saumon, ``Cooling sequences and
  color-magnitude diagrams for cool white dwarfs with hydrogen atmospheres,''
  {\em The Astrophysical Journal}, vol.~543, pp.~216--226, nov 2000.

\bibitem{Haensel2007}
P.~Haensel, A.~Y. Potekhin, and D.~G. Yakovlev, {\em Neutron Stars 1 : Equation
  of State and Structure}.
\newblock Astrophysics and Space Science Library, Springer, 2007.

\bibitem{Sharma2015}
{Sharma, B. K.}, {Centelles, M.}, {Vi\~nas, X.}, {Baldo, M.}, and {Burgio, G.
  F.}, ``Unified equation of state for neutron stars on a microscopic basis,''
  {\em A\&A}, vol.~584, p.~A103, 2015.

\bibitem{InertialFusion2014}
P.~F. Schmit, P.~F. Knapp, S.~B. Hansen, M.~R. Gomez, K.~D. Hahn, D.~B. Sinars,
  K.~J. Peterson, S.~A. Slutz, A.~B. Sefkow, T.~J. Awe, E.~Harding, C.~A.
  Jennings, G.~A. Chandler, G.~W. Cooper, M.~E. Cuneo, M.~Geissel, A.~J.
  Harvey-Thompson, M.~C. Herrmann, M.~H. Hess, O.~Johns, D.~C. Lamppa, M.~R.
  Martin, R.~D. McBride, J.~L. Porter, G.~K. Robertson, G.~A. Rochau, D.~C.
  Rovang, C.~L. Ruiz, M.~E. Savage, I.~C. Smith, W.~A. Stygar, and R.~A. Vesey,
  ``Understanding fuel magnetization and mix using secondary nuclear reactions
  in magneto-inertial fusion,'' {\em Phys. Rev. Lett.}, vol.~113, p.~155004,
  Oct 2014.

\bibitem{InertialFusion2015}
R.~Nora, W.~Theobald, R.~Betti, F.~J. Marshall, D.~T. Michel, W.~Seka,
  B.~Yaakobi, M.~Lafon, C.~Stoeckl, J.~Delettrez, A.~A. Solodov, A.~Casner,
  C.~Reverdin, X.~Ribeyre, A.~Vallet, J.~Peebles, F.~N. Beg, and M.~S. Wei,
  ``Gigabar spherical shock generation on the omega laser,'' {\em Phys. Rev.
  Lett.}, vol.~114, p.~045001, Jan 2015.

\bibitem{Roozehdar2019}
R.~Roozehdar~Mogaddam, N.~Sepehri~Javan, K.~Javidan, and H.~Mohammadzadeh,
  ``Modulation instability and soliton formation in the interaction of x-ray
  laser beam with relativistic quantum plasma,'' {\em Physics of Plasmas},
  vol.~26, no.~6, p.~062112, 2019.

\bibitem{Edwards2019}
M.~R. Edwards, Y.~Shi, J.~M. Mikhailova, and N.~J. Fisch, ``Laser amplification
  in strongly magnetized plasma,'' {\em Physical review letters}, vol.~123,
  no.~2, p.~025001, 2019.

\bibitem{EbelingBook2017}
W.~Ebeling, V.~Fortov, and V.~Filinov, {\em Quantum Statistics of Dense Gases
  and Nonideal Plasmas}.
\newblock Berlin: Springer, 2017.

\bibitem{Savchenko2001}
V.~I. Savchenko, ``Quantum, multibody effects and nuclear reaction rates in
  plasmas,'' {\em Physics of Plasmas}, vol.~8, no.~1, pp.~82--91, 2001.

\bibitem{Salpeter1969}
E.~E. {Salpeter} and H.~M. {van Horn}, ``Nuclear reaction rates at high
  densities,'' {\em Astrophysical Journal}, vol.~155, p.~183, Jan. 1969.

\bibitem{Ichimaru1993}
S.~Ichimaru, ``Nuclear fusion in dense plasmas,'' {\em Rev. Mod. Phys.},
  vol.~65, pp.~255--299, Apr 1993.

\bibitem{Dewitt1999}
H.~Dewitt and W.~Slattery, ``Screening enhancement of thermonuclear reactions
  in high density stars,'' {\em Contributions to Plasma Physics}, vol.~39,
  no.~1-2, pp.~97--100, 1999.

\bibitem{STAROSTIN2002287}
A.~Starostin, A.~Mironov, N.~Aleksandrov, N.~Fisch, and R.~Kulsrud, ``Quantum
  corrections to the distribution function of particles over momentum in dense
  media,'' {\em Physica A: Statistical Mechanics and its Applications},
  vol.~305, no.~1, pp.~287--296, 2002.
\newblock Non Extensive Thermodynamics and Physical applications.

\bibitem{STAROSTIN1}
A.~Starostin, A.~Leonov, and Y.~Petrushevich, ``Quantum corrections to the
  particle distribution function and reaction rates in dense media,'' {\em
  Plasma Phys. Rep.}, vol.~31, pp.~123--132, 2005.

\bibitem{STAROSTIN2}
A.~Starostin, V.~Gryaznov, and Y.~Petrushevich, ``Development of the theory of
  momentum distribution of particles with regard to quantum phenomena,'' {\em
  J. Exp. Theor. Phys.}, vol.~125, pp.~940--947, 2017.

\bibitem{LarkinCPP2016}
A.~S. Larkin, V.~S. Filinov, and V.~E. Fortov, ``Path integral representation
  of the wigner function in canonical ensemble,'' {\em Contributions to Plasma
  Physics}, vol.~56, no.~3-4, pp.~187--196, 2016.

\bibitem{LarkinJAMP2017}
A.~Larkin and V.~Filinov, ``Phase space path integral representation for wigner
  function,'' {\em Journal of Applied Mathematics and Physics}, vol.~5,
  pp.~392--411, 2017.

\bibitem{LarkinCPP2018}
A.~Larkin and V.~Filinov, ``Quantum tails in the momentum distribution
  functions of non-ideal fermi systems,'' {\em Contributions to Plasma
  Physics}, vol.~58, no.~2-3, pp.~107--113, 2018.

\bibitem{LarkinJoP2017}
A.~Larkin, V.~Filinov, and V.~Fortov, ``Peculiarities of the momentum
  distribution functions of strongly correlated charged fermions,'' {\em
  Journal of Physics A: Mathematical and Theoretical}, vol.~51, no.~3,
  p.~035002, 2017.

\bibitem{LarkinCPP2017}
A.~Larkin, V.~Filinov, and V.~Fortov, ``Pauli blocking by effective pair
  pseudopotential in degenerate fermi systems of particles,'' {\em
  Contributions to Plasma Physics}, vol.~57, no.~10, pp.~506--511, 2017.

\bibitem{LarkinTVT2019}
A.~S. Larkin and V.~S. Filinov, ``Monte carlo simulation of the thermodynamic
  properties of hydrogen plasma with the wigner function,'' {\em High
  Temperature}, vol.~57, pp.~651--659, 2019.

\bibitem{Loos2016}
P.-F. Loos and P.~M.~W. Gill, ``The uniform electron gas,'' {\em WIREs
  Computational Molecular Science}, vol.~6, no.~4, pp.~410--429, 2016.

\bibitem{Mahan2000}
G.~Mahan, {\em Many-Particle Physics}.
\newblock Physics of Solids and Liquids, Springer US, 2000.

\bibitem{Yasuhara1976}
H.~Yasuhara and Y.~Kawazoe, ``A note on the momentum distribution function for
  an electron gas,'' {\em Physica A: Statistical Mechanics and its
  Applications}, vol.~85, no.~2, pp.~416--424, 1976.

\bibitem{Kimball1975}
J.~C. Kimball, ``Short-range correlations and the structure factor and momentum
  distribution of electrons,'' {\em Journal of Physics A: Mathematical and
  General}, vol.~8, pp.~1513--1517, sep 1975.

\bibitem{Hunger2021}
K.~Hunger, T.~Schoof, T.~Dornheim, M.~Bonitz, and A.~Filinov, ``Momentum
  distribution function and short-range correlations of the warm dense electron
  gas: Ab initio quantum monte carlo results,'' {\em Phys. Rev. E}, vol.~103,
  p.~053204, May 2021.

\bibitem{LarkinPoP2021}
A.~S. Larkin, V.~S. Filinov, and P.~R. Levashov, ``Single-momentum path
  integral monte carlo simulations of uniform electron gas in warm dense matter
  regime,'' {\em Physics of Plasmas}, vol.~28, no.~12, p.~122712, 2021.

\bibitem{TOUKMAJI199673}
A.~Y. Toukmaji and J.~A. Board, ``Ewald summation techniques in perspective: a
  survey,'' {\em Computer Physics Communications}, vol.~95, no.~2, pp.~73--92,
  1996.

\bibitem{Dornheim2018}
T.~Dornheim, S.~Groth, and M.~Bonitz, ``The uniform electron gas at warm dense
  matter conditions,'' {\em Physics Reports}, vol.~744, pp.~1--86, 2018.

\bibitem{Tatarski1983}
V.~I. Tatarski{\u{\i}}, ``The wigner representation of quantum mechanics,''
  {\em Soviet Physics Uspekhi}, vol.~26, pp.~311--327, apr 1983.

\bibitem{FeynmanHibbs}
R.~Feynman and A.~Hibbs, {\em Quantum Mechanics and Path-Integral}.
\newblock Physics of Solids and Liquids, McGraw-Hill, New York, 1965.

\bibitem{Filinov2003}
V.~S. Filinov, M.~Bonitz, P.~Levashov, V.~E. Fortov, W.~Ebeling, M.~Schlanges,
  and S.~W. Koch, ``Plasma phase transition in dense hydrogen and
  electron{\textendash}hole plasmas,'' {\em Journal of Physics A: Mathematical
  and General}, vol.~36, pp.~6069--6076, may 2003.

\bibitem{Kleinert}
H.~Kleinert, {\em Path Integrals in Quantum Mechanics, Statistics, Polymer
  Physics, and Financial Markets}.
\newblock Path Integrals in Quantum Mechanics, Statistics, Polymer Physics, and
  Financial Markets, World Scientific, 2004.

\bibitem{Hastings1970}
W.~K. Hastings, ``{Monte Carlo sampling methods using Markov chains and their
  applications},'' {\em Biometrika}, vol.~57, pp.~97--109, 04 1970.

\bibitem{allen1988}
M.~Allen and D.~Tildesley, {\em Computer Simulation of Liquids}.
\newblock Clarendon Press, 1988.

\bibitem{Dornheim2016}
T.~Dornheim, S.~Groth, T.~Schoof, C.~Hann, and M.~Bonitz, ``Ab initio quantum
  monte carlo simulations of the uniform electron gas without fixed nodes: The
  unpolarized case,'' {\em Phys. Rev. B}, vol.~93, p.~205134, May 2016.

\end{thebibliography}
\end{document}